\documentclass[twocolumn]{aastex63} 
\usepackage{graphicx} 
\usepackage{natbib}
\usepackage{float}
\usepackage{hyperref}

\newcommand{\enst}{MgSiO$_{3}$}
\newcommand{\forst}{Mg$_{2}$SiO$_{4}$}
\newcommand{\fsed}{$f_{\rm sed}$}
\newcommand{\um}{$\mu{\rm m}$}

\graphicspath{{./}{}}

\received{April 12, 2021}
\revised{July 15, 2021}
\accepted{July 22, 2021}
\submitjournal{ApJ}

\shortauthors{Luna et al.}

\begin{document}

\title{Empirically Determining Substellar Cloud Compositions in the era of \emph{JWST}}

\correspondingauthor{Jessica Luna }
\email{jessicaluna@utexas.edu }

\author[0000-0003-2152-9248]{Jessica L. Luna }
\altaffiliation{NSF Graduate Research Fellow}
\affiliation{Department of Astronomy, The University of Texas at Austin \\
2515 SPEEDWAY, Stop C1402 \\
AUSTIN, TX 78712-1206 \label{utexas} }

\author[0000-0002-4404-0456]{Caroline V. Morley}
\affiliation{Department of Astronomy, The University of Texas at Austin \\
2515 SPEEDWAY, Stop C1402 \\
AUSTIN, TX 78712-1206 \label{utexas} }

\begin{abstract}

Most brown dwarfs have atmospheres with temperatures cold enough to form clouds. A variety of materials likely condense, including refractory metal oxides and silicates; the precise compositions and crystal structures of predicted cloud particles depend on the modeling framework used and have not yet been empirically constrained. \emph{Spitzer} has shown tentative evidence of the silicate feature in L dwarf spectra and \emph{JWST} can measure these features in many L dwarfs. Here, we present new models to predict the signatures of the strongest cloud absorption features. We investigate different cloud mineral species and determine how particle size, mineralogy, and crystalline structure change spectral features. We find that silicate and refractory clouds have a strong cloud absorption feature for small particle sizes ($\leq$ 1 \um). Model spectra are compared to five brown dwarfs that show evidence of the silicate feature; models that include small particles in the upper layers of the atmosphere produce a broad cloud mineral feature, and that better match the observed spectra than the \citet{AM01} cloud model. We simulate observations with the MIRI instrument on \emph{JWST} for a range of nearby, cloudy brown dwarfs, demonstrating that these features could be readily detectable if small particles are present. Furthermore, for photometrically variable brown dwarfs, our predictions suggest that with \emph{JWST}, by measuring spectroscopic variability inside and outside a mineral feature, we can establish silicate (or other) clouds as the cause of variability. Mid-infrared spectroscopy is a promising tool to empirically constrain the complex cloud condensation sequence in brown dwarf atmospheres.

\end{abstract}



\defcitealias{AM01}{AM01}

\section{Introduction \label{sec:intro}} 
Brown dwarfs are substellar objects that are too small in mass to fuse hydrogen. Their cool atmospheres are analogous in both temperature and composition to gas giant planets. Brown dwarfs cool as they age and their temperatures determine the species that can condense to form clouds in their atmospheres. For L dwarfs ( $>$ 1300 K), layers of thick silicate and refractory clouds likely form \citep{Tsuji96,Allard01,Marley02,burrows06, Cushing08}. As brown dwarfs cool, these clouds appear to clear significantly at the L/T transition ( $\approx$ 1200 K). For late-T dwarfs ( $<$ 1000 K), sulfide and salt clouds can condense \citep{Visscher06,Morley12}. For the coldest Y dwarfs, water ice and ammonia clouds can form \citep{Burrows03,Morley14a}. Clouds have a substantial impact on brown dwarf spectra, and, 25 years after the discoveries of the first known brown dwarfs, modeling these clouds accurately in detail remains one of the largest uncertainties in brown dwarf astrophysics \citep{Marley13, marley15}.

\subsection{Approaches to Cloud Modeling}
We will briefly review the 3 major classes of brown dwarf cloud models and highlight the cloud species predicted by each. These models can differ significantly in the clouds that form, their masses, number densities, size distributions, and the resulting impact on the spectra. 

One straight-forward approach, ``rainout equilibrium" \citep{Chabrier00,Allard01}, adopts equilibrium chemistry to predict condensation. This approach has the advantage of not requiring knowledge of microphysical processes that are challenging to model from first principles with many uncertain physical quantities. Rainout equilibrium calculations use a ``bottom-up" approach to the chemistry of cloud formation, assuming that the atmosphere is in thermochemical equilibrium but removing material from the gas phase as materials condense \citep{Lodders06, Marley13}. The cloud locations and particle sizes are determined by balancing the downward transport of particles by sedimentation with the upward mixing of vapor and condensate \citep[hereafter \citetalias{AM01}]{Saumon08,Stephens09,Morley12,AM01}. It is yet to be determined if every possible species does condense in the predicted sequence. Some species may have barriers to their formation (eg. slow nucleation timescales) that are not modeled in this framework, which assumes that all vapor in excess of the saturation vapor pressure condenses. 

A second approach to cloud modeling uses grain chemistry by treating cloud formation as a kinetic process using a ``top-down" approach \citep{Helling08a, Helling08b}. They start with TiO$_2$ seed particles at the top of the atmosphere and follow their growth as they fall downward, modeling the heterogeneous chemical reactions occurring on their surfaces. These seed particles accrete condensate material and grow, which results in ``dirty grains" with mixed compositions. The resulting clouds can be different in composition, location and, particle size distributions than rainout equilibrium. For example, \citet{Helling&Woitke06} predict that SiO$_2$ condenses, which is not predicted from equilibrium chemistry calculations for a solar composition atmosphere.

A microphysical framework developed for the solar system but recently used for exoplanets and brown dwarfs is CARMA. CARMA attempts to model microphysical processes from first principles using a bin scheme approach to fully resolve particle size distributions \citep{Turco1979, Toon1988,Ackerman1995,  Gao18a,Gao18,Powell18, Powell19}. It treats the microphysical processes of nucleation (heterogeneous and homogeneous), condensation, evaporation, and coagulation, and vertical transport by atmospheric mixing and gravitational settling.

Modeling microphysics is challenging due to the plethora of unknowns that are required as inputs. They require assumptions about properties that are unknown, including vertical mixing rates, material properties like surface tension, and the detailed knowledge of formation pathways. The resulting clouds are broadly similar to equilibrium condensation clouds but differ in particle sizes and locations \citep{Gao18a}.

\subsection{Directly Observing Cloud Properties}
With different models resulting in a range of cloud properties, and some models not requiring clouds at all, a key next step is to empirically test our approaches by measuring cloud properties directly for an ensemble of brown dwarfs. Some tentative evidence from \emph{Spitzer} IRS suggests that silicate features are present in L dwarf spectra but with smaller particle sizes than produced by \citetalias{AM01} \citep{Cushing06, Helling&Woitke06,Looper08}. Further evidence also suggests that some brown dwarfs have small (sub-micron) grain in their photospheres: \citet{Hiranaka16} showed that small sub-micron-sized grains above the main cloud deck can explain the NIR colors of red L dwarfs. They conclude that brown dwarf models should include both large and small (sub-micron) sized particles to reproduce the spectra of red L dwarfs. 

The MIRI instrument on the James Webb Space Telescope will allow us to measure cloud mineral spectral features in many L dwarfs, allowing us to investigate cloud compositions, particle sizes, and mineral structures, as well as the impact of these clouds on variability. These observations will provide the strongest direct evidence of cloud formation and allow us to determine which cloud modeling frameworks best capture the physics of brown dwarf atmospheres.  

In this paper, we investigate the effect clouds have on the observed thermal emission spectra of brown dwarf atmospheres. In particular, we focus on clouds present in L dwarfs, where the atmospheres are dominated by silicate and other refractory clouds. We present new theoretical models for brown dwarf atmospheres motivated by observational results, including small particles to produce cloud spectral features. We discuss the properties of our model in Section \ref{sec:methods}. In Section \ref{sec:results}, we show the results of our model thermal emission spectra and in Section \ref{sec:matchingobs} we fit models to \emph{Spitzer} IRS observations and simulate \emph{JWST} observations. Finally in Section \ref{sec:discussion} we discuss the implications of our new cloud models and in Section \ref{sec:conclusion} we conclude.

%METHODS

\section{Methods} \label{sec:methods}
We used a series of 1D models to investigate the effect that clouds can have on thermal emission spectra of unirradiated substellar objects. 

\begin{figure}
\centering
    \includegraphics[width=1\linewidth]{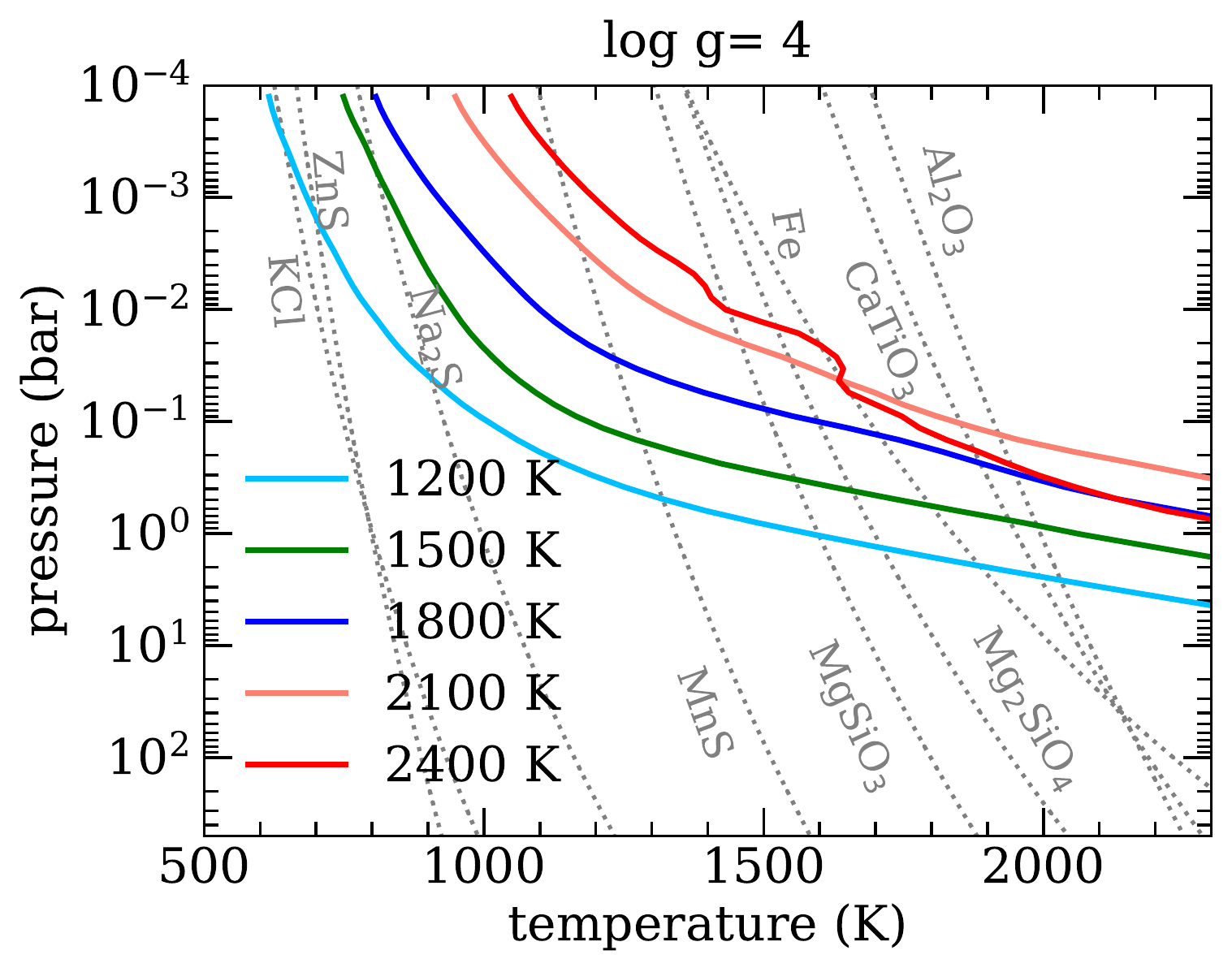}
    \includegraphics[width=1\linewidth]{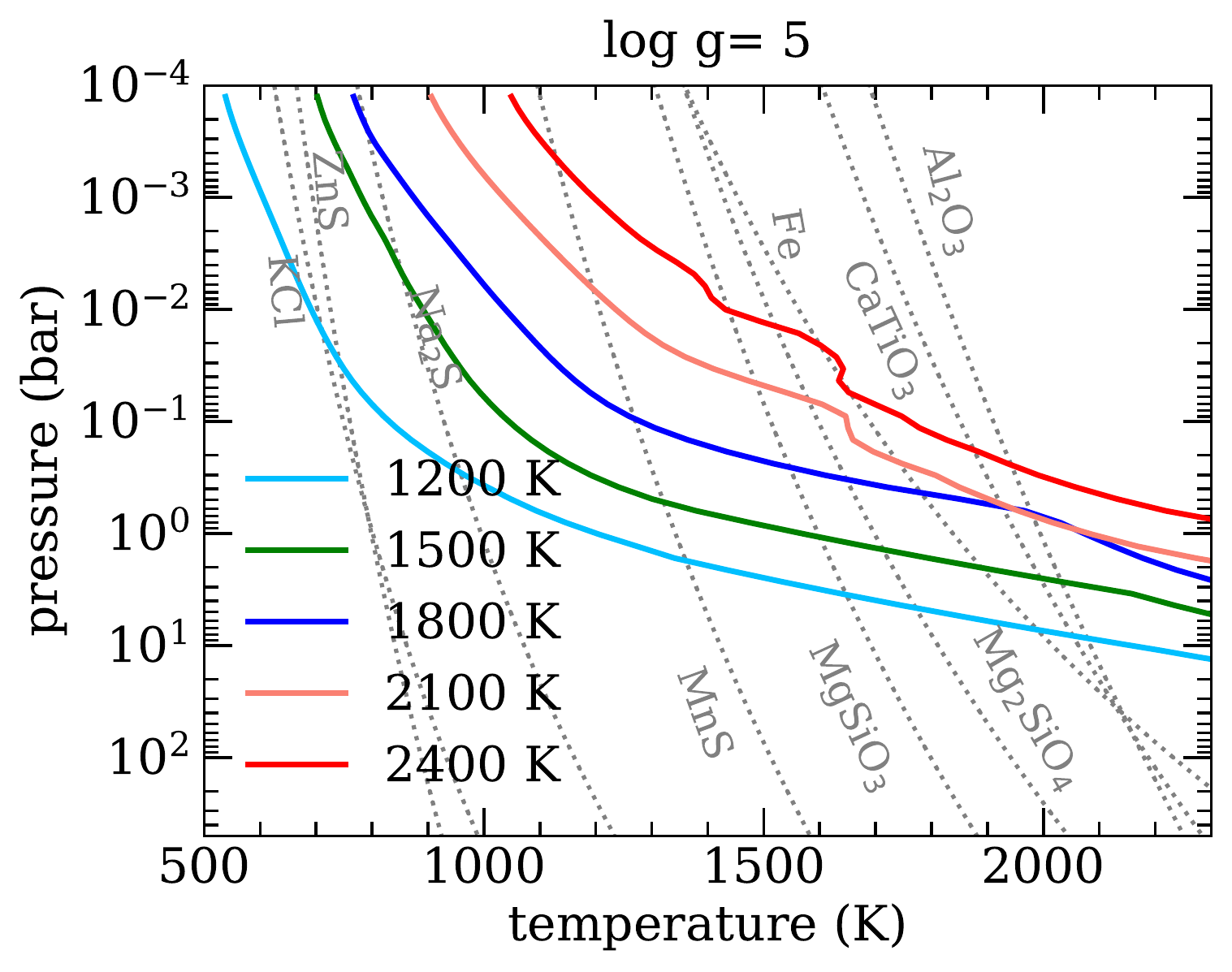}
\caption{Pressure-Temperature profiles for models with log g= 4 (top) and log g= 5 (bottom). We compute these atmosphere models assuming radiative-convective equilibrium. }
\label{fig:PTprofs}
\end{figure}

\subsection{Atmosphere Models}
We calculate the temperature structures of model atmospheres assuming radiative-convective equilibrium. The atmosphere models are described in detail by \citet{Mckay89}, \citet{Marley96, Marley99, Marley02}, \citet{Burrows97}, \citet{Fortney08b}, \citet{Saumon08} and \citet{Morley12, Morley14a}. The opacity database for gases is described in \citet{Freedman08,Freedman14}. We ran a small grid to explore a range in effective temperatures of 1200 K to 2400 K and log g= 4 and log g= 5 shown in Figure~\ref{fig:PTprofs}. These models include  ``standard" brown dwarf clouds (forsterite, iron, and corundum) using the \citetalias{AM01} cloud to determine cloud locations and particle sizes, assuming a sedimentation efficiency \fsed= 2.

\subsection{\text{Ad hoc Cloud Model}\label{sec:adhoccloud}}
Since the standard brown dwarf cloud models do not include the small particles required to match observations we invoke an ad hoc cloud layer to control the size, locations, and number density of particles. We insert our ad hoc cloud layer at a pressure level where we expect these silicate clouds could be present, at lower pressures than the \citetalias{AM01} cloud.

\begin{figure}
\centering
   \includegraphics[width=1\linewidth]{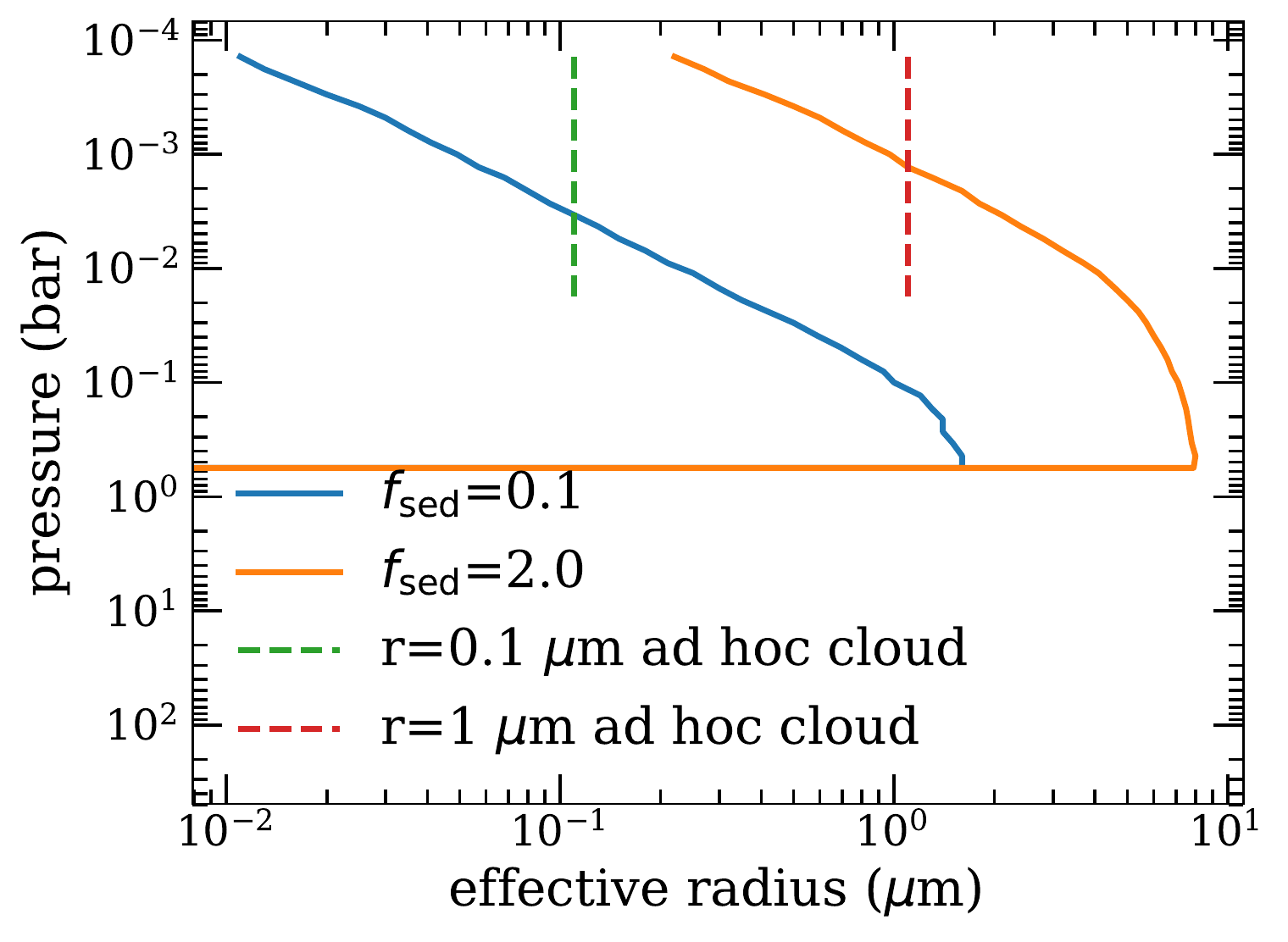}
   \includegraphics[width=1\linewidth]{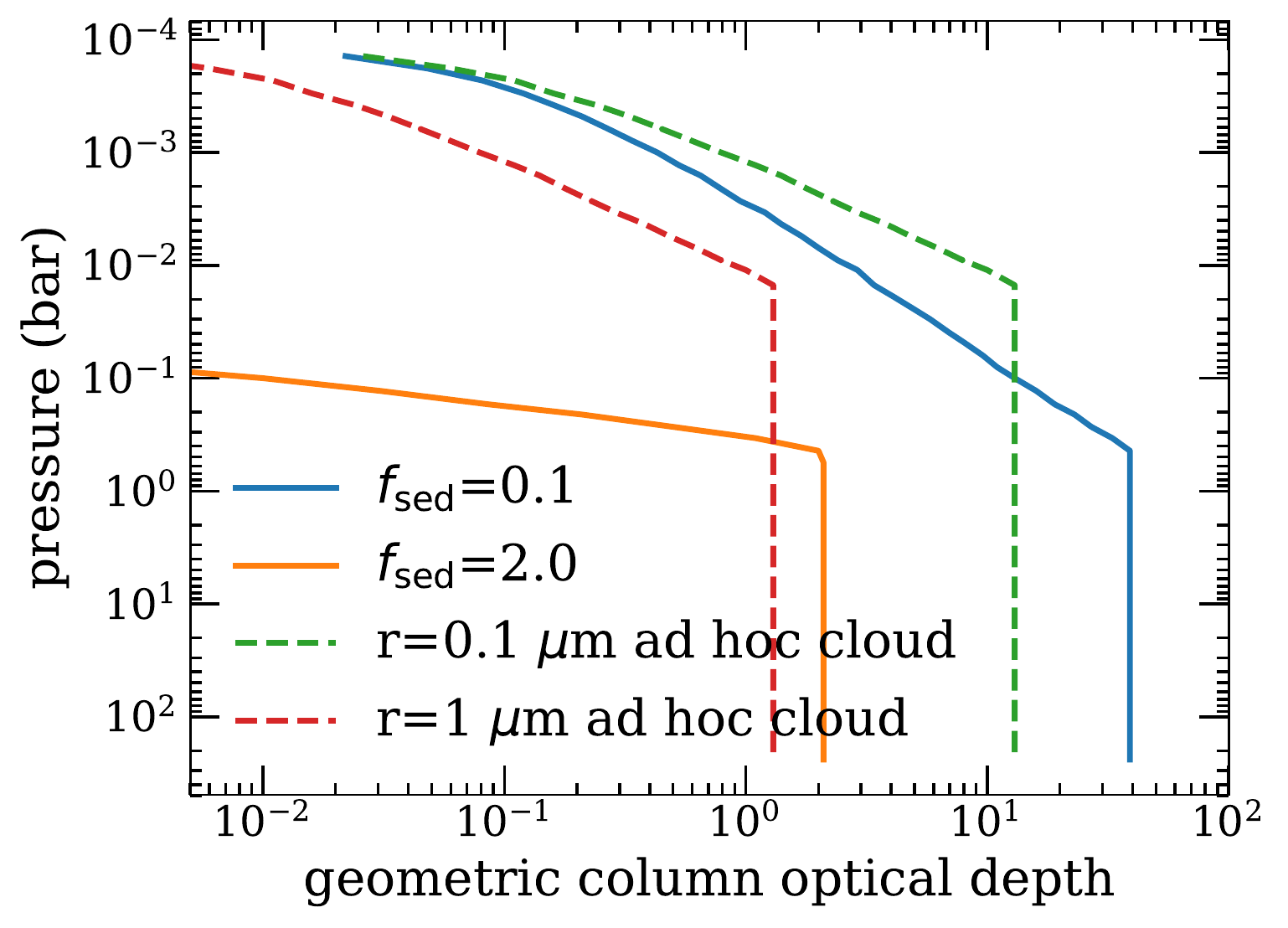}
\caption{ Using forsterite as an example, we compare our ad hoc cloud model (red and green) with the \citetalias{AM01} cloud model for \fsed~ values of 0.1 and 2. The top panel shows the effective radius at each pressure level and the bottom shows the geometric column optical depth. The ad hoc cloud falls in the parameter space of the \citetalias{AM01} cloud model. The ad hoc cloud is marginally optically thick ($\tau \sim 2/3$) with small particles.}
\label{fig:AMo1comp}
\end{figure}

As in the \citetalias{AM01} cloud model, we use a log-normal size distribution of particles defined as,
\begin{equation}
    \frac{dn}{dr} = \frac{N}{r \sqrt{2\pi}~ \ln\sigma} ~\text{exp} \bigg[-\frac{\ln^2(r/r_g)}{2\ln^2\sigma} \bigg], \label{eqn:lognormdist}
\end{equation}
where $N$ is the total number of particles, $\sigma$ is the standard deviation of the log-normal distribution, and $r_g$ is the geometric mean radius. $\sigma$ represents the width of the log-normal distribution of particles. For this study, we use a narrow distribution of particle sizes ($\sigma=1.2$). As $\sigma$ approaches 1, the distribution of particles sizes decreases until $r$ equals the geometric mean radius and Equation~\ref{eqn:lognormdist} reduces to the Dirac-delta function. By setting $\sigma$= 1.2, we average over unphysical Mie scattering effects for a single particle size while ensuring that the majority of particles will be close to the mode radius. 

We assume a constant number density of ad hoc cloud particles in each layer by setting the ratio of cloud particle number density, $n_{\rm cloud}$, to gas number density, $n$, in each atmospheric layer as a constant ($c= n_{cloud}/n$). The constant, $c$, is defined as,
\begin{equation}
     c =  \frac{\tau_{\rm cloud}}{\pi ~r_{\rm eff}^2 \sum n\Delta z},
\end{equation}
where $r_{\rm eff}$ is the effective (area-weighted) radius, $\tau_{\rm cloud}$ is the ad hoc cloud optical depth summed over all layers, and $n$ and $\Delta z$ are respectively the number density of molecules and vertical height of an atmospheric layer. The effective radius\footnote{\citet{Hansen1974} derive the effective radius for a log-normal size distribution (Equation 2.53).} is defined as, 
\begin{equation}
 r_{\rm eff}= r_g ~ \text{exp}\bigg[ {5\ln^2\sigma/2} \bigg] .
\end{equation}

The number density of molecules, $n$, in each layer is found using the pressure and temperature of each atmospheric layer assuming the ideal gas law. We create a marginally optically thick ad hoc cloud ($\tau_{\rm cloud}=2/3$), where we set the total cloud optical depth for 1 \um~particles to $\tau_{\rm cloud,~1\mu m}=0.67$ for amorphous particles. For crystalline particles, we scaled down the optical depth to  $\tau_{\rm cloud,~1\mu m}= 0.2$ to better match the amplitude of features observed in the \emph{Spitzer}/IRS spectra. To determine the number density of cloud particles for larger and smaller particle sizes we scale the number density for 1 \um~particles by,

\begin{equation}
 n_{\rm scaled}= n ~\bigg(\frac{r_{\rm v}}{r_{\rm v}(1 \mu \rm m)}\bigg)^{-3},
\end{equation}
where $r_{\rm v}$ is the volume averaged radius defined as,
\begin{equation}
         r_{\rm v}= r_g ~ \text{exp}\bigg[ {3\ln^2\sigma/2} \bigg].
\end{equation}

We keep the total mass of cloud material constant across different particle sizes with the mass density of the cloud material, $M$, defined as,
\begin{equation}
M = (4/3) ~\pi ~r_{\rm v}^3~  \rho_{\rm material}~  n_{\rm cloud}.
\end{equation}
Here $\rho_{\rm material}$, is the density of the cloud particle. For an 1800K, log g$=$5 model atmosphere, the mass of a 1 \um~amorphous ad hoc forsterite cloud is $M=5.3\times 10^{-4}$~g/cm$^2$, which is 16$\%$ of the total forsterite cloud.

We compare the properties of the ad hoc cloud with the \citetalias{AM01} model. We compute two models setting the sedimentation efficiency parameter, \fsed, to 0.1 and 2. A smaller \fsed~leads to vertically taller clouds with smaller particles and a larger \fsed~results in thin clouds with larger particles. We show the effective radius and optical depth at each pressure level for the 3 cloud models in Figure~\ref{fig:AMo1comp}. Although the \fsed=0.1 model produces small particles in the upper layers of the atmosphere, the resulting cloud becomes optically thick at 1.6 mbar, leading to models with blackbody-like spectra in the near-infrared that do not match observations of L dwarfs.

\begin{figure}
\centering
  \includegraphics[width=1\linewidth]{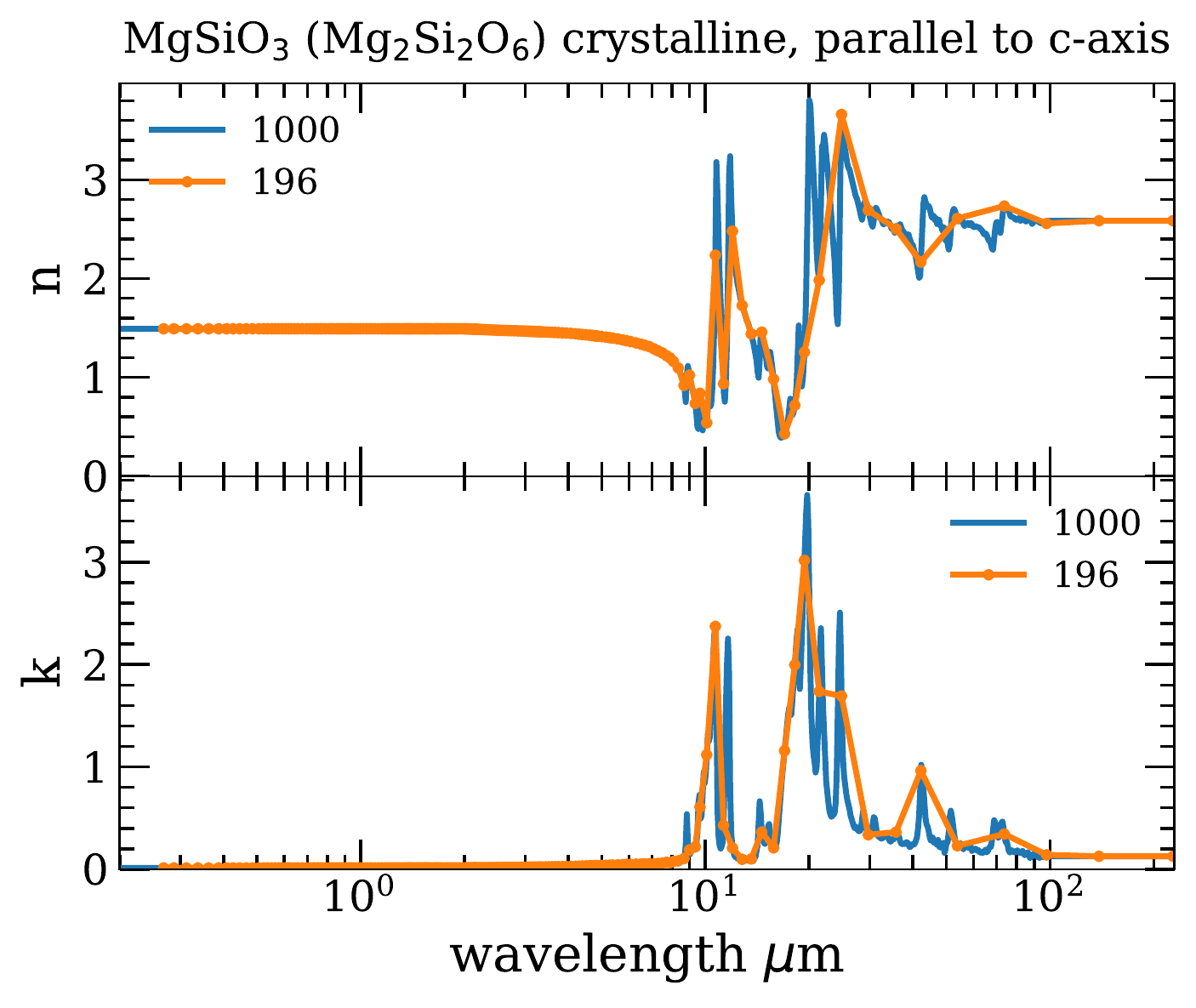}
\caption{The optical properties for crystalline MgSiO$_3$ (Mg$_2$Si$_2$O$_6$) shown with the 196 and 1000 point wavelength grid.  }
\label{fig:cloudresolution}
\end{figure}

\subsection{Mie Scattering \label{sec:miescattering}}
We use Mie scattering theory to calculate the absorption and scattering coefficients of particles in the atmosphere for each cloud species independently. We calculate Mie scattering coefficients for particle sizes ranging from 0.001 $\mu$m to 10 $\mu$m.

\begin{figure}
    \centering 
    \includegraphics[width=1\linewidth]{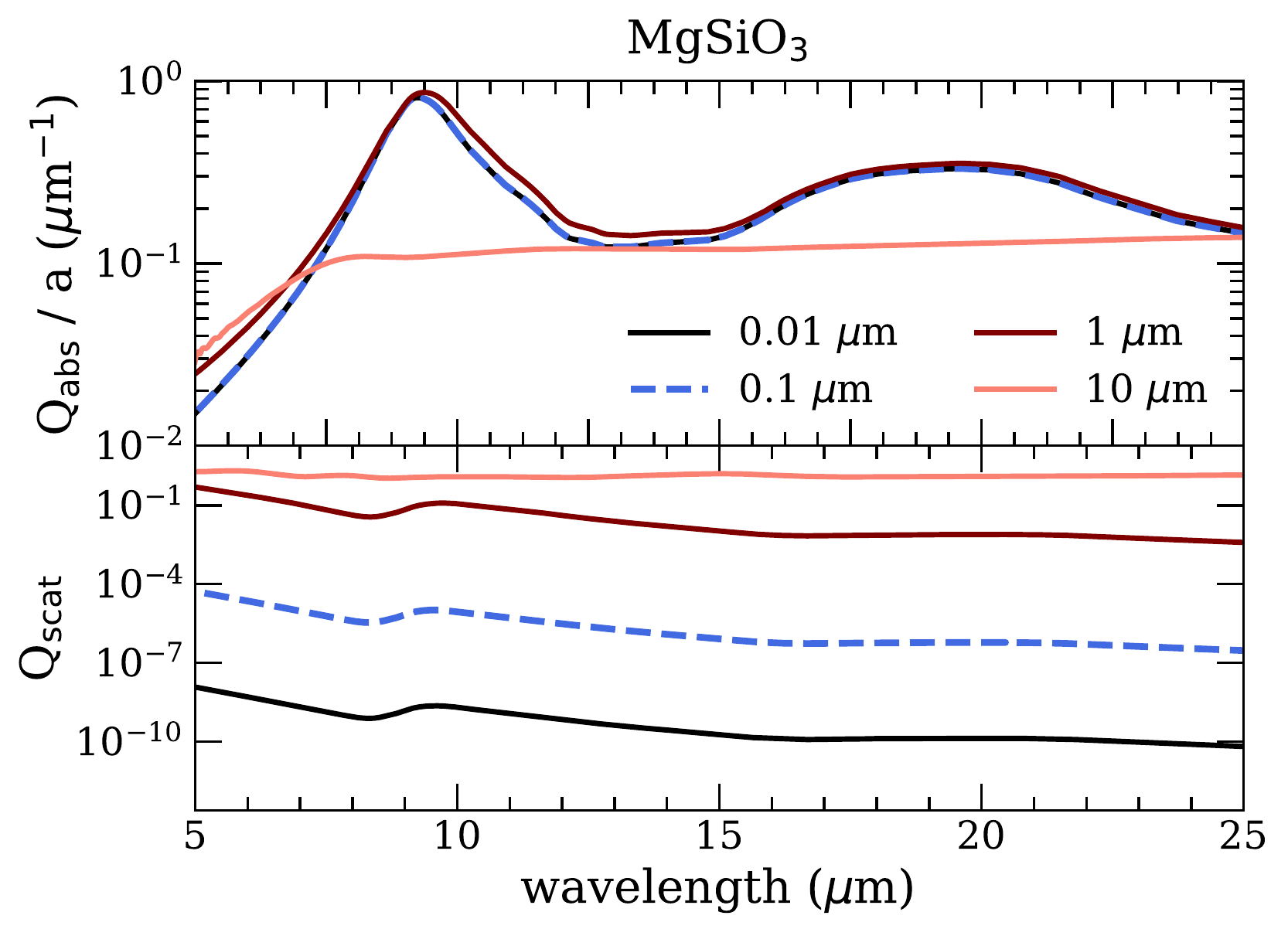}
    \caption{The optical properties for amorphous enstatite (MgSiO$_3$) for a range of particle sizes. We see a peak in absorption around the silicate feature for particles $\sim$ 1$\mu$m or less, and nearly featureless absorption for 10 $\mu$m particles.}
    \label{fig:Qabsparticlesizes}
\end{figure}

\begin{figure*}[t]
\gridline{\fig{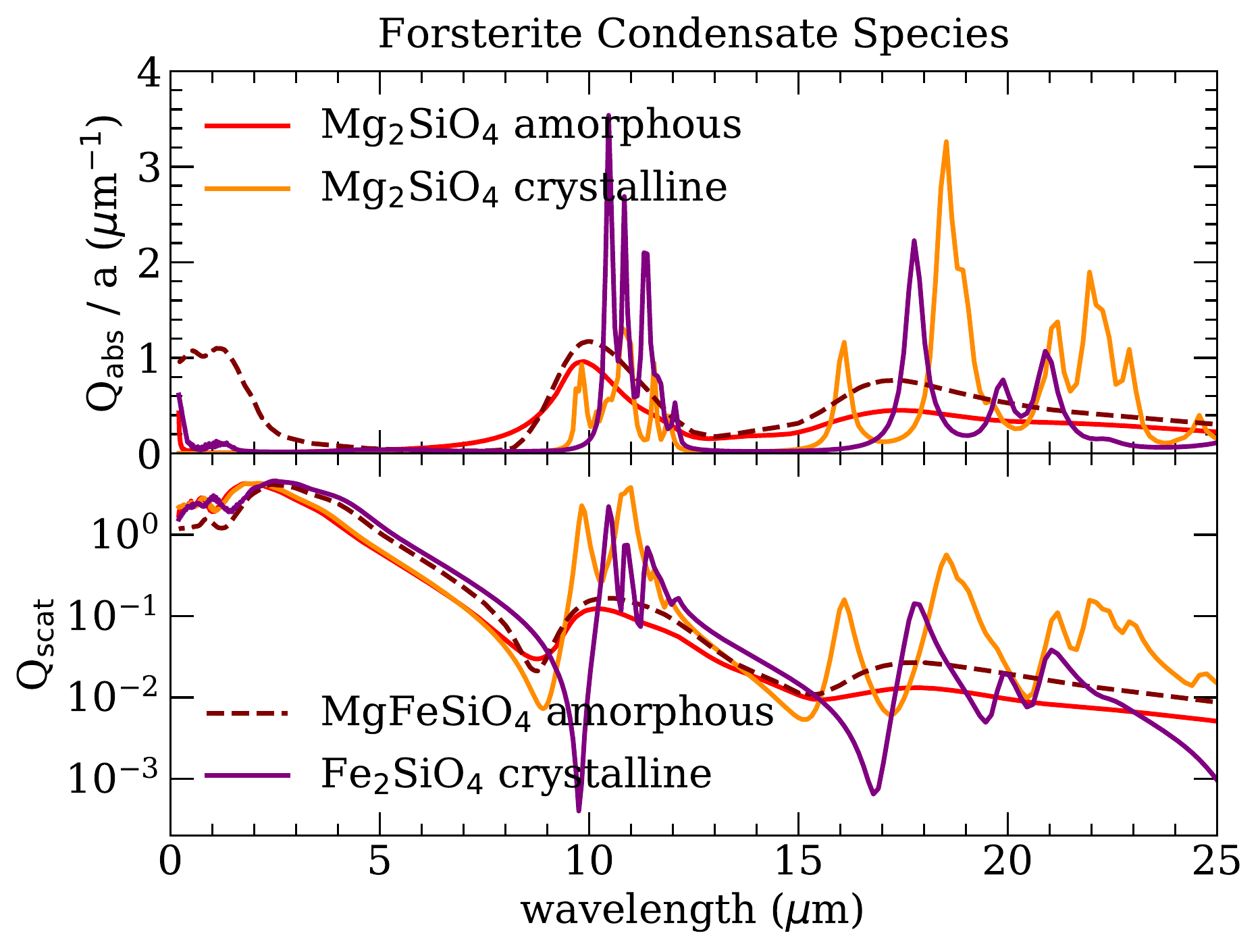}{0.5\linewidth}{}
         \fig{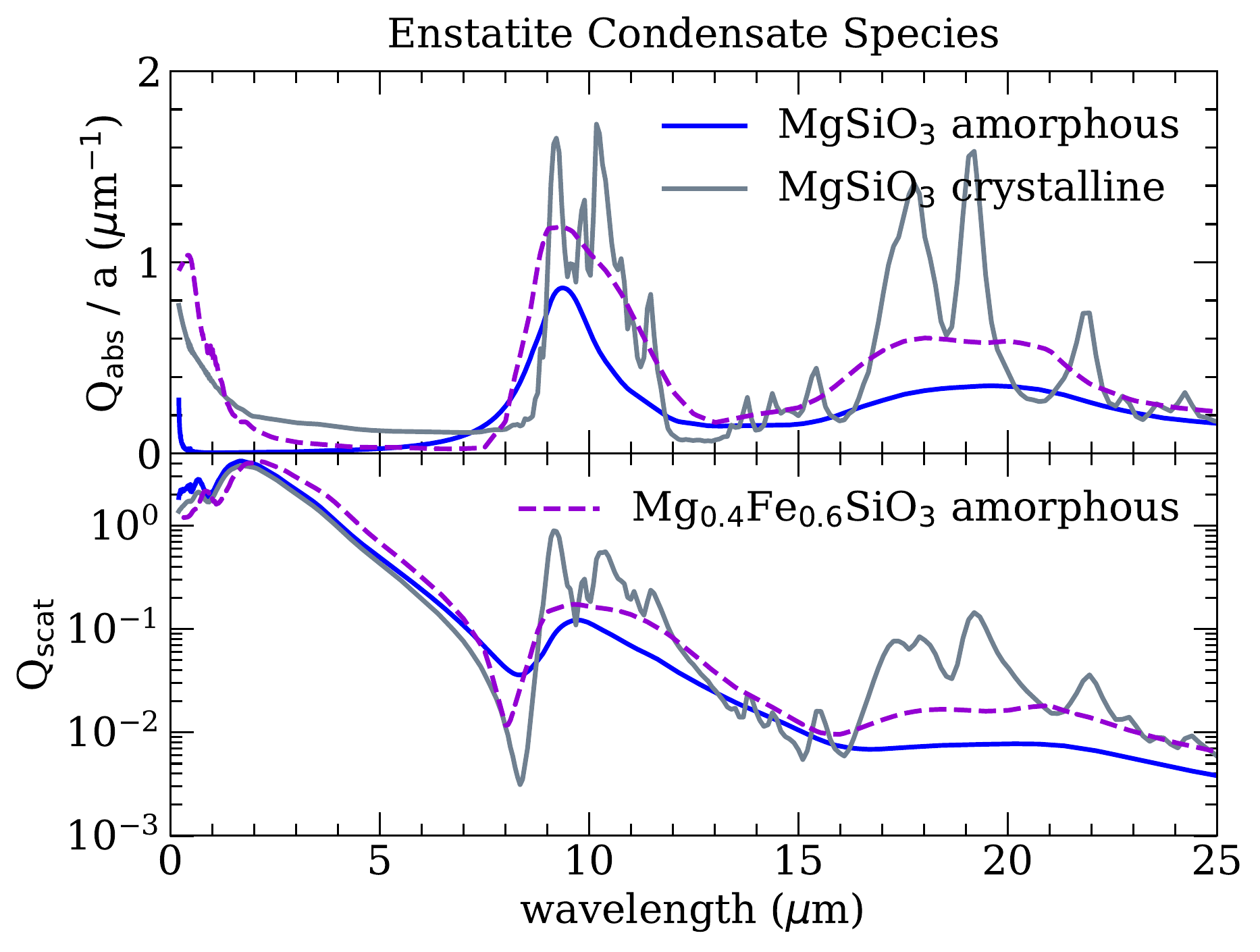}{0.5\linewidth}{}   }
            \vspace{-6mm}
\gridline{ \fig{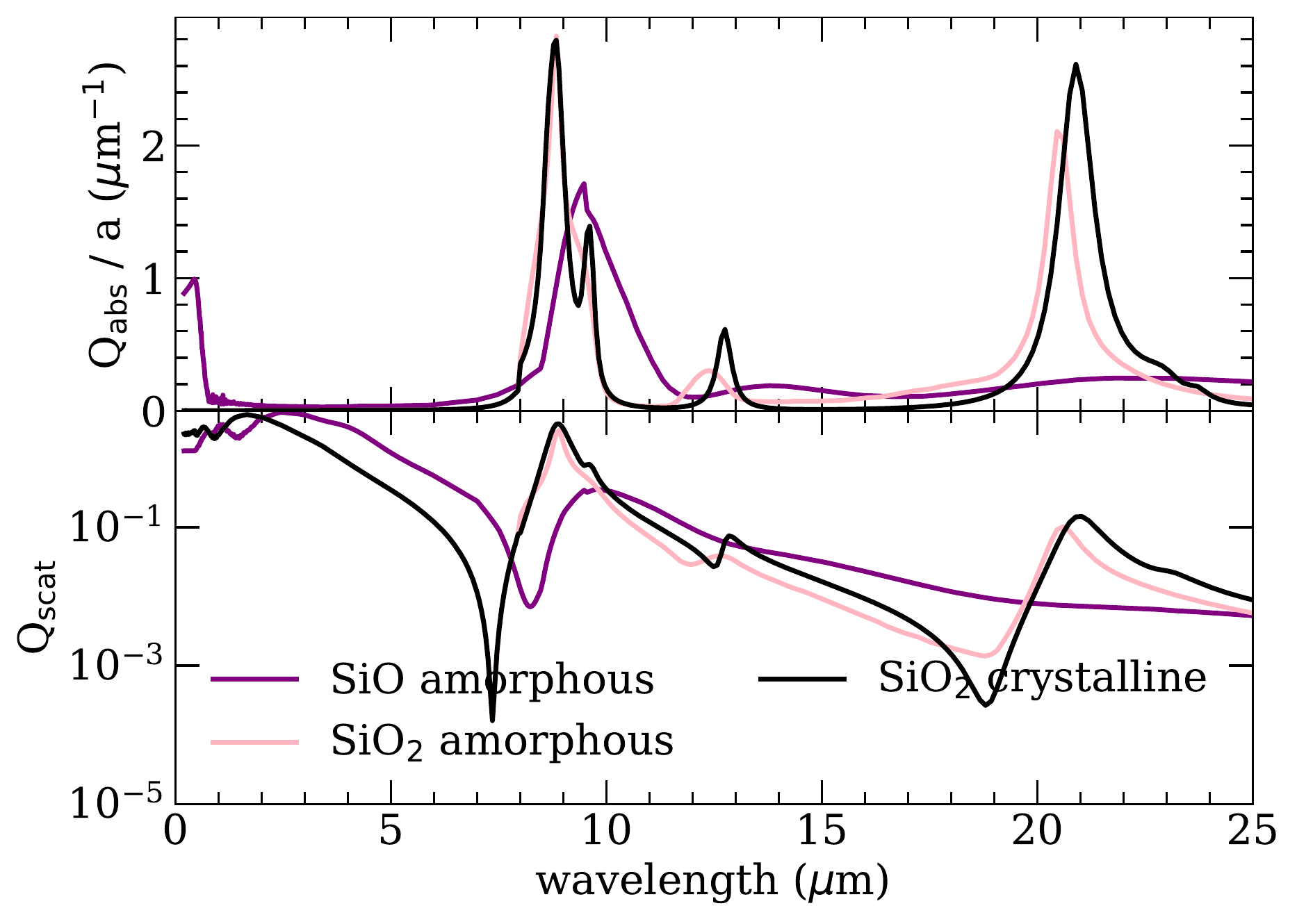}{0.5\linewidth}{}
         \fig{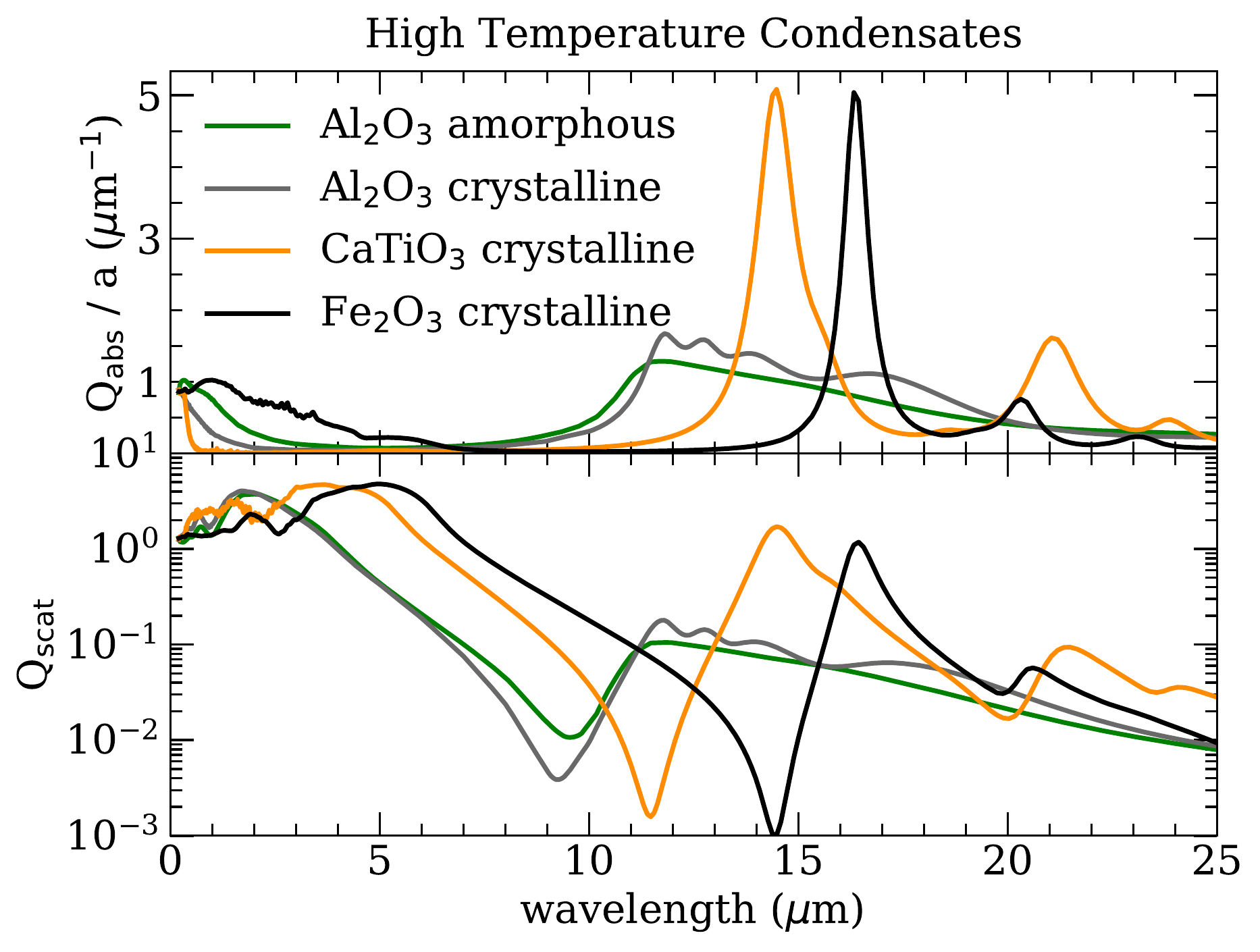}{0.5\linewidth}{}}   
         \vspace{-5mm}
\caption{The absorption and scattering efficiencies  for a range of condensates. The top panels show the forsterite and enstatite condensate species, where iron silicate variants are shown for both amorphous and crystalline structures. The bottom left panel shows silicon oxide species. The bottom right panel represents the high temperature condensates that form in hot L dwarfs and super-hot Jupiters.
\label{fig:Qabs}}
\end{figure*}

\begin{deluxetable*}{ccccc}
\tablecolumns{5}
\tablecaption{Compilation of optical constants for potential condensates in brown dwarf and gas giant exoplanets \label{tab:opticalconstants}}
\tablehead{
\colhead{Condensate} \vspace{-0.2cm} & \colhead{Name} & \colhead{Comments} &\colhead{Reference} &\colhead{$\lambda$ (\um)}\\ \vspace{-0.2cm}}
\startdata
    Al$_2$O$_3$  & corundum & amorphous; 873K & \citet{Begemann97} & 7.8 - 500  \\
    Al$_2$O$_3$  & corundum & crystalline & \citet{Koike95} & 0.2 - 400  \\
    CaTiO$_3$ & perovskite & crystalline  & \citet{Ueda98} & 0.02 - 2 \\ 
     &  &   & \citet{Posch03} & 2 - 5843 \\ 
    Fe$_2$O$_3$ & hematite  &  crystalline & A.H.M.J. Triaud$^{\rm b}$ & 0.1 - 1000 \\
    \forst  & forsterite & amorphous (sol-gel) & \citet{Jaeger03} & 0.2 - 500 \\
    \forst$^{\rm a}$ & forsterite  & crystalline; 295K & \citet{Mutschke19} & 45 - 500 \\
     MgFeSiO$_4$ & olivine & amorphous (glassy)  &   \citet{Dorschner95} & 0.19 - 500  \\
    Mg$_{0.8}$Fe$_{1.2}$SiO$_4$ & olivine & amorphous (glassy)  & \citet{Dorschner95} & 0.19 - 500     \\
    Fe$_{2}$SiO$_4$ & fayalite  & crystalline   &    \citet{Fabian01}$^{\rm b}$ & 2 - 200 \\
    \enst  & enstatite & amorphous (sol-gel) & \citet{Jaeger03}  & 0.2 - 500 \\
    \enst  & enstatite & amorphous (glassy) &  \citet{Dorschner95} & 0.19 - 500 \\
    \enst(Mg$_{2}$Si$_{2}$O$_{6}$)$^{\rm a}$  & enstatite & crystalline & \citet{Jaeger98} & 2 - 100  \\ 
    Mg$_{0.8}$Fe$_{0.2}$SiO$_3$ & pyroxene & amorphous (glassy)  &   \citet{Dorschner95} & 0.19 - 500     \\
    Mg$_{0.5}$Fe$_{0.5}$SiO$_3$ & pyroxene & amorphous (glassy)  & \citet{Dorschner95} & 0.19 - 500   \\
    Mg$_{0.4}$Fe$_{0.6}$SiO$_3$ & pyroxene & amorphous (glassy)  &   \citet{Dorschner95} & 0.19 - 500  \\
    Mg$_{0.92}$Fe$_{0.09}$SiO$_3^{\rm  a}$ & orthoenstatite & crystalline; 10 K, 300 K, 928 K  &   \citet{Zeidler15} & 5 - 60   \\
    SiO &  & amorphous  &   \citet{Palik85}  & 0.05 - 0.8   \\    
     &  &   &   \citet{Wetzel13}  & 2 - 100   \\    
    SiO$_2$ & quartz & amorphous; 300 K  & \citet{Palik85}  & 0.05 - 8.4     \\ 
     &  &  & \citet{Henning97}  & 2 - 500   \\    
    SiO$_2$ & quartz & crystalline; 928 K  &  \citet{Palik85} & 0.05 - 8.4 \\  
     &  &  &  \citet{ Zeidler13} & 5 - 50     \\  
\enddata
\tablecomments{$^{\rm a}$Computed by averaging absorption and scattering efficiencies from 3 crystallographic axes \\
    $^{\rm b}$Database of Optical Constants for Cosmic Dust, Laboratory Astrophysics Group of the AIU Jena (\url{http://www.astro.uni-jena.de/Laboratory/OCDB/index.html}}
\end{deluxetable*}

We use optical properties of each cloud species from the literature summarized in Table~\ref{tab:opticalconstants}. A few condensates have optical properties taken from multiple sources (SiO, amorphous SiO$_2$ and, crystalline SiO$_2$). The amorphous and crystalline SiO$_2$ data sets were combined as described in Section~4 of \citet{Kitzmann18}. To combine the data, the overlapping IR data is replaced with the most recent publication, and the Kramers-Kronig relation is used to combine the IR and short-wavelength data. For SiO, Section~4.1 in \citet{Wetzel13} describes how the UV-visible data is taken from \citet{Palik85} ($\lambda<$ 0.8 \um), and from 0.8 \um~- 8 \um~a Brendel-oscillator model is fit to experimental results.

To capture sharper crystalline optical constants, we change the resolution of the optical properties from the 196 wavelength point grids used in prior \citetalias{AM01} models (eg. \citet{Saumon08,Morley12}) to a new 1000 wavelength point grid that ranges from 0.2 $\mu$m  to 230 $\mu$m. In Figure~\ref{fig:cloudresolution} we show the optical properties for crystalline enstatite (Mg$_2$Si$_2$O$_6$) to illustrate the resolution of these models compared to prior works.

Using the complex refractive indices, we calculate the absorption and scattering efficiency for a range of particle sizes which we show in Figure~\ref{fig:Qabsparticlesizes}. These coefficients show which wavelengths strongly exhibit Mie scattering features. In Figure~\ref{fig:Qabs}, we show the absorption and scattering efficiencies for a broad range of condensates that can form in brown dwarf and giant exoplanets. Crystalline structures have refractive indices that are different for each crystallographic axis. To combine information from the 3 axes into our 1D cloud model, we calculate the absorption and scattering efficiencies for each crystallographic axes separately, then averaged the efficiencies to combine the information \citep{Steyer1974,Jaeger98}, shown in Figure~\ref{fig:Qabs}. We find that the numerical calculation approaches the analytic expression for r $\ll$ $\lambda$: at r $\leq$ 0.1 \um~\citep{VandeHulst1957}.

\subsection{Thermal Emission Spectra}
Using the cloud model and the pressure-temperature profile as inputs, we use the radiative transfer model developed in \citet{Morley15} to calculate thermal emission spectra. This model calculates the intensity and fluxes in multiple-scattering and emitting layered atmospheres. It takes the optical depth, single scattering albedo, and asymmetry parameter from the cloud model and calculates the flux at a given wavenumber using the C version of the open-source radiative transfer code \texttt{disort} \citep{Stamnes88, Buras11}. The radiative transfer code implements the discrete-ordinate method for unpolarized light through a vertically inhomogeneous media.

%%%%--------------
%%%%------RESULTS SECTION-------------------------------------------------
\section{Results} \label{sec:results}
In the following section, we discuss the impacts of silicate and refractory clouds on model spectra of brown dwarfs. Then, we discuss the observability of these features in the near future using \emph{JWST} simulations and determine the best targets for future observations. 

\subsection{Model Spectra}
\begin{figure*}[t]
\gridline{\fig{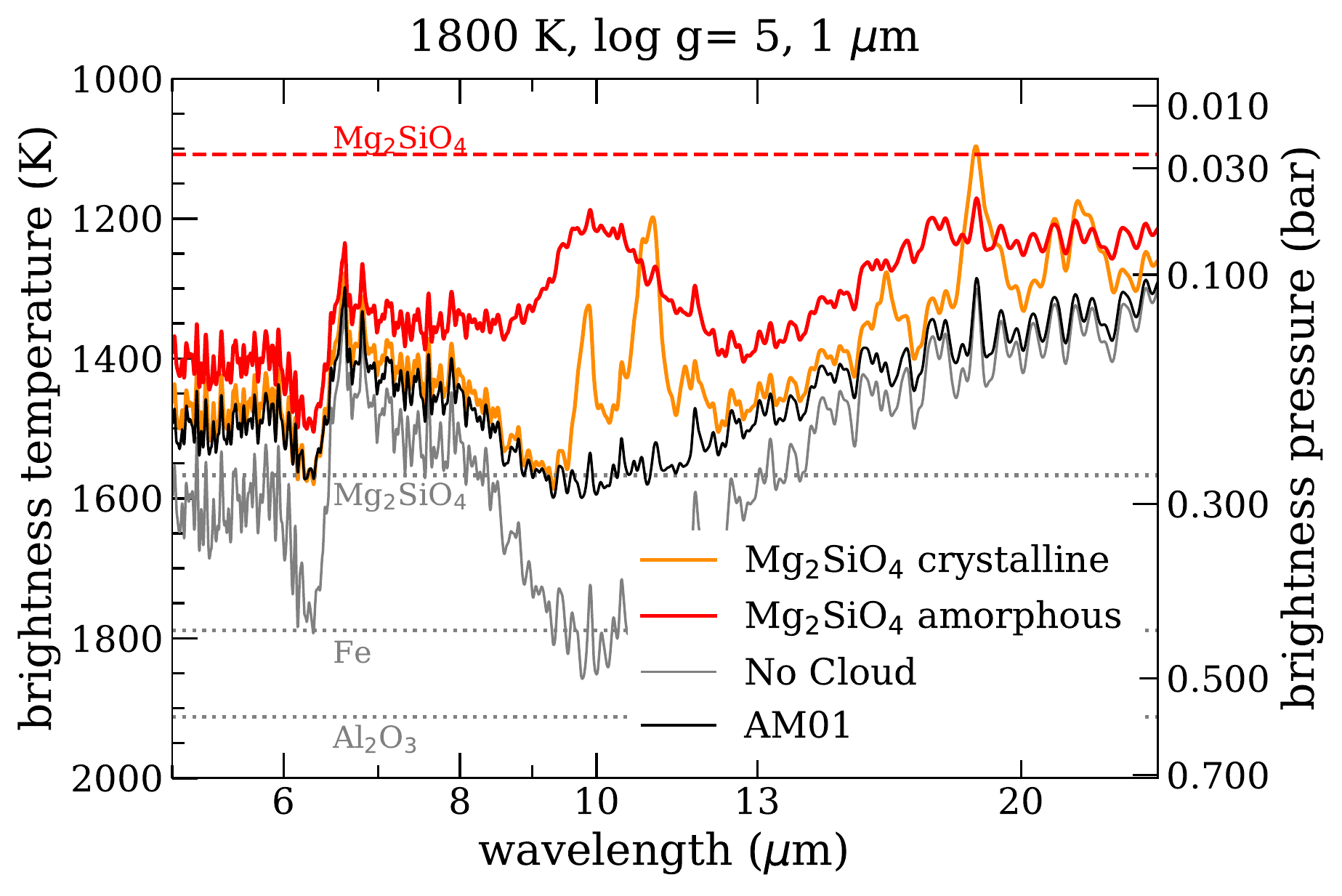}{0.5\linewidth}{}
         \fig{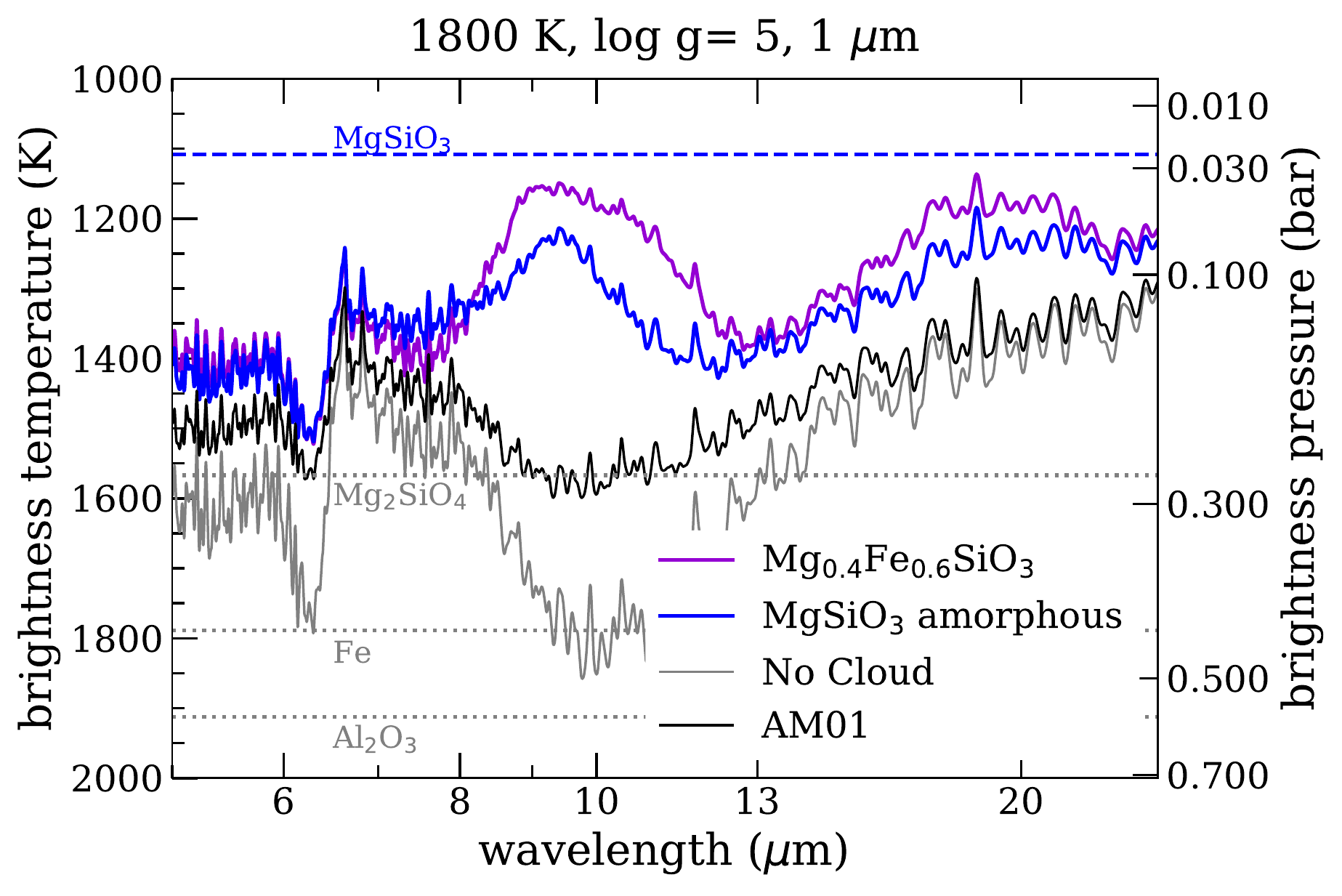}{0.5\linewidth}{}   }
      \vspace{-6mm}
\gridline{ \fig{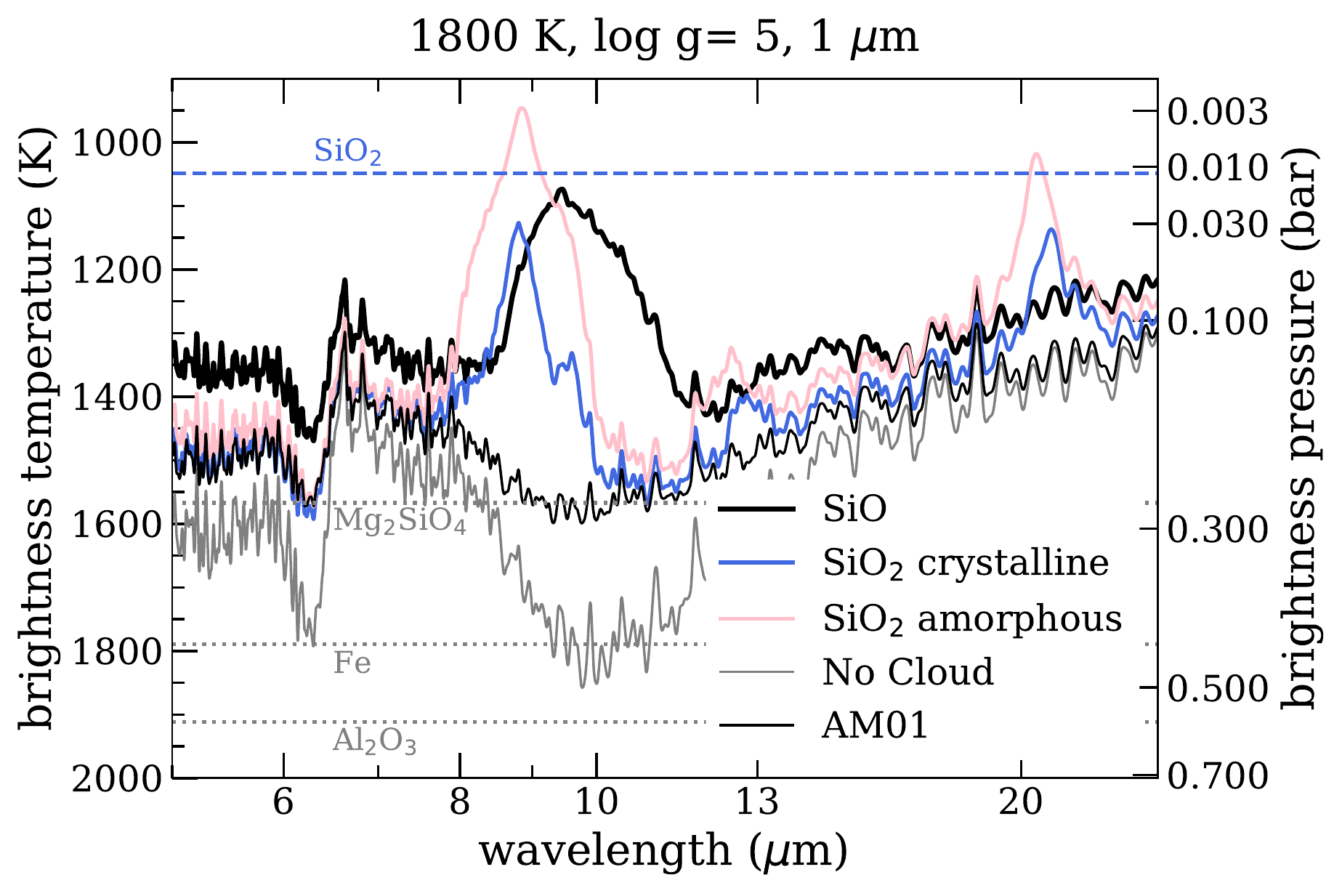}{0.5\linewidth}{}
         \fig{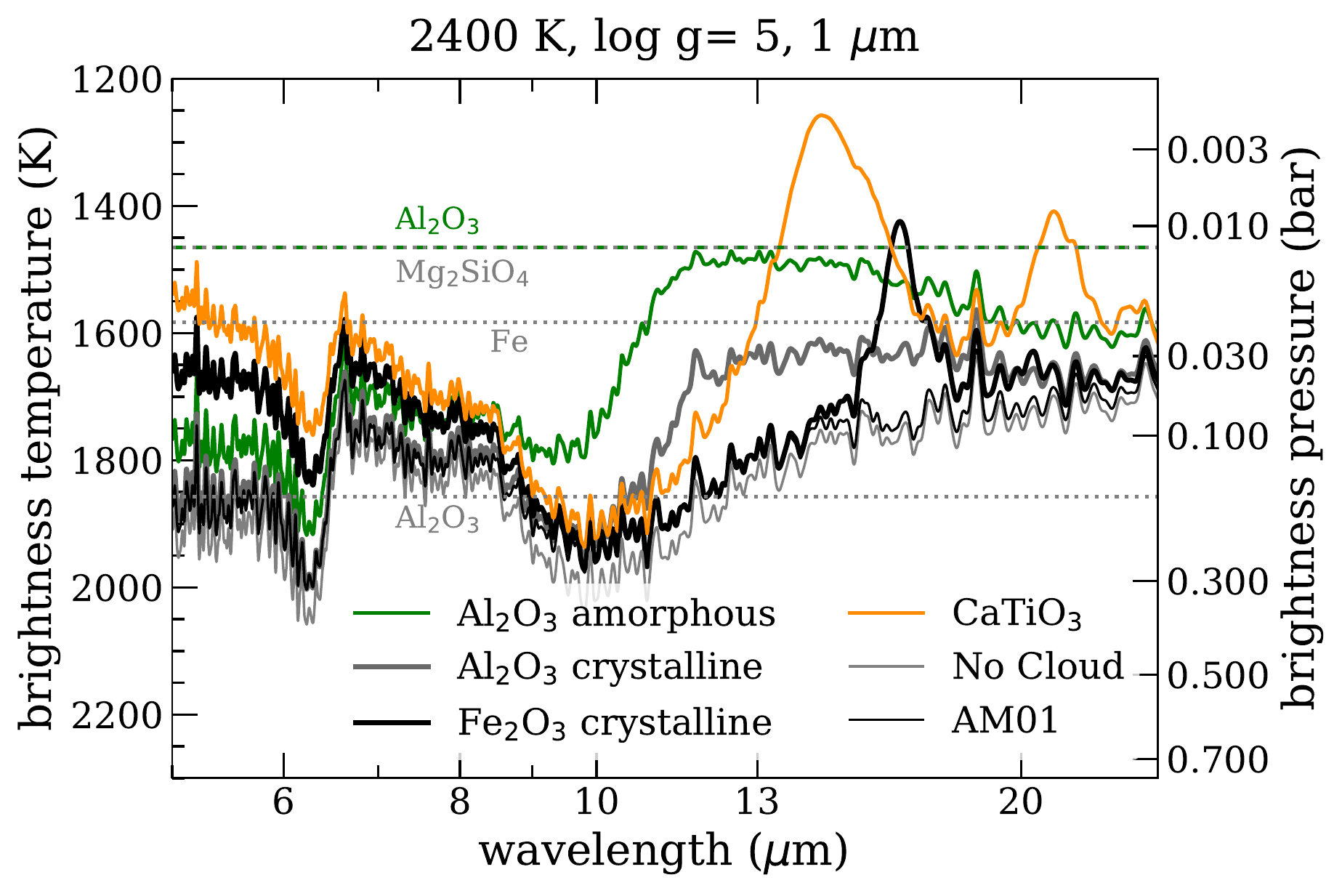}{0.5\linewidth}{}} 
         \vspace{-5mm}
\caption{We show the brightness temperature for forsterite silicate species (top left), enstatite silicate species (top right), quartz species (bottom left) and high-temperature species (bottom right). The pressure along the pressure-temperature profile corresponding to each brightness temperature is shown on the right y-axis In each panel, we also show a cloud free (gray) and an \citetalias{AM01}, \fsed= 2 (black) model. The horizontal dashed gray lines represent the location of the standard \citetalias{AM01} clouds.  \label{fig:ptbright} The flux from each silicate or refractory features come from the top, cooler, part of the atmosphere (horizontal colored dashed lines). This is in contrast to the cloudless and \citetalias{AM01} models, where the flux in those regions come from the deeper, hotter part of the atmosphere. }
\end{figure*}

\begin{figure*}[t]
\gridline{\fig{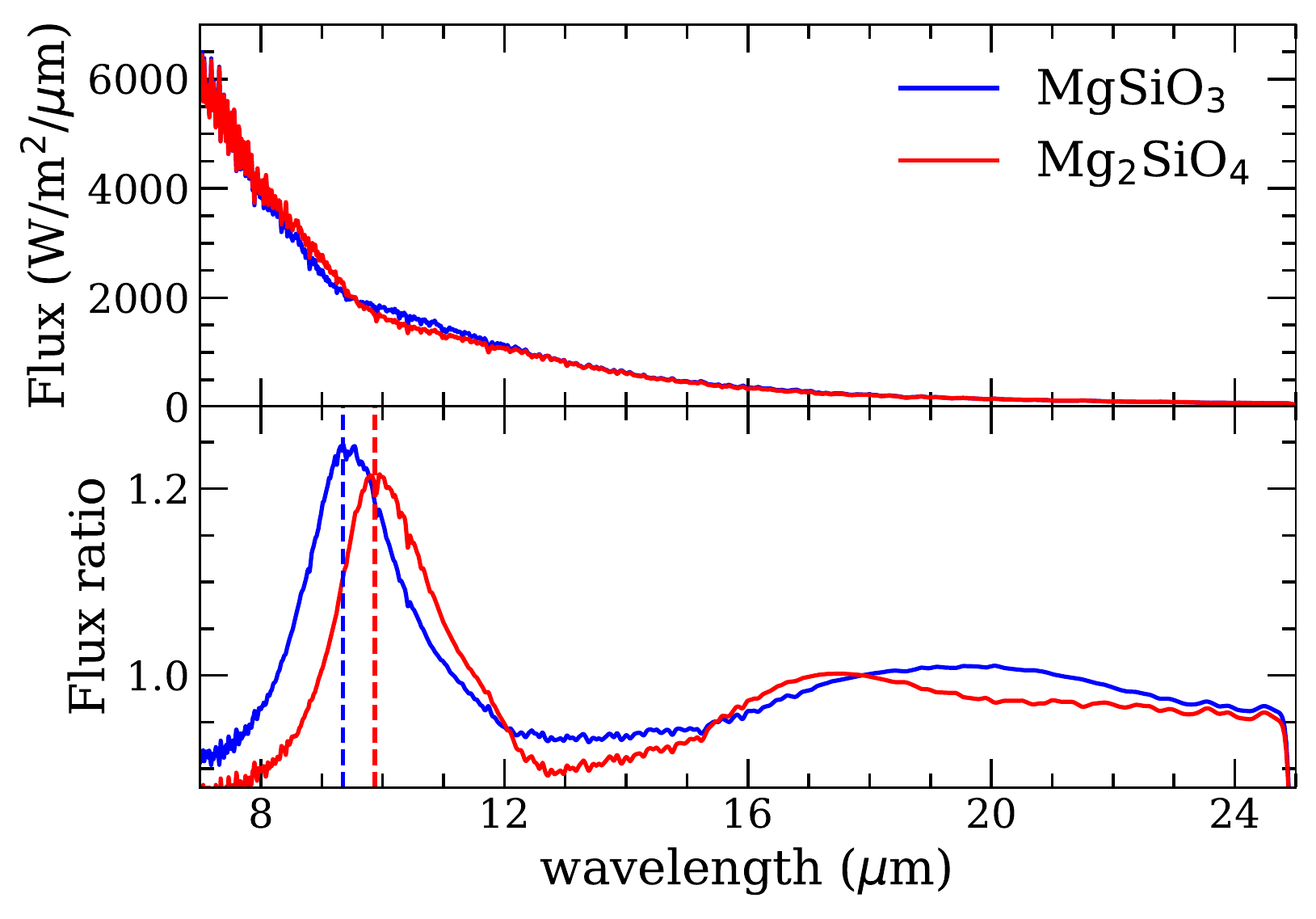}{0.5\linewidth}{}
         \fig{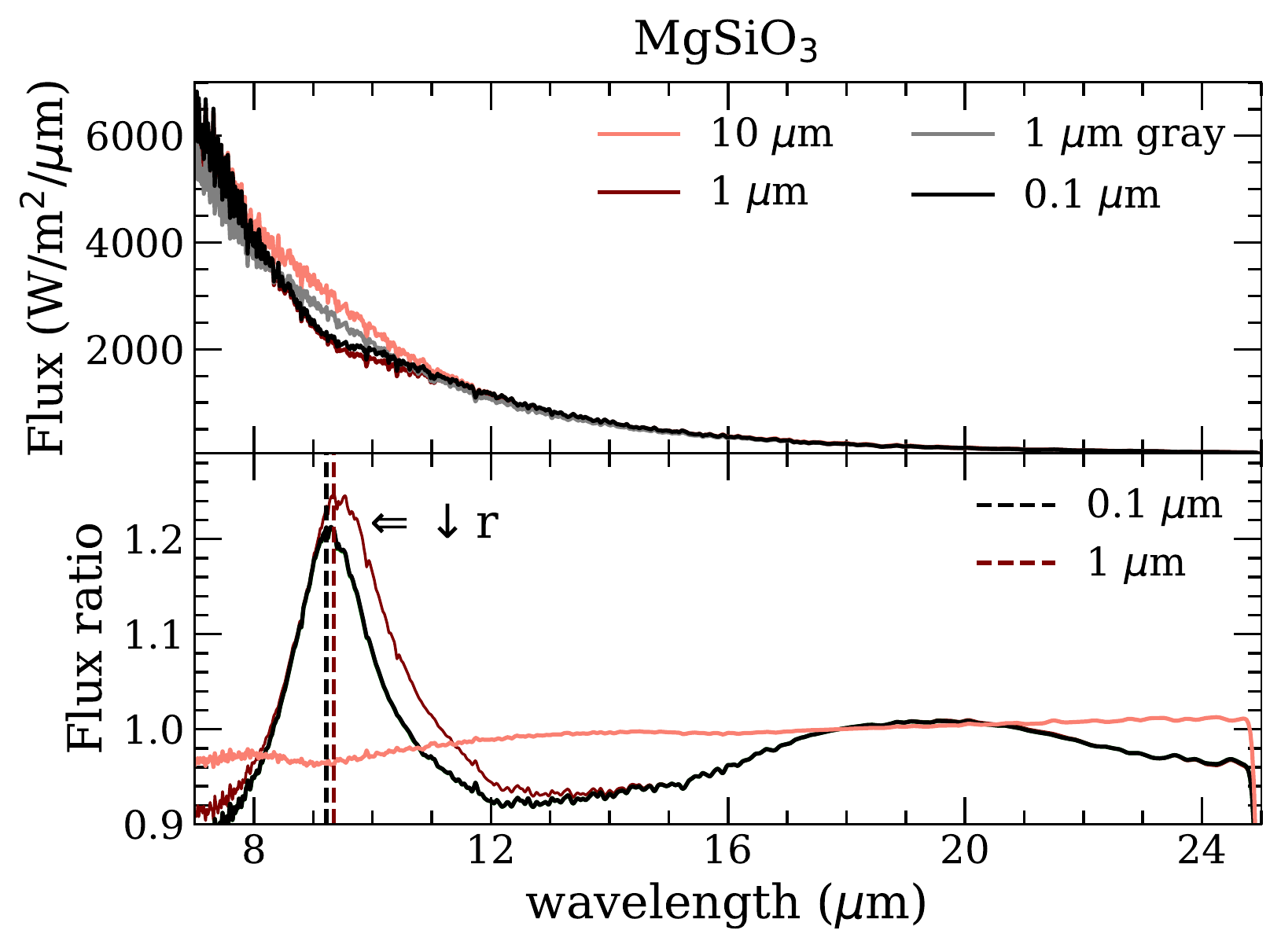}{0.5\linewidth}{}   }
   \vspace{-6mm}
\gridline{ \fig{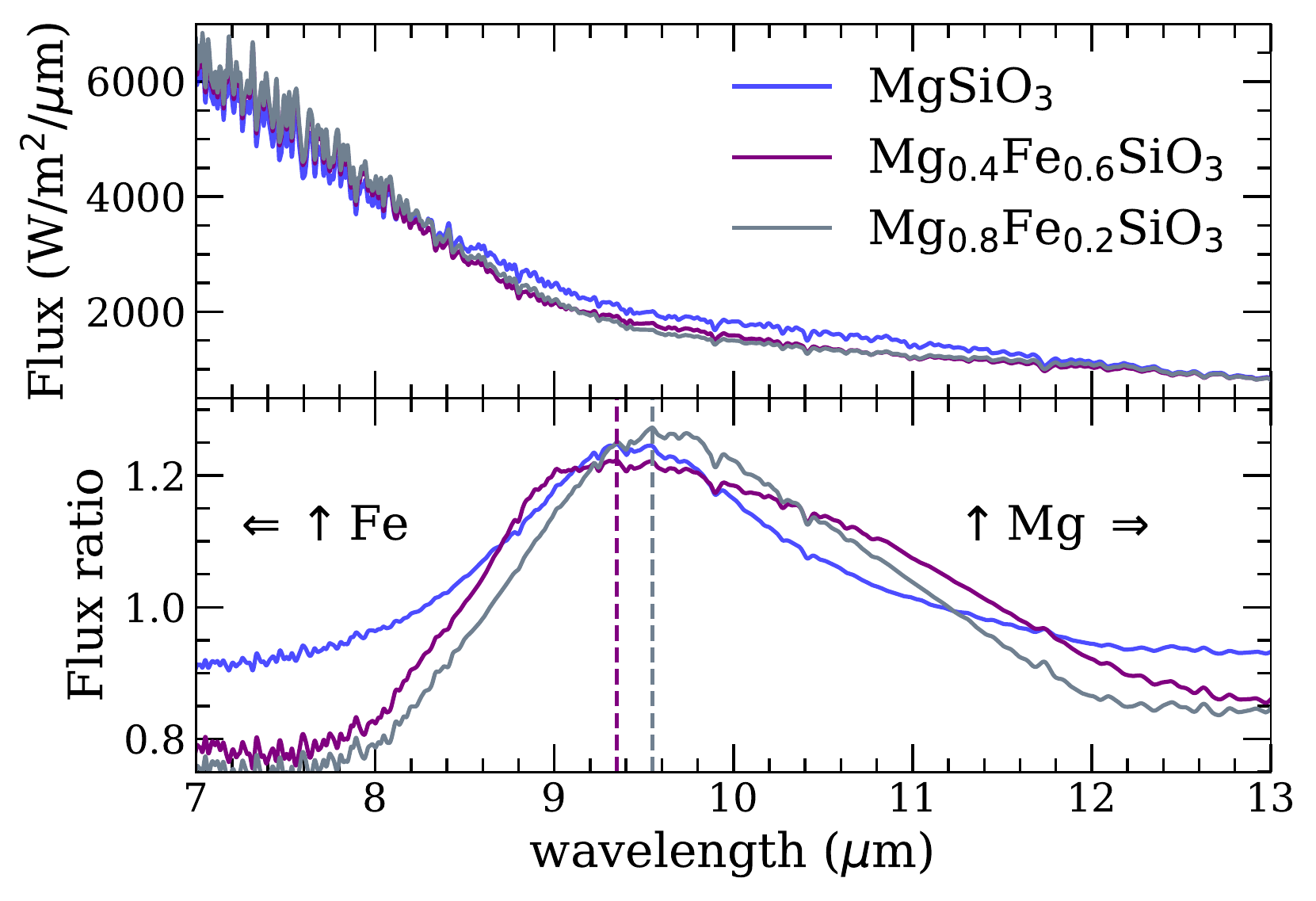}{0.5\linewidth}{}
 \fig{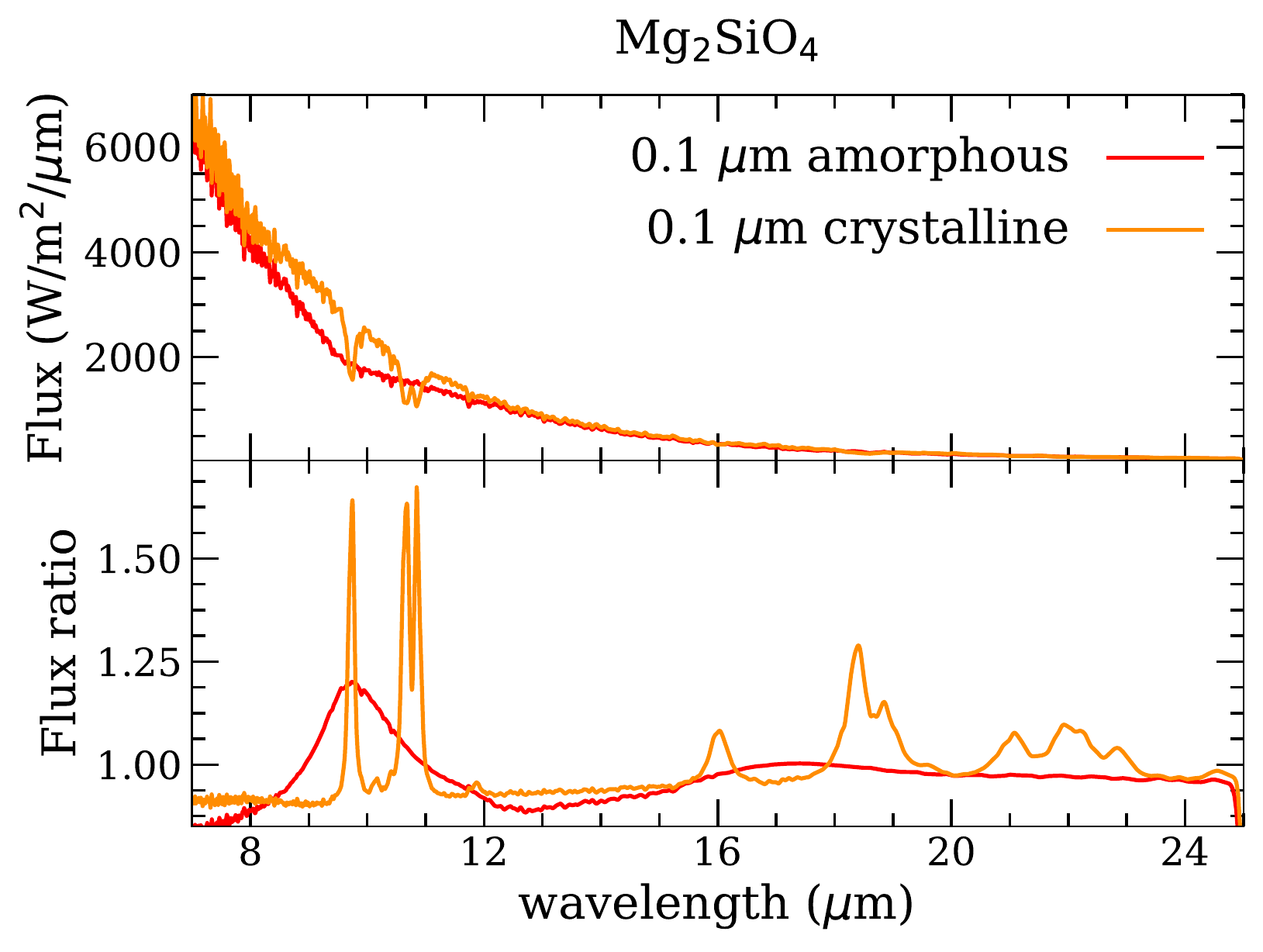}{0.5\linewidth}{}
        }    
    \vspace{-6mm}
\gridline{ \fig{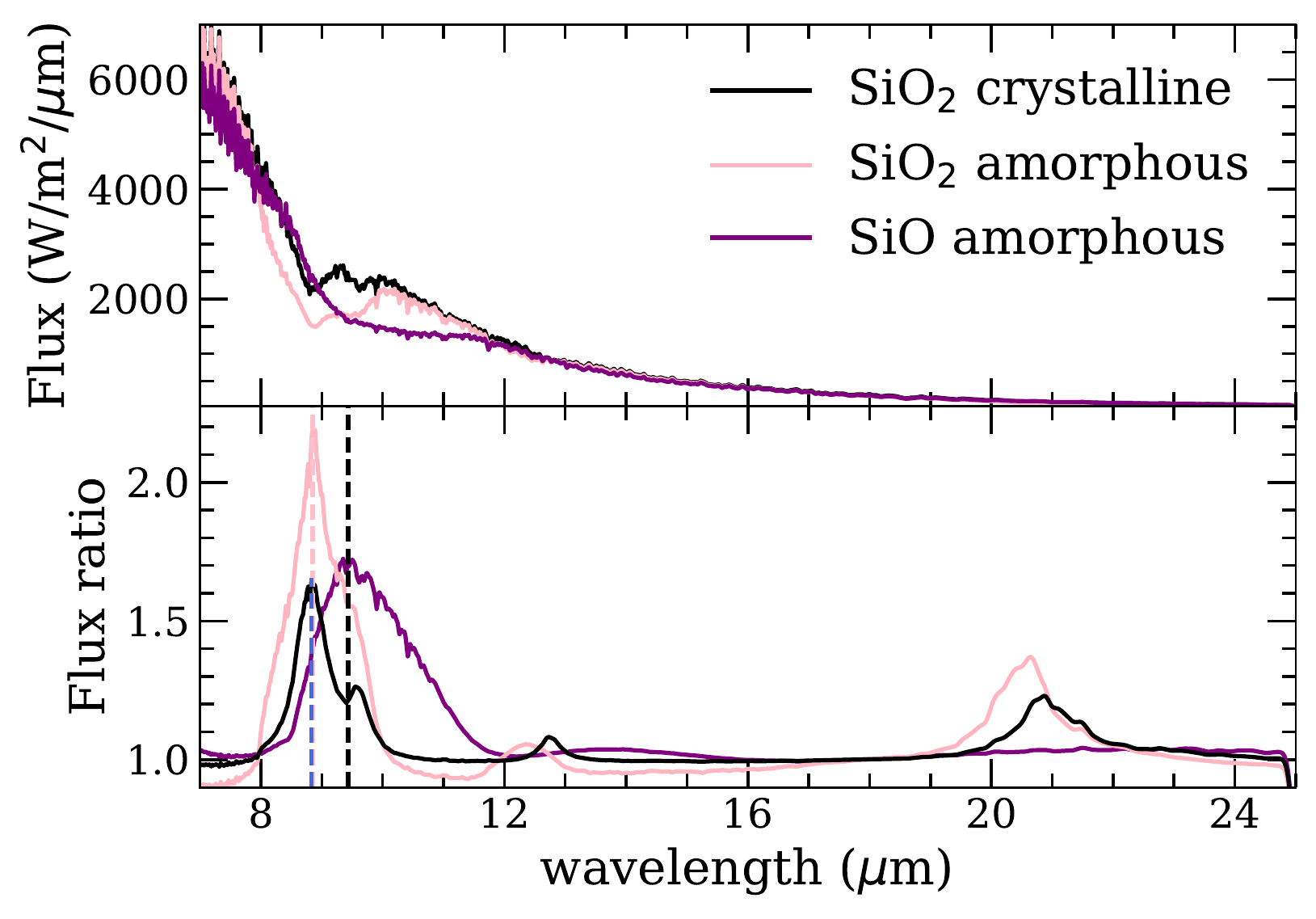}{0.5\linewidth}{}
         \fig{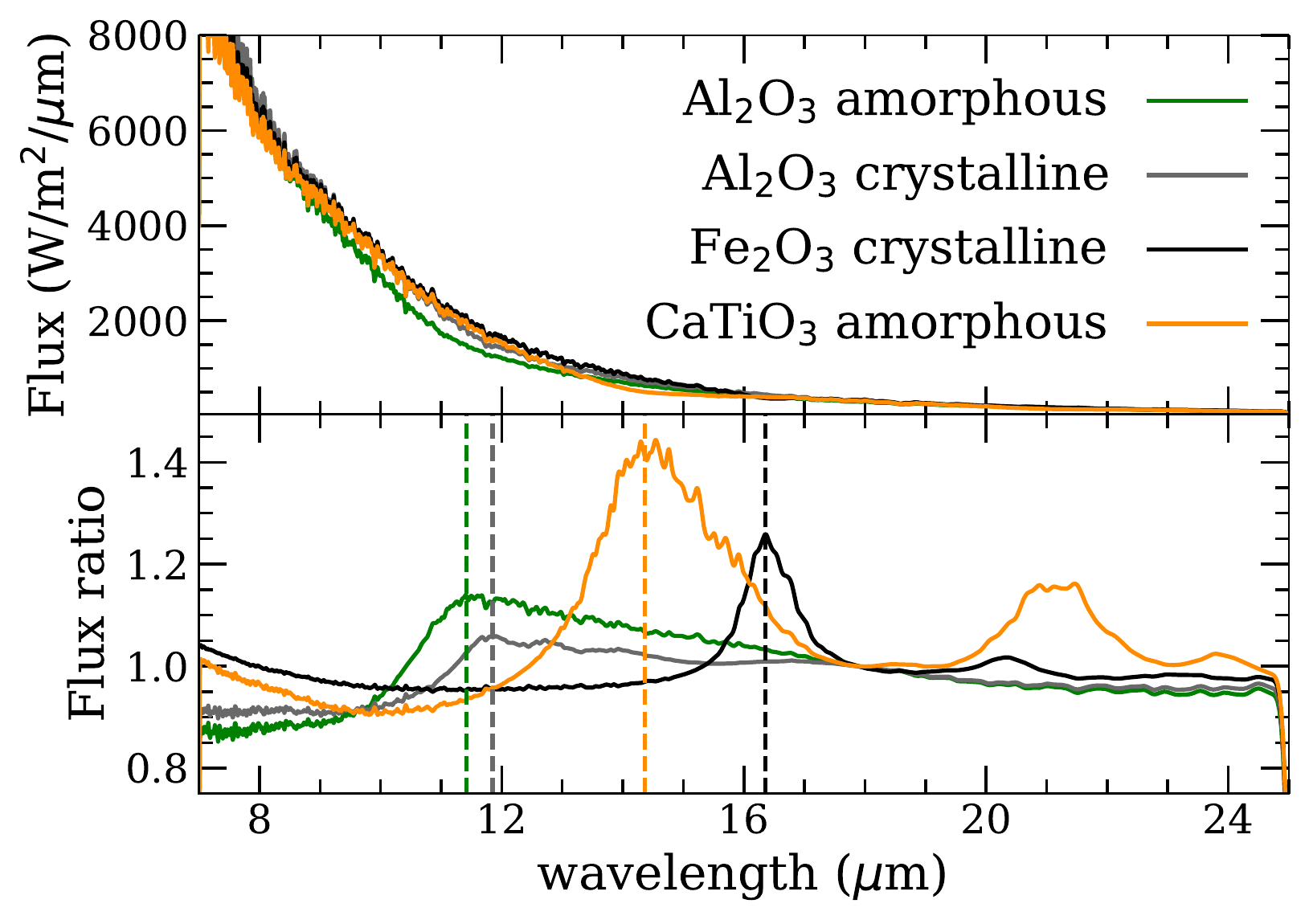}{0.5\linewidth}{}}   

         \vspace{-5mm}         
\caption{Thermal emission spectra and flux ratios for a range of particle sizes, compositions and crystalline structures to demonstrate how the cloud mineral absorption feature changes. 
\label{fig:fluxratios}We use a 2400 K, log g= 5, 1 $\mu$m model spectrum for the high-temperature clouds (bottom right). All other panels use an 1800 K, log g= 5, 1 $\mu$m model spectrum.
}
\end{figure*}

In Figures~\ref{fig:ptbright} and~\ref{fig:fluxratios}, we show spectra for each of our cloud mineral species. We show an 1800 K, log g= 5 model spectrum for the silicate and quartz species and a 2400 K, log g= 5 model spectrum for the high-temperature species as representative examples. In Figure~\ref{fig:ptbright}, we show the wavelength-dependent brightness temperature and pressure. This figure allows us to estimate which average pressure is probed at each wavelength. We compare to a cloud-free model and a standard cloudy (\fsed= 2, \citetalias{AM01}) model. In models with a low-pressure ad hoc cloud, the mid-IR flux is emitted from lower pressures in the atmosphere where the silicate clouds are placed. We show the mid-IR thermal emission spectra in Figure~\ref{fig:fluxratios} for the same set of models. 

To more easily visualize the impact of each cloud species on its corresponding thermal emission spectrum, we define two quantities: flux ratio and amplitude difference. 

\subsection{Flux Ratios \label{sec:fluxratio}}
To determine how the varying cloud optical depth with wavelength changes with spectra, for each model atmosphere, we run a corresponding model, $F_{\rm gray}$, where the cloud optical depth $\tau(\lambda)$, single scattering albedo $\omega_o$ and asymmetry parameter $g_o$ is taken to be the average optical depth $\overline{\tau}$, average single scattering albedo $\overline{\omega_o}$ and average asymmetry parameter $\overline{g_o}$ from 8 \um~ to 18 \um. This wavelength range covers the cloud mineral absorption features for silicates and refractory condensates. 
We plot the ratio of the cloudy model to the model with the gray cloud, 
\begin{equation}
 F_{\rm ratio}= \frac{F_{\rm gray}}{F_{\rm cloud}}.    
\end{equation}
The flux ratio allows us to determine the wavelengths of the strongest non-gray mineral features and systematically identify how that feature changes with respect to particle size and composition. 

\begin{figure}[h]
    \centering
    \includegraphics[width=1\linewidth]{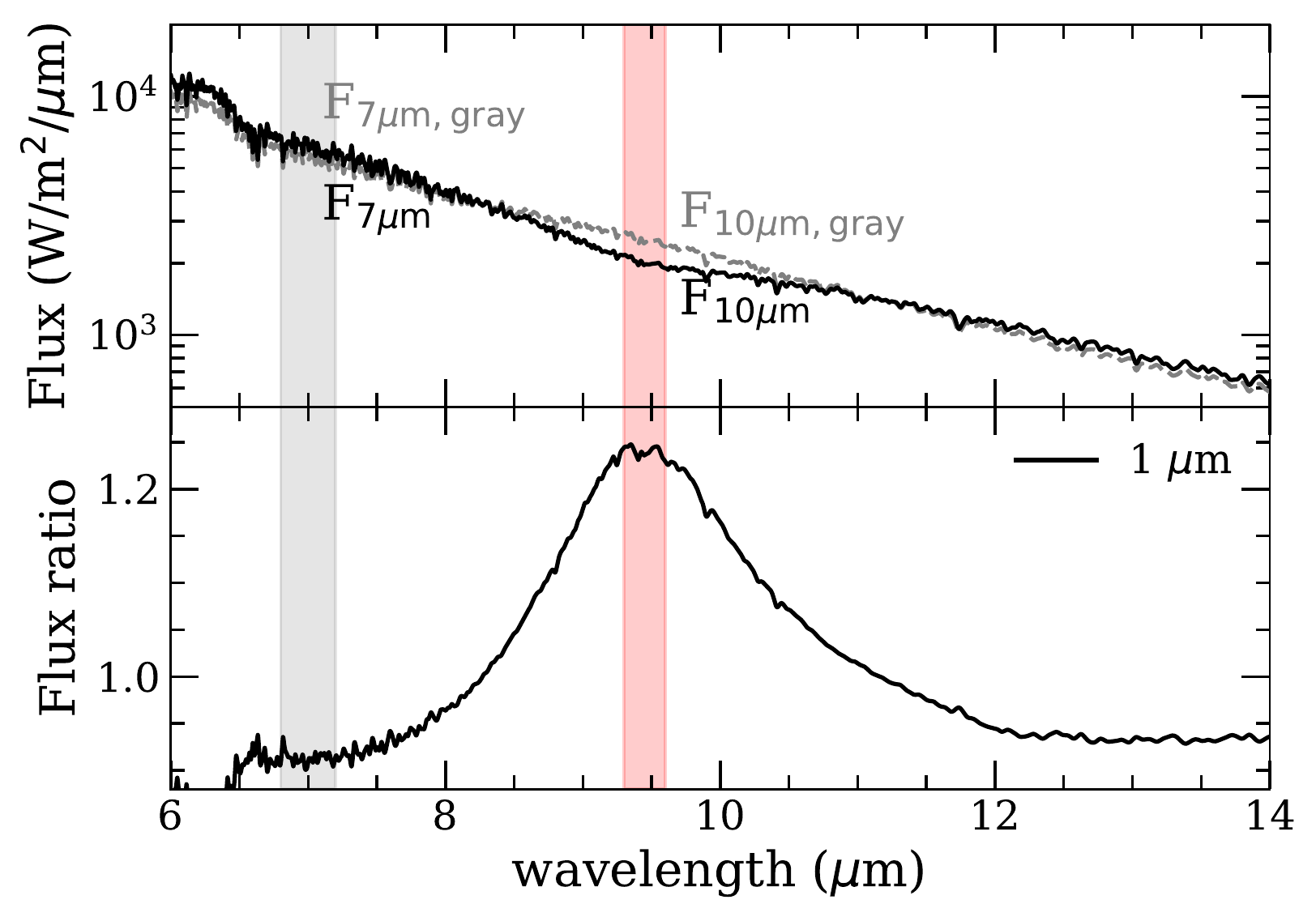}
    \caption{We demonstrate how we calculate the amplitude difference. The amplitude difference is simply the ratio of flux at two wavelengths: inside a mineral feature (shaded pink region) and outside (shaded gray region) where little variability is expected.}
    \label{fig:ampdiffex}
\end{figure}

Each cloud-forming mineral produces an absorption feature at a unique and particle size-dependent wavelength, potentially allowing us to distinguish between cloud particle size and compositions directly. In Figure~\ref{fig:fluxratios}, we show thermal emission spectra and flux ratio plots for a range of cloud compositions, sizes, and crystalline structures. We show a few representative examples to demonstrate the trends we find when changing the composition, size, and crystalline structure of the cloud model.

\subsection{The Amplitude Difference}
To quantify the strength of the cloud absorption features, we calculate the amplitude difference. We compute the difference in the ratios of flux inside and outside a mineral feature to the same model with the wavelength dependence averaged out (gray cloud), as described in Section \ref{sec:fluxratio}. We define the amplitude difference as, 
\begin{equation}
   {\rm amplitude~ difference} = \frac{F_{7~ \mu \rm m}}{F_{9-10 ~\mu \rm m}}-\frac{F_{7 ~\mu \rm m,~ gray}}{F_{9-10 ~\mu \rm m, ~gray}}. \label{eqn:ampdiff}
\end{equation}
We choose 7 $\mu$m as an appropriate reference wavelength where there should be little cloud impact. Figure~\ref{fig:ampdiffex}, shows a graphic representation of Equation~\ref{eqn:ampdiff}, with an 1800 K, log g= 5, 1 $\mu$m particle enstatite model. We also show the same model, with the wavelength dependence taken out (gray). 
The bottom panel of Figure~\ref{fig:ampdiffex}, shows the flux ratio inside and outside the wavelength windows of interest. 

\begin{figure}[h]
    \centering
    \includegraphics[width=1\linewidth]{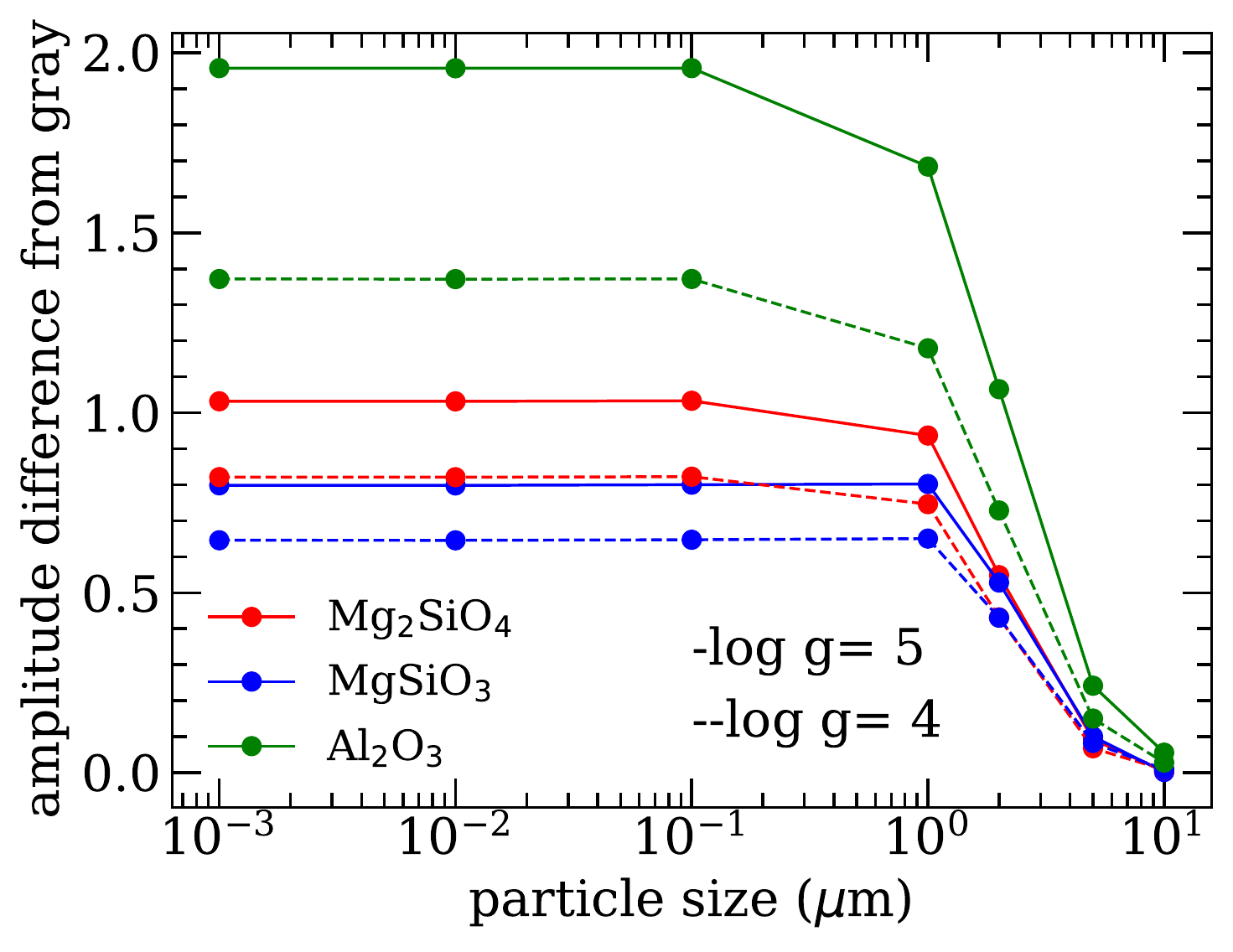}
    \caption{The amplitude difference of different particle sizes for forsterite, enstatite for an 1800 K model, and 2400 K for corundum. Particles that are 1 \um m in size or less, produce a strong silicate feature. When particles are larger than a few microns, the Mie scattering features disappear and yield gray absorption.}
    \label{fig:ampdiff}
\end{figure}

We calculate the amplitude difference for a range of particle sizes between 0.001 $\mu$m to 10 $\mu$m, shown in Figure~\ref{fig:ampdiff}, for enstatite, forsterite, and corundum. This figure summarizes the particle size dependence, where the Mie scattering effects occur only for particles smaller than a few microns.

\subsection{Particle Size Dependence}
We investigated the effect that particle size and composition have on thermal emission spectra. The top right panel in Figure~\ref{fig:fluxratios} shows how different sized particles can significantly change the spectrum in the mid-IR. This figure shows the spectra and flux ratios for a range of particle sizes from 0.001 $\mu$m to 10 $\mu$m.
We find that small particles 1 $\mu$m or less, produce the silicate feature, while particle sizes $\leq$ 0.1 $\mu$m produce the strongest feature. Once particles become larger than a few microns, the cloud becomes gray. Thus non-gray mineral features seen in observations are likely due to absorption from small particles.

\subsection{Enstatite versus Forsterite}
Rainout equilibrium calculations predict that forsterite forms a more massive cloud deeper in a brown dwarf atmosphere, with a less massive enstatite cloud at higher altitudes. The top left panel in Figure~\ref{fig:fluxratios} shows an amorphous 1 $\mu$m forsterite (red) and enstatite (blue) cloud model. Amorphous enstatite and forsterite have a distinct single absorption feature at $\sim$ 9.35 $\mu$m and $\sim$ 9.87 $\mu$m respectively. Since they are well separated in wavelength space, future mid-IR observations of brown dwarfs could distinguish between forsterite or enstatite.

\subsection{Amorphous versus Crystalline}
It has been typical in the past to use optical properties of amorphous structures (\citetalias{AM01}; \citealt{Morley12,Kitzmann18}), thus most minerals investigated are that of amorphous glass or measured using the sol-gel method \citep{Jaeger03}. Amorphous and crystalline structures produce distinct spectral features. Crystalline structures produce deeper, narrower absorption features than amorphous structures, assuming the same particle size distributions and number density. We currently do not know the crystalline structure of cloud particles in brown dwarfs. However, it has been assumed that the silicate particles are in an amorphous state and that they may be the precursors to crystalline particles \citep{Jaeger03}. If a cloud particle rises and sinks in the atmosphere, annealing could happen to amorphous particles and cause them to crystallize. If this process occurs, mid-IR observations of brown dwarfs with small particles can reveal the crystalline structure.

\subsection{Iron-rich Silicates}
Standard brown dwarf cloudy models include iron droplets, but recent microphysical studies have shown that iron clouds do not readily form \citep{2021gao}. Iron has a higher surface energy that creates a nucleation energy barrier preventing cloud formation \citep{2021gao}. However, FeH disappears from the spectra of late L-dwarfs \citep{Kirkpatrick05} suggesting that iron is indeed removed from the gas phase. We explore the hypothesis that the iron, instead of condensing into droplets, is instead incorporated into silicate clouds. When iron is introduced to enstatite or forsterite they become pyroxene or olivine, respectively. We show model spectra for pyroxene compared to enstatite in the middle left panel of Figure~\ref{fig:fluxratios} and demonstrate the trend we see for iron-rich silicate clouds. Increasing the iron content of either pyroxene or olivine pulls the absorption feature to bluer wavelengths. 

\subsection{Silicon oxides and Refractory Clouds}
The bottom two panels in Figure~\ref{fig:fluxratios} show the silicon oxide species and high-temperature refractory cloud species for 1 \um~particles. SiO uniquely absorbs at $\sim$ 9.43 \um~and crystalline and amorphous SiO$_2$ at $\sim$  8.83 \um~and $\sim$ 8.84 \um, respectively. The potential importance of these clouds is discussed further in Section \ref{sec:physicsofclouds}.

Refractory clouds are expected to form in low-mass M dwarfs and super-hot Jupiters. Corundum is often included in ``standard" brown dwarf cloud models \citep{AM01,Saumon08} but, a variety of different refractory species could potentially condense. The initial Al-bearing and Ca-bearing condensates depend on the pressure and temperature of the atmosphere with different condensates being favored at different pressures \citep{Lodders02b,Wakeford17}. For example, in a solar composition gas perovskite (CaTiO$_3$) and corundum (Al$_2$O$_3$) are favored at low-pressures, Ca$_4$Ti$_3$O$_{10}$ and hibonite (CaAl$_{12}$O$_{19}$) at intermediate pressures, and Ca$_3$Ti$_2$O$_7$ and grossite (CaAl$_4$O$_7$) at high pressures \citep{Burrows99,Lodders02b,Wakeford17}. Corundum has a distinct absorption feature at $\sim$ 11.42 \um~for amorphous corundum and $\sim$ 11.85 \um~for crystalline corundum.  Fe$_2$O$_3$ absorbs at $\sim$ 14.36 \um~and CaTiO$_3$ absorbs at $\sim$ 16.35 \um. These high-temperature refractory clouds can potentially be seen in hot L dwarf spectra. However, these clouds do not form as abundantly as silicate condensates since the limiting gas species for each condensate is less abundant.

\begin{figure*}[t]
\gridline{\fig{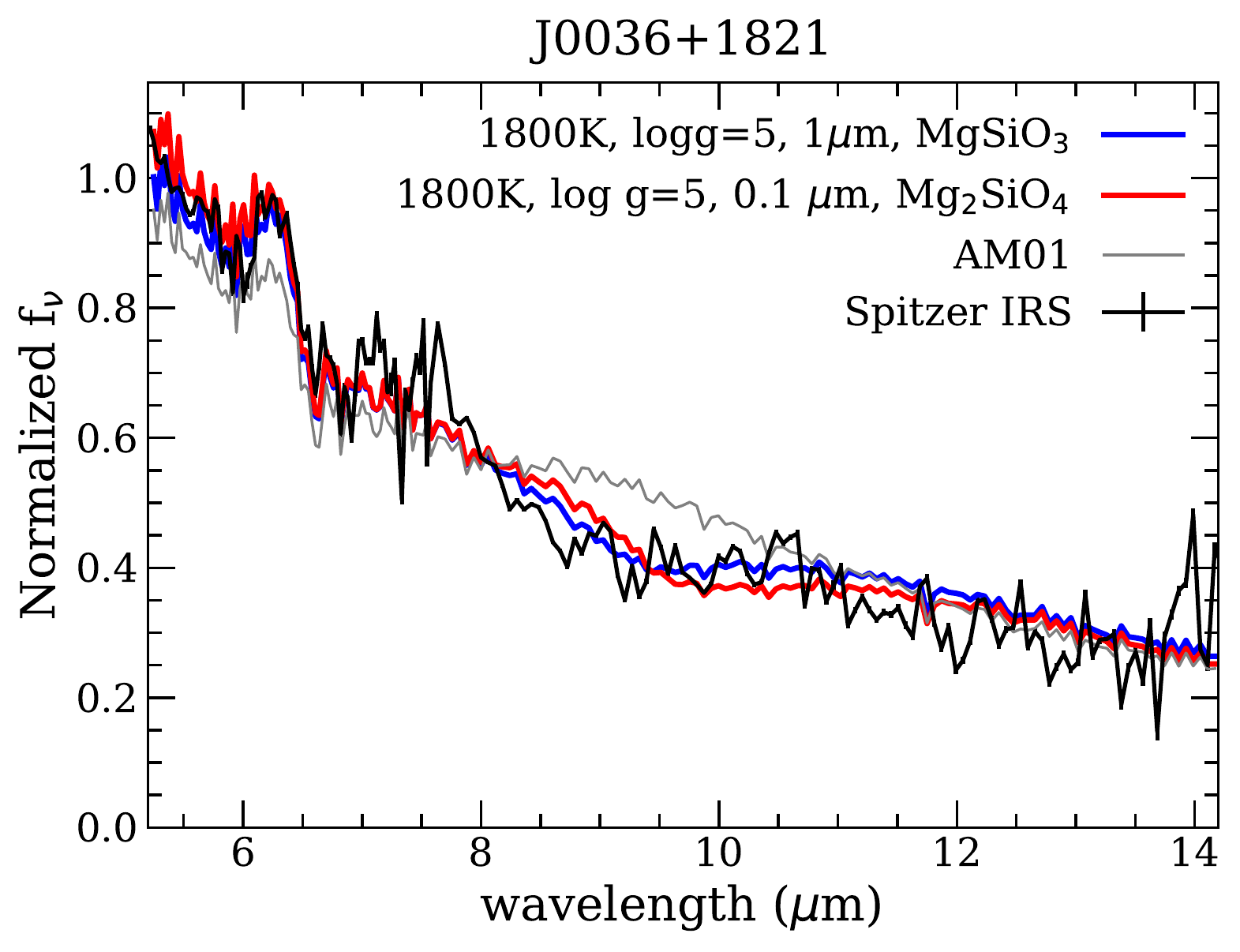}{0.5\linewidth}{}
         \fig{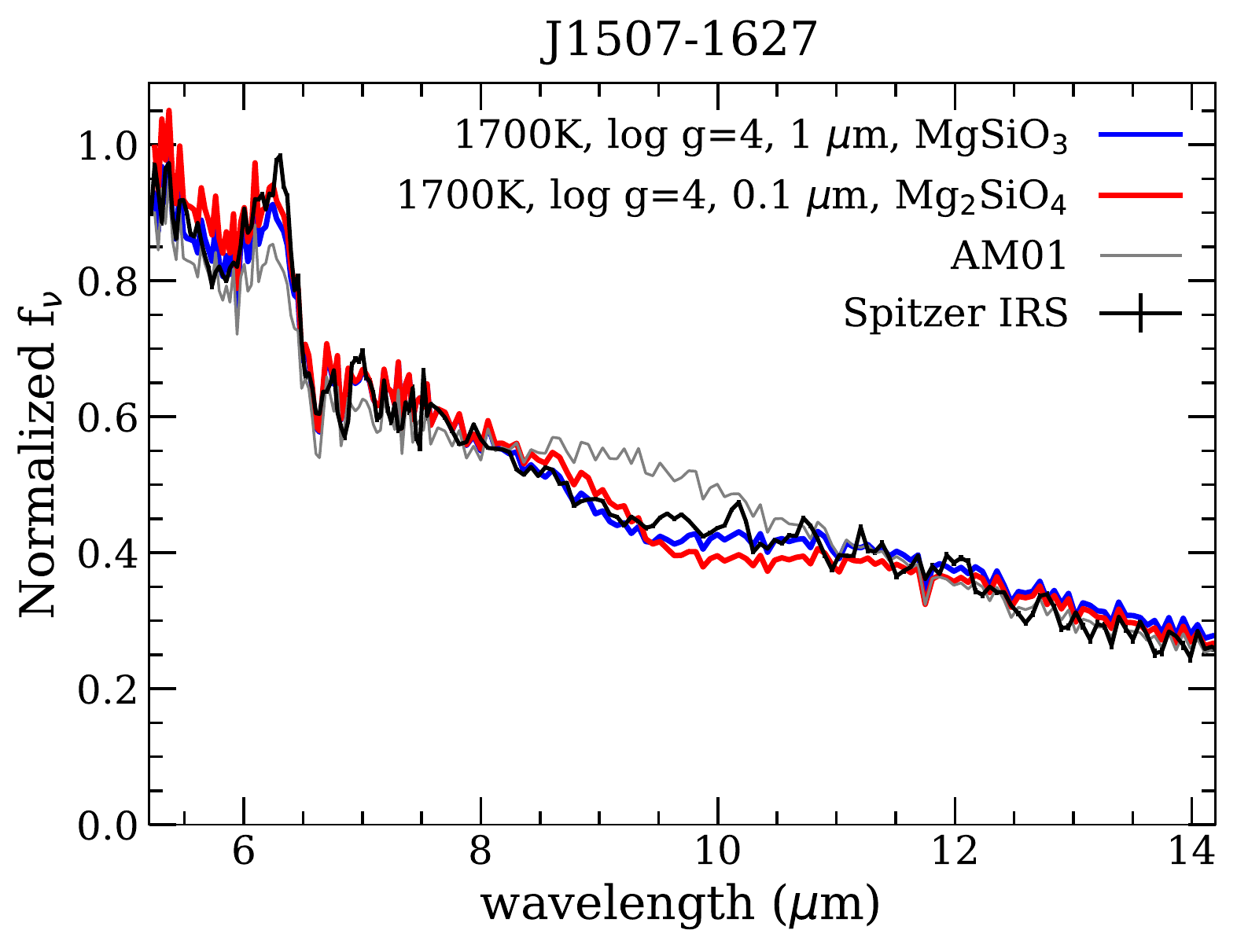}{0.5\linewidth}{}   }
   \vspace{-7mm}
\gridline{ \fig{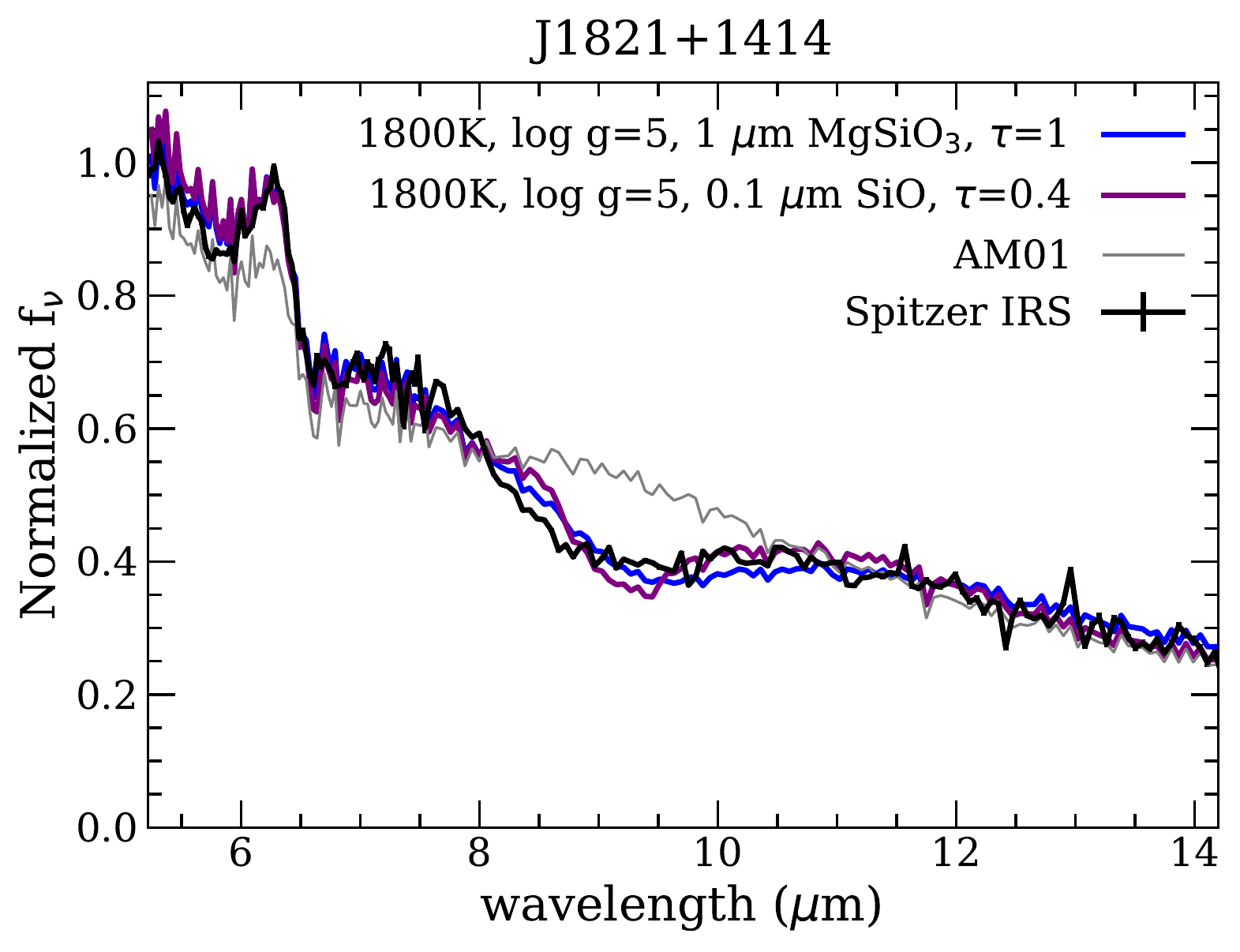}{0.5\linewidth}{}
         \fig{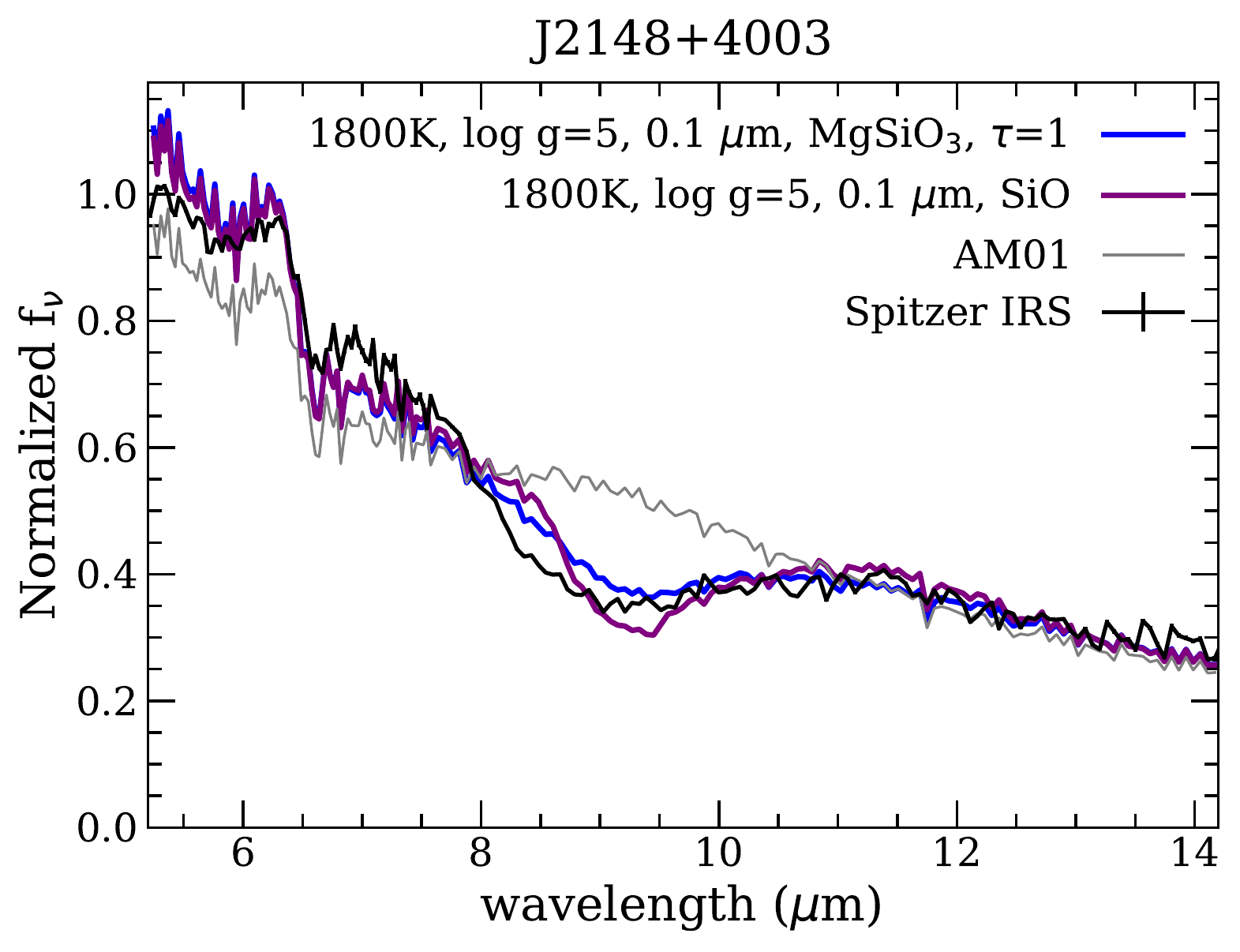}{0.5\linewidth}{}}   
         \vspace{-7mm}
  \centering  \includegraphics[width=0.5\linewidth]{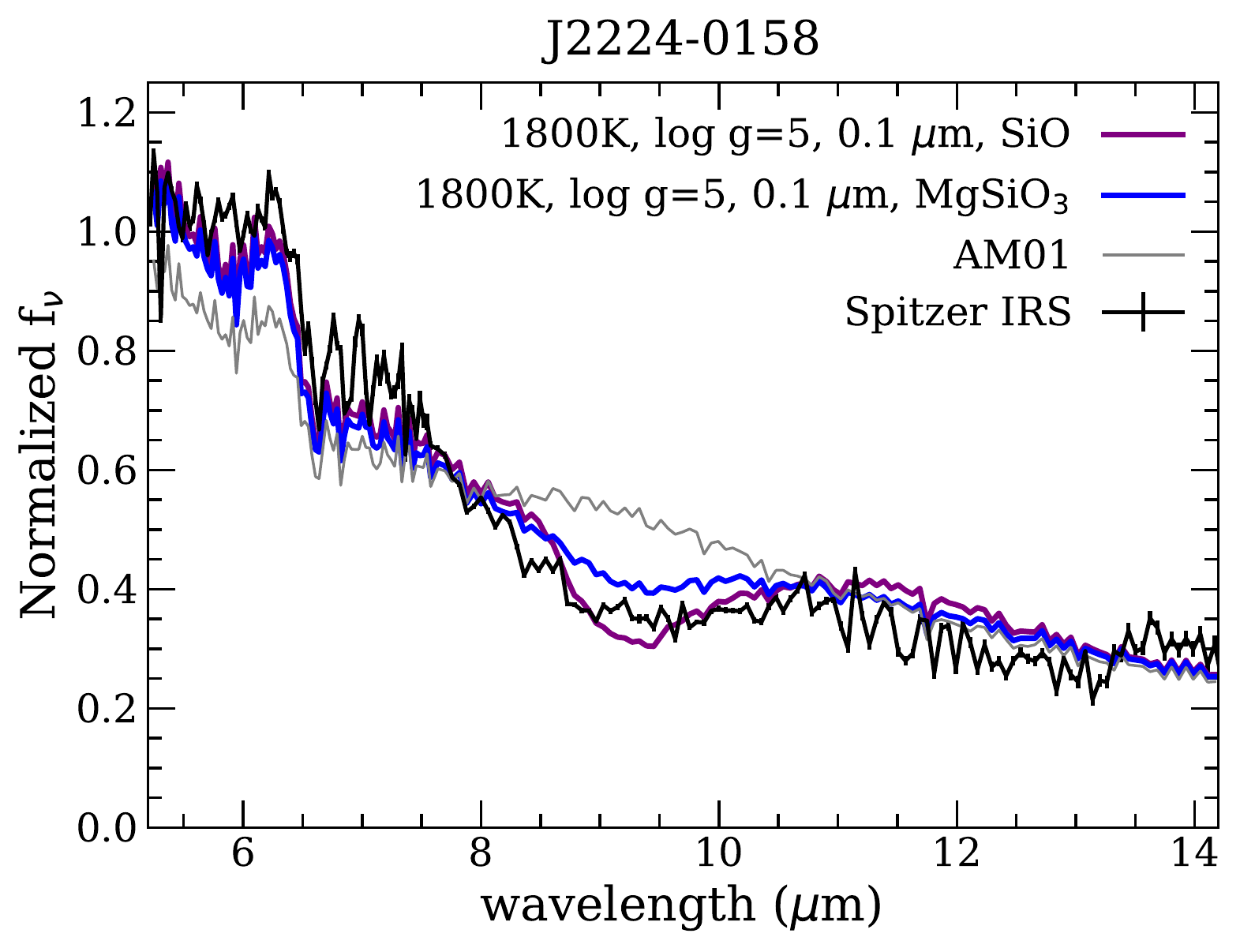}
\caption{Brown dwarf observations taken with the \emph{Spitzer} IRS  that tentatively show the silicate feature \citep{Cushing06,Looper08} at $\lambda / \Delta \lambda \sim$90. The gray line represents a standard \citetalias{AM01} cloud model for an \fsed= 2 with the temperature and surface of the corresponding best fitting model. The models are smoothed to the resolving power of the \emph{Spitzer} IRS spectra. The black line represents the \emph{Spitzer} IRS data with the observed errors. We also show the second best fitting model and list the $\chi ^2$ goodness of fit values in Table \ref{tab:Gkvalues}. Each cloud type is represented by a different color: enstatite in blue, forsterite in red and SiO in purple.}
\label{fig:compareobs_newmodels}
\end{figure*}

\section{Matching Observations to Models \label{sec:matchingobs}}
\subsection{Fitting {\it Spitzer} IRS Data with Models \label{sec:fittingobs}}
There are currently five brown dwarfs in the literature with a tentative silicate feature detection in their spectra using \emph{Spitzer} IRS (listed in Table~\ref{tab:targets}). \citet{Cushing06} used \emph{Spitzer} IRS to observe a sequence of low-mass objects and three objects show a tentative silicate feature. \citet{Looper08} observed two more peculiar brown dwarfs that both show broad 8-10 \um~absorption. One of these brown dwarfs, 2MASS J2148+4003, will be observed in \emph{JWST} cycle 1 under program ID: 2288 (PI: Joshua Lothringer, Co-PI: Jeff Valenti). 

We find the best fitting model for each brown dwarf with a tentative \emph{Spitzer} IRS silicate feature. We compare the model spectra to the data by computing the $\chi^2$ between the data and model defined as, 

\begin{equation}
    \chi^2 = \sum_{i=1}^n \left( \frac{f_i- F_{k,i}}{\sigma_i}\right)^2,
\end{equation}
The model that corresponded to the smallest $\chi^2$ was chosen as the best-fitting model.
 
The \emph{Spitzer}/IRS spectra have $\lambda / \Delta \lambda \sim$ 90 and each model spectrum is smoothed to this resolving power. In the Appendix, in Table~\ref{tab:Gkvalues}, we list the chi-squared value for all models fit to each brown dwarf. As a simple comparison between models, we also include whether the change in goodness of fit value is large enough to statistically reject models with worse fits than our best-fit model for each object. We highlight which models we can reject at the 3-$\sigma$ level, assuming 184 degrees of freedom (189 data points in each IRS spectrum; 5 model parameters fit), though we note that no models provide perfect fits to the data. We note that this is a crude assessment using a coarse grid of models; future work could use more rigorous retrieval techniques to assess a more finely-sampled set of model parameters as pioneered by \citet{burningham21} for cloudy L dwarfs. However, this simple and fast technique allows us to quickly assess how well different cloud species can fit the observed data. 

\textit{2MASS J0036+1821} was best fit by an 1800 K, log g= 5, 0.1 $\mu$m enstatite cloud. We show the best fitting model in the top left panel of Figure~\ref{fig:compareobs_newmodels} and an \citetalias{AM01} model with 1800 K, log g= 5, \fsed= 2. 

\textit{2MASS J1507-1627}  was best fit by an 1700 K, log g= 4, 1 $\mu$m enstatite cloud. We show the best fitting model in the top right panel of Figure~\ref{fig:compareobs_newmodels} and an \citetalias{AM01} model with 1700 K, log g= 4, \fsed= 2. 

\textit{2MASS J1821+1414}  was best fit by an 1800 K, log g= 5, 1 $\mu$m enstatite cloud with a total cloud optical depth (somewhat higher than the standard $\tau_{\rm cloud}=$ 0.67 set for the bulk of our models) $\tau_{\rm cloud}=$ 1 . We show the best fitting model in the middle left panel of Figure~\ref{fig:compareobs_newmodels} and an \citetalias{AM01} model with 1800 K, log g= 5, \fsed= 2.

\textit{2MASS J2148+4003}  was best fit by an 1800 K, log g= 5, 1 $\mu$m enstatite cloud with a total cloud optical depth of $\tau_{\rm cloud}=$ 1. We show the best fitting model in the middle right panel of Figure~\ref{fig:compareobs_newmodels} and an \citetalias{AM01} model with 1800 K, log g= 5, \fsed= 2. \citet{Looper08} concluded that 2MASS J2148+4003 is possibly metal-rich, and the larger discrepancy between the model and data for this object may be attributed to the assumed solar composition atmosphere. Future studies should aim to compute higher-metallicity models to better capture the feature. 

\textit{2MASS J2224-0158} was best fit by an 1800 K, log g= 5, 0.1 $\mu$m SiO cloud. We show the best fitting model in the bottom panel of Figure~\ref{fig:compareobs_newmodels} and an \citetalias{AM01} model with 1800 K, log g= 5, \fsed= 2. Similarly to J2148+4003, this brown dwarf may be metal-rich \citep{burningham21}.

Interestingly, we find that 4 of the 5 objects studied are better fit by an amorphous enstatite cloud than a forsterite cloud, which equilibrium rainout models predict to be more massive. This lends credence to the idea that because enstatite condenses at lower pressures, it may dominate spectroscopically. 

\subsection{Simulating Spectra with \emph{JWST}/MIRI}
We simulate spectroscopic observations of brown dwarfs with a tentative \emph{Spitzer} silicate detection using the \emph{James Webb Space Telescope} (\emph{JWST}). We use the publicly available \emph{JWST} exposure time calculator (hereafter, \emph{JWST} ETC) that calculates the signal-to-noise ratio for all \emph{JWST} observing modes \citep{Pontoppidan16} (\href{JWST/ETC}{https://jwst.etc.stsci.edu}). We simulate spectroscopic observations with the MIRI instrument, in MRS mode, which is capable of observing the full wavelength range of mineral features found in L dwarfs. 

To determine the signal-to-noise ratio for each object of interest we use the best fitting model from Section \ref{sec:fittingobs} as inputs to the \emph{JWST} ETC. The model spectra give the flux as a function of wavelength at the surface of a brown dwarf. To obtain the flux at Earth, the best fitting model spectra must be multiplied by $(R/D)^2$, where $R$ is the radius of the brown dwarf and $D$ is the distance to the brown dwarf. We use the radius and distance for each brown dwarf from literature estimates (listed in Table~\ref{tab:targets}).

\begin{figure*}
  \centering  \includegraphics[width=1.3\linewidth,angle=90]{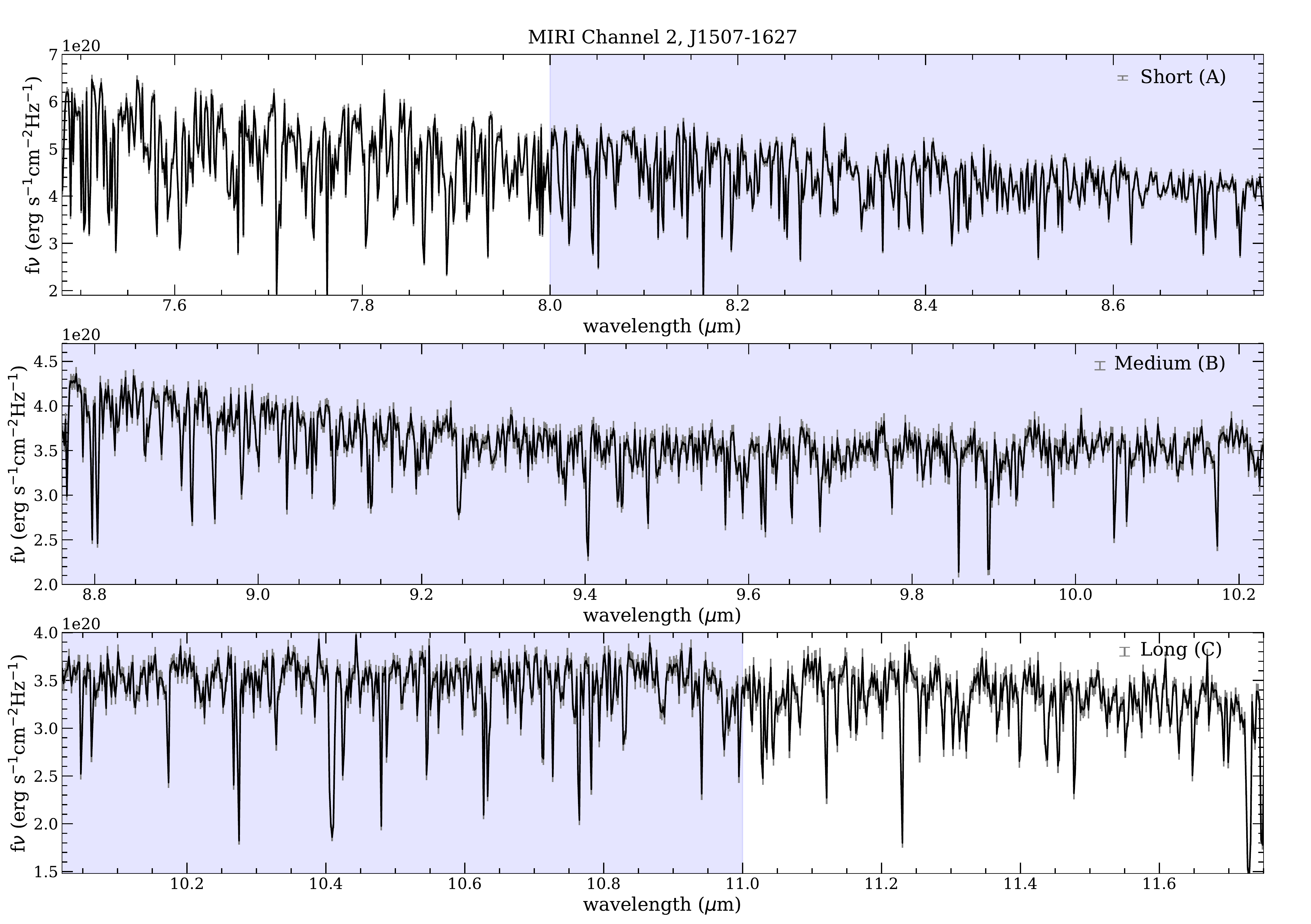}
\caption{\emph{JWST}/MIRI Channel 2 spectra for 2MASS J1507-1627 for a 837s exposure. The error bars are plotted in gray and we show an example errorbar to scale in the upper right, before the channel names. The area of the spectrum affected by cloud opacity from \enst~is shaded and this region of the spectrum is largely shaped by gas opacity from H$_2$O.
} \label{fig:jwstmiri_ch2}
\end{figure*}

We simulated data based on MIRI MRS with the first three channels. For each best-fitting model spectrum, we simulate data using 2 exposure times for the first 3 Channels in MIRI MRS: 837s and 2220s. In Figure \ref{fig:jwstmiri_ch2}, we show a simulated \emph{JWST}/MIRI Channel 2 spectrum for 2MASS J1507-1627 for a 837s exposure. The area of the spectrum affected by cloud opacity from \enst~is shaded. This region of the spectrum is largely shaped by the gas opacity from H$_2$O but CH$_4$, SiO, CO$_2$ and H$_2$S can also be detectable in the MIRI bandpasses if present in the atmosphere. In each panel we also plot the errorbar to scale in the upper right, before the channel names.

The results presented in Section \ref{sec:fittingobs} and \citet{burningham21} demonstrate the \emph{Spitzer}/IRS-like resolution ($\sim$90) and S/N (typically $\sim$ 40 per resolution element) are adequately high for characterizing the broad features of silicate clouds. Observations with \emph{JWST} will provide several advantages that we will briefly discuss here. 

\emph{JWST}/MIRI can achieve the same signal to noise (SNR) per spectral element as \emph{Spitzer}/IRS in roughly the same integration time, but with 30 times the spectral resolution. For example, in a 16 minute exposure on either \emph{JWST}/MIRI (at 9.4\um) or \emph{Spitzer}/IRS yields a signal to noise of $\sim$40 per spectral element. The increase in spectral resolution will potentially allow us to detect trace species like SiO, CH$_4$, CO$_2$, and H$_2$S at mid-infrared wavelengths (see Figure \ref{fig:jwstmiri_ch2}). 

Binning \emph{JWST}/MIRI observations to \emph{Spitzer}/IRS-like resolutions, would enable us to characterize more distant objects. To achieve the same S/N$\sim$40 in 16 minutes, we could observe objects $\sim5.5\times$ further away. For context, the most distant mid-L dwarf with a silicate feature observed with \emph{Spitzer} was 2MASS~J2224438-015852 with a distance of 11.5 parsec \citep{Stephens09}. JWST will enable more observations of fainter and more distant brown dwarfs. We can potentially observe dozens more brown dwarfs at high signal-to-noise, allowing us to target interesting and more rare targets, such as very young brown dwarfs, or those with high or low metallicities, at a range of viewing geometries. 

Lastly and perhaps obviously, \emph{Spitzer} stopped operating cryogenically in 2009. Many brown dwarfs have been discovered since then, and many interesting fields of brown dwarf research have emerged or matured in that time, including studies of their variability \citep{Metchev15, vos19, vos20}, obliquities \citep{Bryan20}, and characteristics of youth \citep{vos19,vos20}. Directly studying the cloud optical properties will inform, and be informed by these fields.

\subsection{Simulating Time-Series Spectra with JWST/MIRI}
Four brown dwarfs (2MASS J0036+1821, 2MASS J1507-1627, 2MASS J1821+1414, and 2MASS J2148+4003) with a tentative \emph{Spitzer} silicate detections are also variable, suggesting that their atmospheres are inhomogeneous, potentially with patchy clouds \citep{Bailer-Jones01,Gelino02,Radigan14,Metchev15}. One way to determine the cause of brown dwarf variability is by studying the variability of the silicate feature itself. 

\citet{Metchev15} used the [3.6] and [4.5] \emph{Spitzer} IRAC  channels to determine the amplitudes and periods of variability for field brown dwarfs (including the brown dwarfs in Section \ref{sec:fittingobs}). We use the amplitudes at [3.6] and [4.5] and periods to guide simulations of variability in the mid-IR.

We simulate time-series observations of variable brown dwarfs with a tentative silicate detection on the \emph{James Webb Space Telescope} (\emph{JWST}). We use the signal-to-noise estimates from Section 4.2 for the short exposure time (837s) to simulate time-series observations with the MIRI instrument.

It is unknown which clouds are causing the brightness variations seen with \emph{Spitzer}. There are two scenarios that we explore: one cloud changing and causing brightness variations or both clouds changing. 

\begin{itemize}
    \item One cloud varying: In this case, the deep \citetalias{AM01} cloud is static while the ad hoc cloud in the upper layers of the atmosphere is undergoing weather. 
    \item Two clouds varying: In this case, both the deep \citetalias{AM01} cloud and the ad hoc cloud in the upper layers are undergoing weather and causing brightness variations. 
\end{itemize}

To model one cloud varying, we calculate and sum the flux from two cloudy models. The first is an \fsed= 2 \citetalias{AM01} cloud that becomes optically thick at 0.4 bar. The second includes an ad hoc cloud model above at 0.011 bar, as described in Section \ref{sec:adhoccloud}. We linearly combine the two cloudy models, varying the fraction of the \citetalias{AM01} and ad hoc cloud models. If one hemisphere is cloudier than the other, we will observe varying brightness as the two cloudy regions come in and out of view. We model the total flux from the two cloudy components as follows,
\begin{equation}
    F_{\rm total}= {\rm f} ~F_{\rm (ad~ hoc~+~AM)} + (1-{\rm f})~F_{\rm AM}
\label{eqn:ftotal}
\end{equation}
where f is the fractional cloud coverage, $F_{\rm AM}$ is the flux from the \citetalias{AM01} cloud, and $ F_{\rm (ad~ hoc~+~AM)}$ is equal to the flux from the ad hoc cloud. 

\begin{figure}[h]
\centering
    \includegraphics[width=1\linewidth]{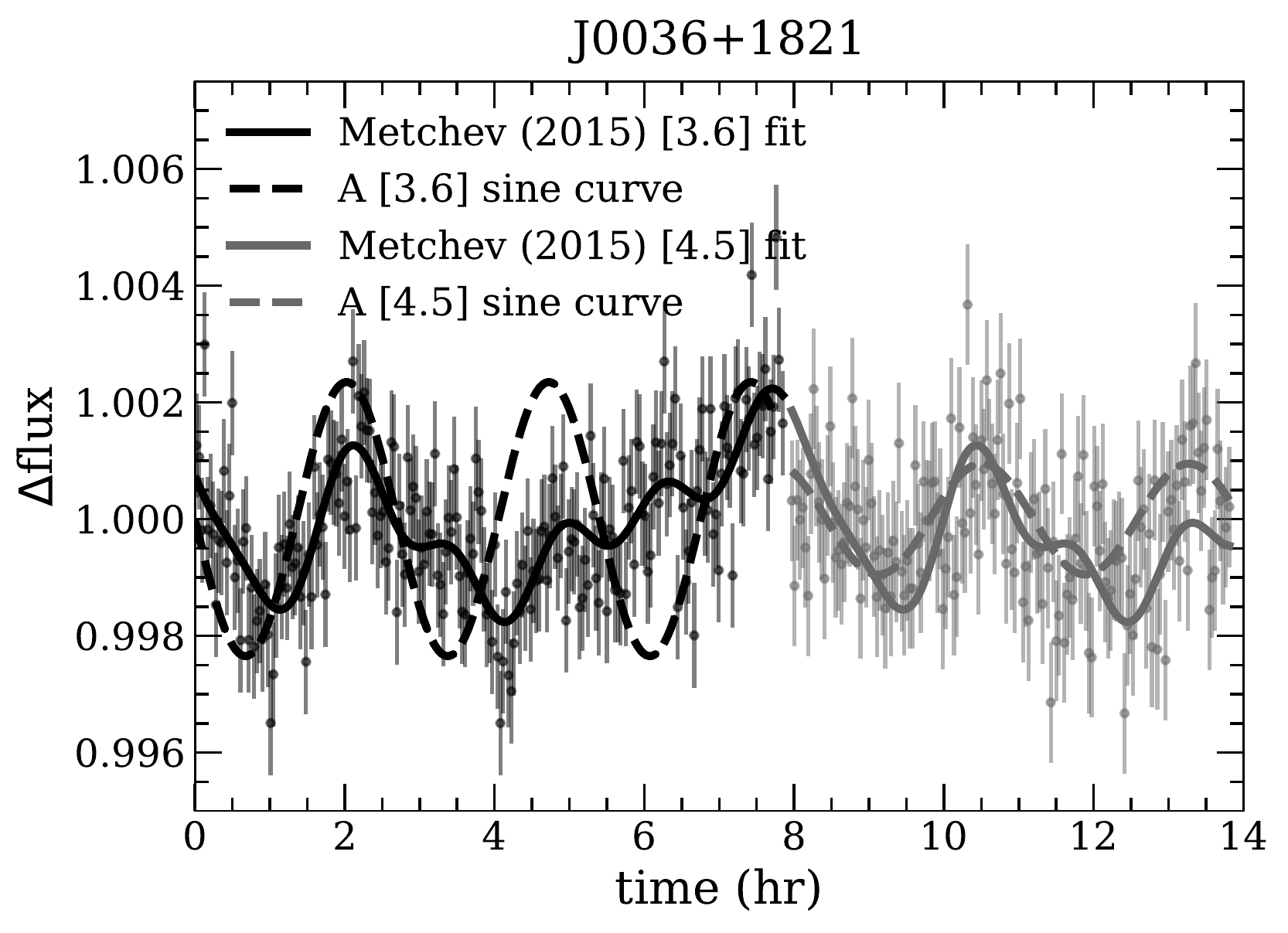}
    \includegraphics[width=1\linewidth]{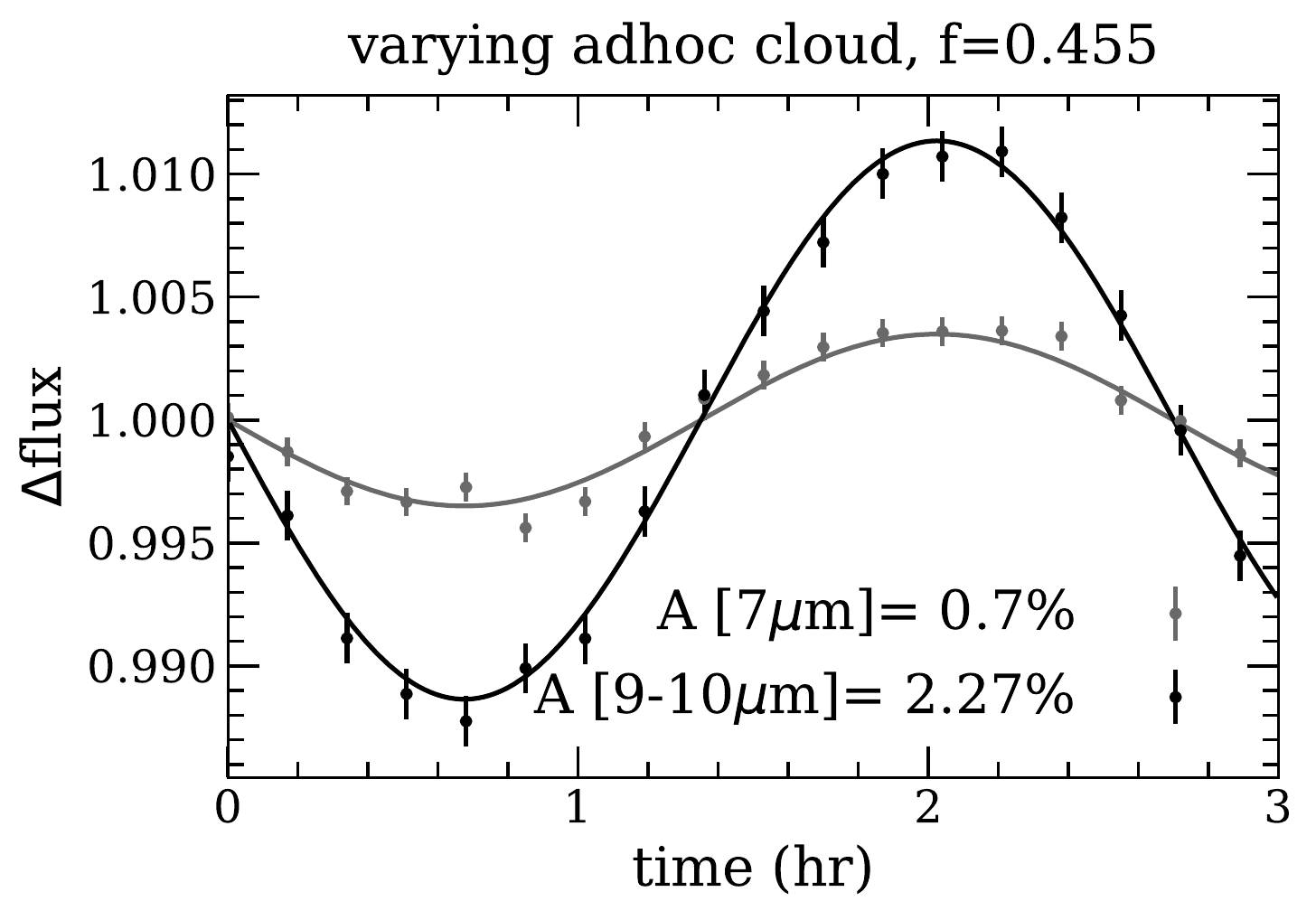}
    \includegraphics[width=1\linewidth]{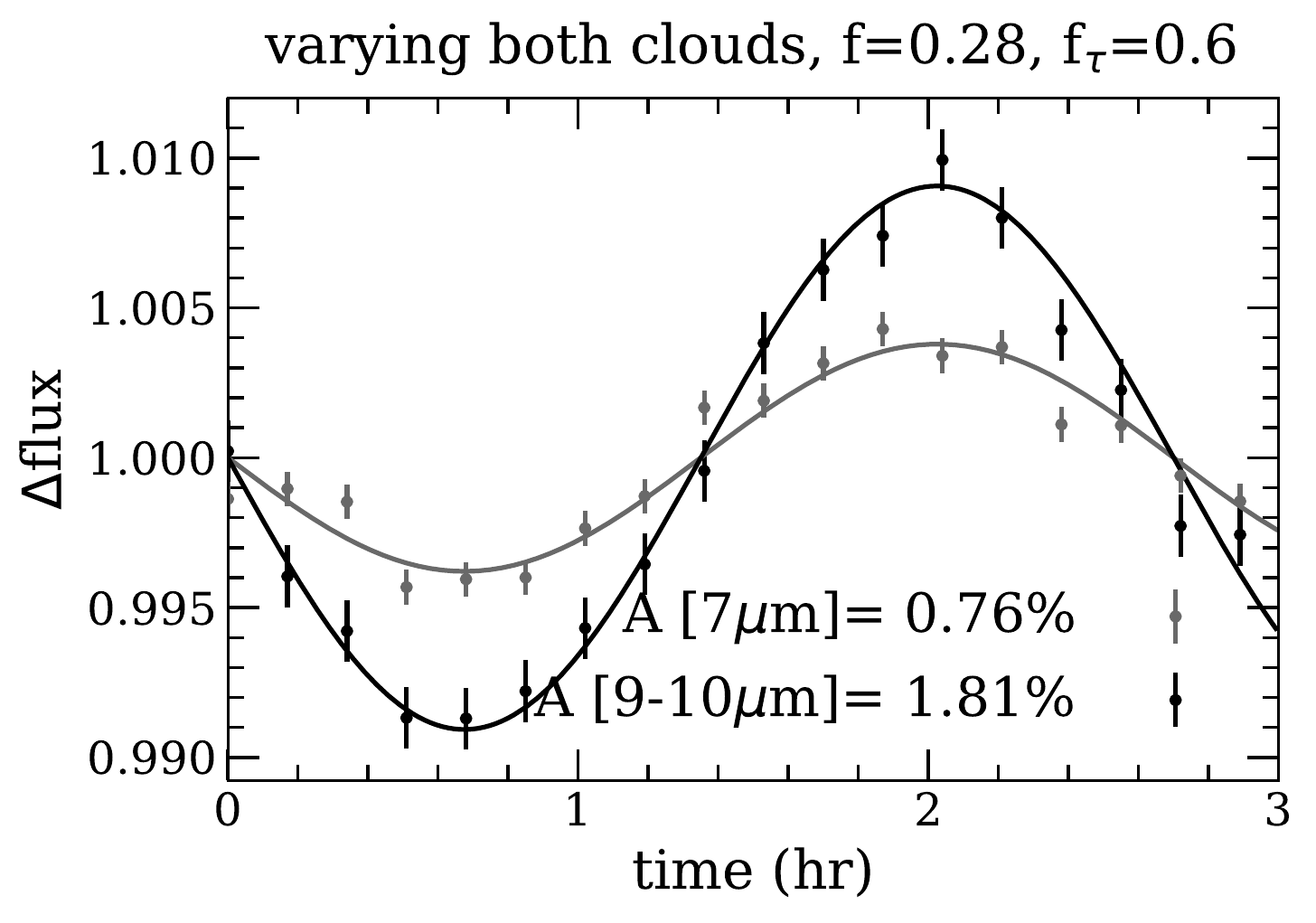}
    \caption{
    \label{fig:J0036fluxvstime}\emph{Spitzer} light curves from \citet{Metchev15}
   for J0036+1821 (top). The IRAC [3.6] points are in black and IRAC [4.5] points are in gray with the corresponding fit. We also show in black (dashed) and gray(dashed), the simplistic sine curve model that use the period and amplitude in Table~\ref{tab:targets}. We show the predicted mid-IR variability for J0036+1821 by varying the ad hoc cloud (middle) and by varying both clouds (bottom). }
\end{figure}

To model two clouds varying, we assume both the ad hoc and \citetalias{AM01} cloud varies. To model both components as varying, we replace $F_{\rm AM}$, in Equation~\ref{eqn:ftotal}, with an alternative $F(\tau \times {\rm f}_\tau)_{ \rm ad~ hoc}$, where the cloud optical depth in each layer is multiplied by some fraction, f$_\tau$. In this way, we are varying both the \citetalias{AM01} and the ad hoc cloud model. 
We use the fraction that can best match the observed \emph{Spitzer} amplitudes of variability to simulate the variability in the mid-IR. We use a toy model sine curve to simulate the brightness variations we can expect from the mid-IR using the MIRI instrument on \emph{JWST}. This is a simplistic version of the modeling by \citet{Metchev15} and non-sinusoidal variations in the amplitude and phase are to be expected. We show the light curves and fit from \citet{Metchev15} in the top panel of Figures~\ref{fig:J0036fluxvstime} and~\ref{fig:J1507fluxvstime}. To illustrate our simple toy model, we over plot the simulated [3.6] and [4.5] amplitudes of variability. This sine curve simply uses the period and amplitude in Table~\ref{tab:targets}.

We now describe the steps to simulate the mid-IR variability. 
\begin{itemize}
    \item  Find the fraction, f, that best reproduces the observed \emph{Spitzer} amplitudes of variability. 
    \item  If possible find the fractions, f, and f$_\tau$, that can reproduce the observed \emph{Spitzer} amplitudes of variability. 
    \item Using the fraction(s), we calculate the amplitudes of variability inside and outside the silicate feature. 
\end{itemize}

We start by finding the correct fraction, f, that reproduces the observed amplitudes of variability in IRAC channels 1 and 2 from \citet{Metchev15}. To do this we use the best fitting model from Section \ref{sec:fittingobs} and a standard \fsed= 2, \citetalias{AM01} cloud model. Using Equation~\ref{eqn:ftotal}, we combine the flux from the cloudy models using a range of fractions f, 0.1 - 0.5 in steps of 0.005. We then integrate the flux in the first two IRAC filter channels [3.6] and [4.5] and compute the amplitudes in the two channels. We then choose the fraction that yielded the smallest $\chi^2$ to the observed \emph{Spitzer} amplitudes of variability and associated errors listed in Table~\ref{tab:targets}. In the case where both clouds are varying, we chose the combination of f and f$_\tau$ that yielded the smallest $\chi^2$. We explored a range of f$_\tau$'s, 0.1 - 0.9 in steps of 0.1. 

Next, we describe the steps we use to simulate the variability in the mid-IR. We start by modeling the clouds as varying using the fraction that best reproduces the observed \emph{Spitzer} amplitudes of variability. Using this best fraction, we calculate the amplitudes of variability from 9 \um~- 10 \um~(A [9-10 $\mu$m]) and from 6.5 \um~to 7.5 \um~(A [7 $\mu$m]). We then attempt the same process, but by varying both clouds. We calculate the A [3.6 $\mu$m] and A [4.5 $\mu$m] for each combination of fractions f (0.1 - 0.5 in steps of 0.005) and f$_\tau$ (0.1 - 0.9 in steps of 0.1). Using the fractions, f and f$_\tau$, that gives the smallest $\chi^2$, we calculate A [7 $\mu$m] and A [9-10 $\mu$m]. We should expect that the variability inside the silicate feature (9-10 $\mu$m) to be larger than the variability at a reference wavelength, which we choose as 7 $\mu$m. 

\begin{deluxetable*}{cccccccc}
\tablecolumns{8}
\tablewidth{0pt}
\tablecaption{ Bright L dwarfs with a tentative silicate detection. \label{tab:targets}}
\tablehead{
\colhead{Name} & \colhead{SpT} & \colhead{J} &\colhead{ R } & \colhead{D }  &\colhead{P} & \colhead{A[3.6]} &\colhead{A[4.5]}\\ \vspace{-0.2cm}
& &  \vspace{-0.2cm}  & \colhead{R$_\odot$} &  \colhead{pc} &  \colhead{hrs} & \colhead{$\%$}& \colhead{$\%$}  \vspace{-0.2cm}\\ }
\startdata
2MASS J0036159$+$182110  & L3.5 & 12.47 & 0.11  & 8.8 $\pm$ 0.1 &2.7 & 0.47 $\pm$ 0.05 &  0.19 $\pm$ 0.04  \\
2MASS J1507476-162738 & L5   & 12.83 & 0.08 & 7.3 $\pm$ 0.03 &2.5  & 0.53 $\pm$ 0.11& 0.45 $\pm$ 0.09\\
2MASS J18212815+1414010 & L4.5   & 13.43 & 0.09$^{\rm a}$ & 9.3568 $\pm$ 0.022$^{\rm a}$ & 4.2  & 0.54 $\pm$ 0.05 & 0.71 $\pm$ 0.14\\
2MASS J21481628+4003593 & L6   & 14.15 & 0.10$^{\rm b}$ & 7 $\pm$ 1$^{\rm c}$ & 19  &  1.33 $\pm$ 0.07 & 1.03 $\pm$ 0.10 \\
2MASS J2224438-015852  & L4.5   & 14.07 & 0.12 (0.11 - 0.13) & 11.5 $\pm$ 0.1 & - & - & -\\
\enddata
\tablecomments{Periods and amplitudes of variability for the 3.6 $\mu$m and 4.5 $\mu$m from \citet{Metchev15}.
\\Radius and Distance from \citet{Stephens09} unless noted otherwise. 
\\2MASS J2224438-015852 is not variable. 
\\{\bf References:} (a) \cite{sebastian2021}; (b) \cite{vos20}; (c) \cite{faherty2009}.
}
\end{deluxetable*}

\subsubsection{2MASS J0036+1821} 
We were able to reproduce the observed amplitudes of variability for this object by varying one or two clouds. 
The fraction f $=$ 0.455 gives A [3.6 $\mu$m]$=$ 0.52$\%$ and A [4.5 $\mu$m]$=$ 0.16$\%$. The predicted variability in the mid-IR is shown in the top panel of Figure \ref{fig:J0036fluxvstime}. The variability inside the silicate feature is $\sim$ 4 times the variability outside the feature. 

The combination of fractions f $=$ 0.28 and f$_\tau=0.6$ gives A [3.6 $\mu$m]$=$  0.46$\%$ and A [4.5 $\mu$m]$=$ 0.20$\%$. The predicted variability in the mid-IR is shown in the bottom panel of Figure~\ref{fig:J0036fluxvstime}. Although the variability is larger in the silicate feature, the difference in amplitudes at each wavelength decreases.

\subsubsection{2MASS J1507-1627} 
\begin{figure}[h]
\centering
    \includegraphics[width=1\linewidth]{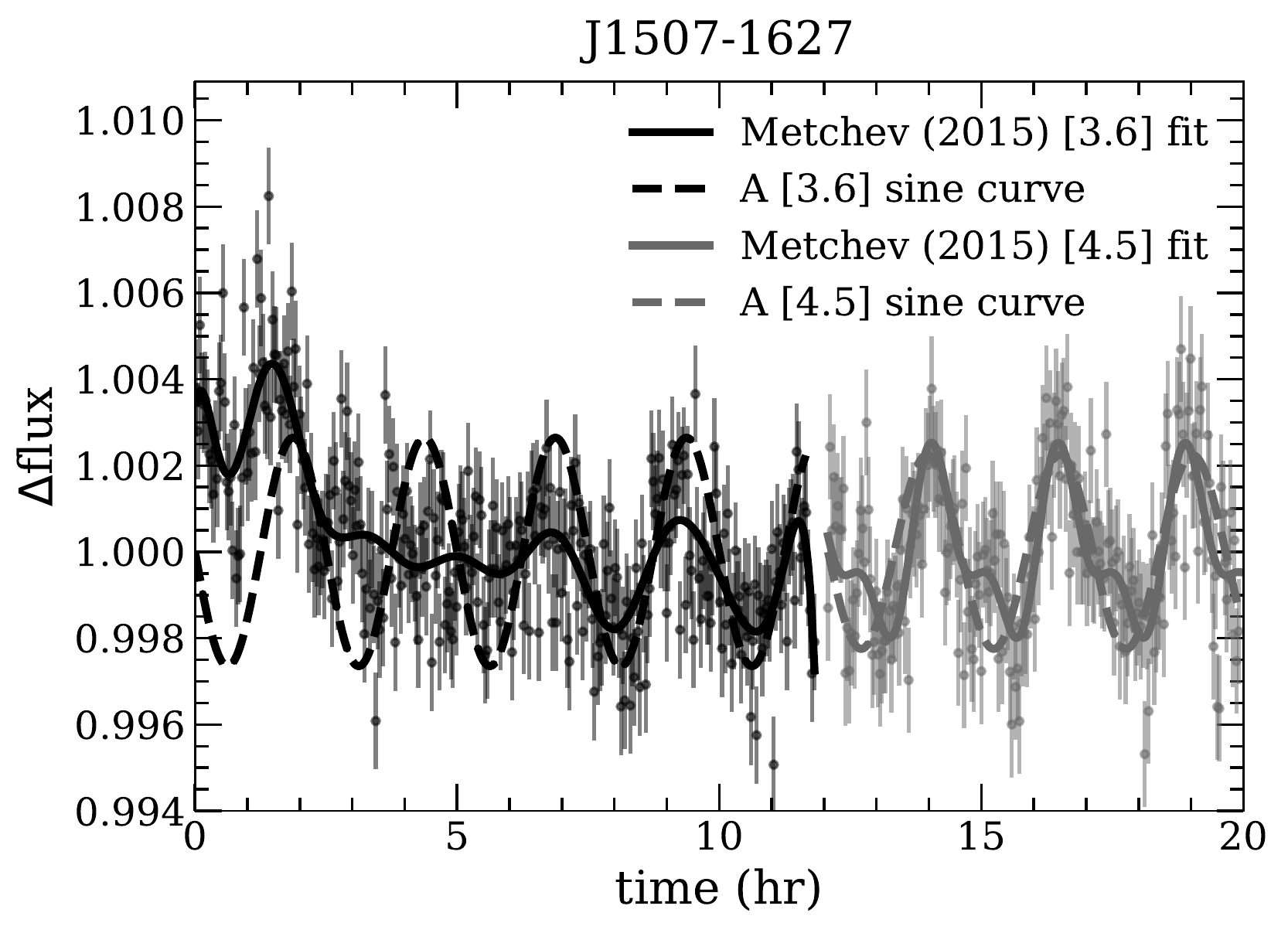}
    \includegraphics[width=1\linewidth]{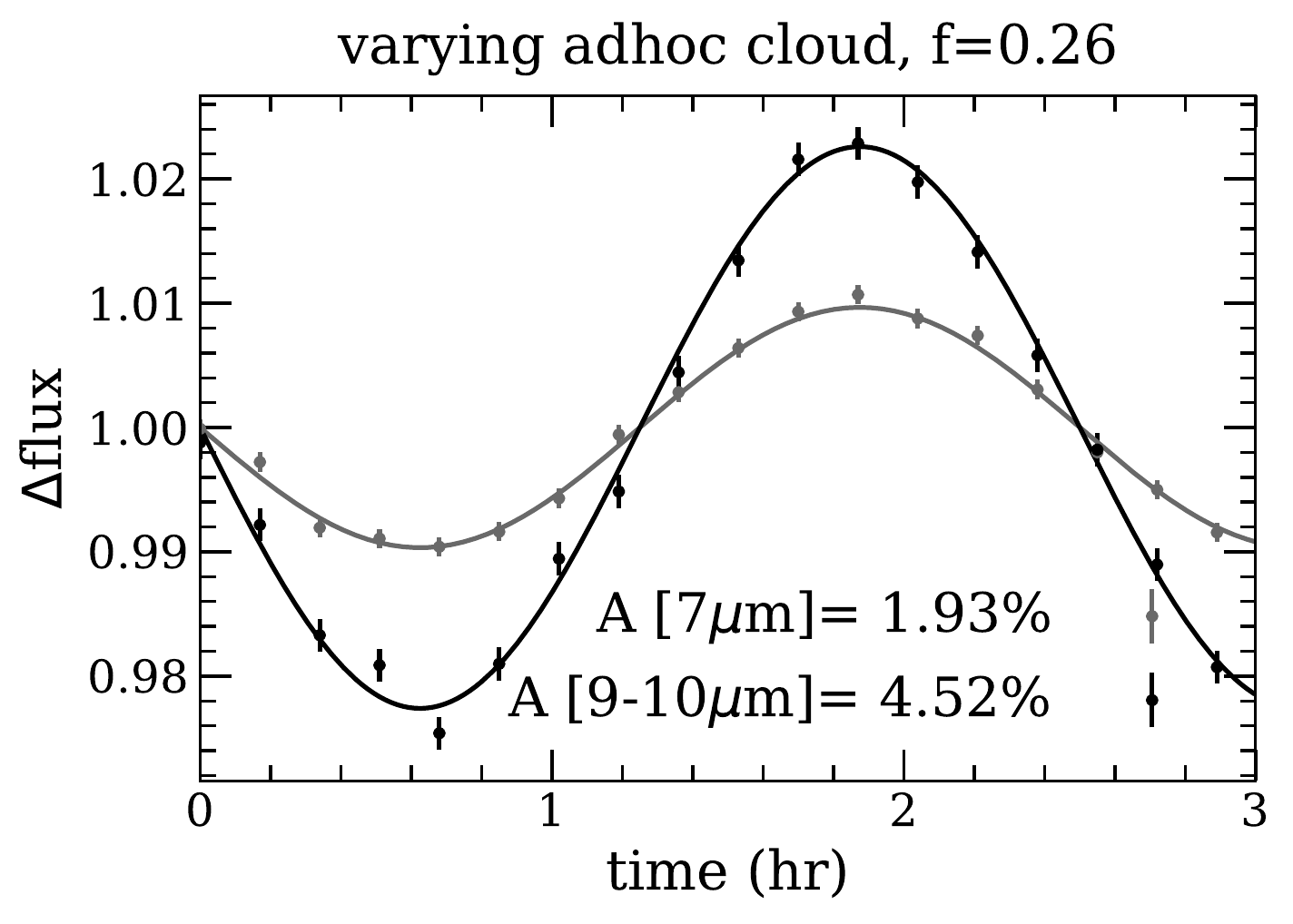}
    \caption{\label{fig:J1507fluxvstime} 
    \emph{Spitzer} light curves from \citet{Metchev15}
   for J1507-1627 (top). The IRAC [3.6] points are in black and IRAC [4.5] points are in gray with the corresponding fit. We also show in  black (dashed) and gray(dashed), the simplistic sine curve model that use the period and amplitude in Table~\ref{tab:targets}. We show the predicted mid-IR variability for J1507-1627 by varying the ad hoc cloud (bottom).
    }
\end{figure}

We were able to reproduce the observed amplitudes of variability for this object only by varying the ad hoc cloud alone. The fraction f $=$ 0.26 gives A [3.6 $\mu$m]$=$ 0.51$\%$ and A [4.5 $\mu$m]$=$ 0.47$\%$. We show the predicted mid-IR variability in the bottom panel of Figure~\ref{fig:J1507fluxvstime}. The variability inside the silicate feature is $\sim$ 2 times the variability outside the feature. 

\subsubsection{2MASS J1821+1414} 
\begin{figure}[h]
\centering
    \includegraphics[width=1\linewidth]{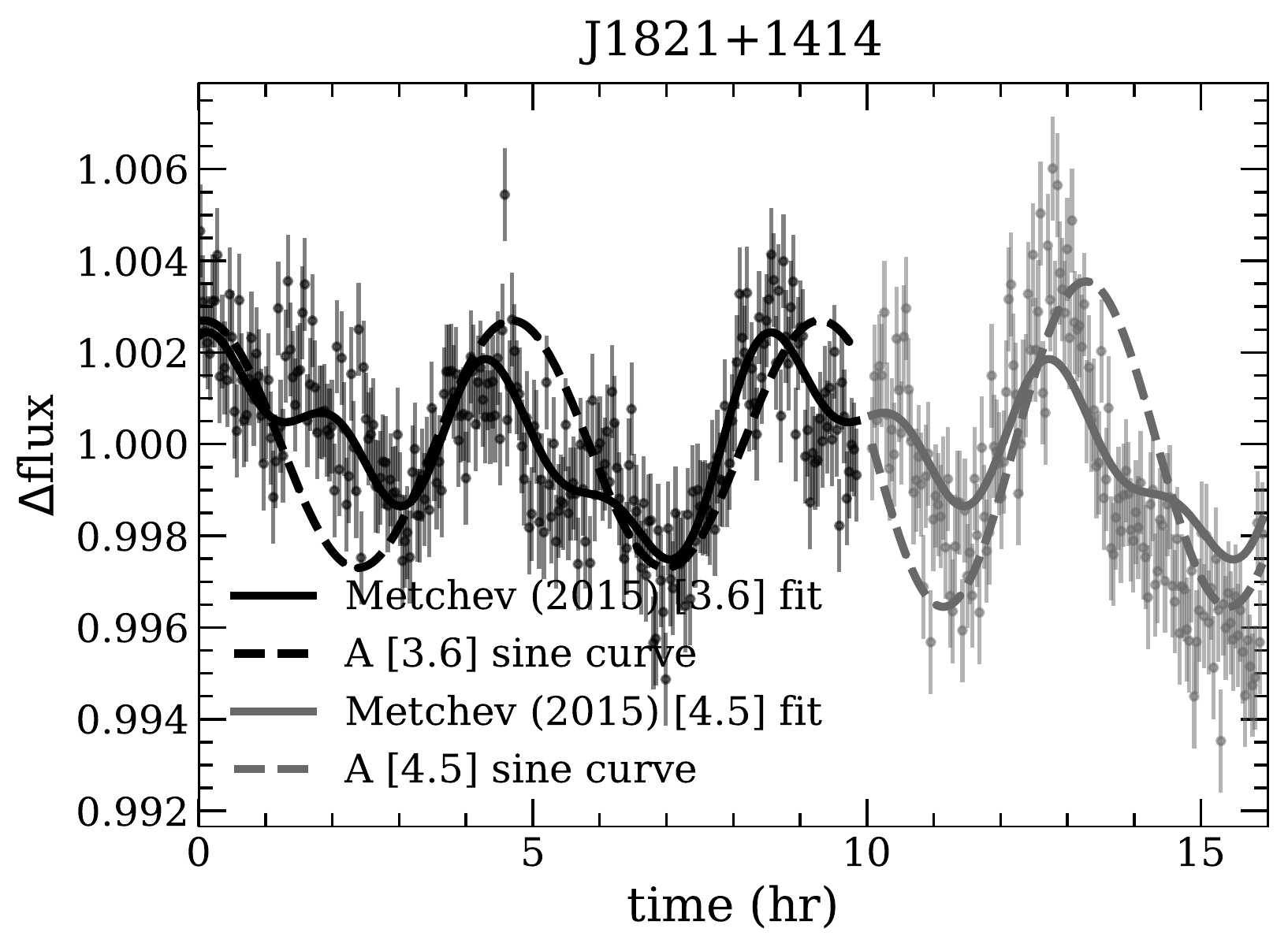}
    \includegraphics[width=1\linewidth]{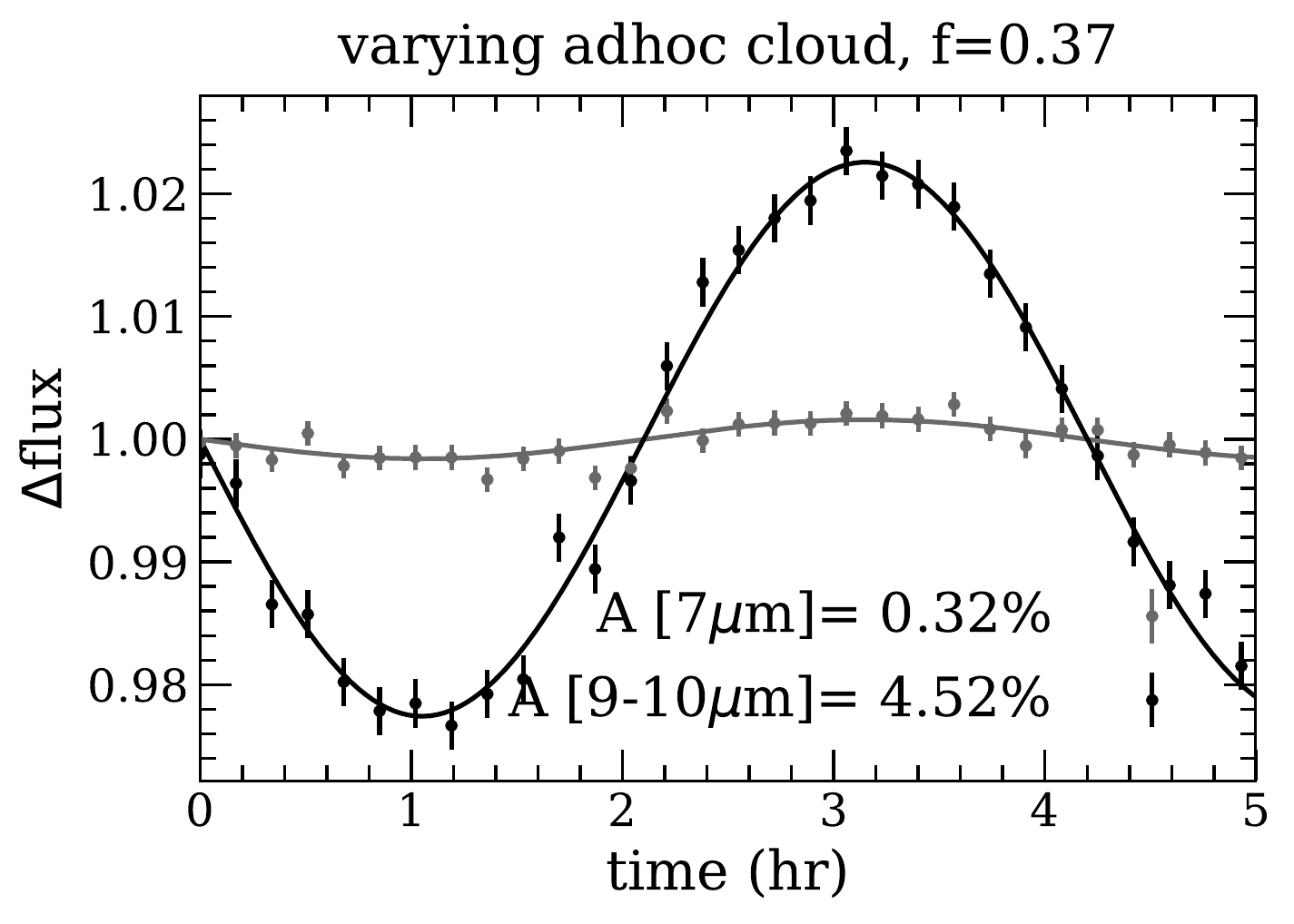}
    \caption{Same as Figure~\ref{fig:J1507fluxvstime} for J1821+1414. \label{fig:J1821fluxvstime} }
\end{figure}

We were able to reproduce the observed amplitudes of variability for this object only by varying the ad hoc cloud alone. The fraction f $=$ 0.37 gives A [3.6 $\mu$m]$=$ 0.62$\%$ and A [4.5 $\mu$m]$=$ 0.56$\%$. We show the predicted mid-IR variability in the bottom panel of Figure~\ref{fig:J1821fluxvstime}. The variability inside the silicate feature is $\sim$ 11 times the variability outside the feature.

\subsubsection{2MASS J2148+4003} 
\begin{figure}[h]
\centering
    \includegraphics[width=1\linewidth]{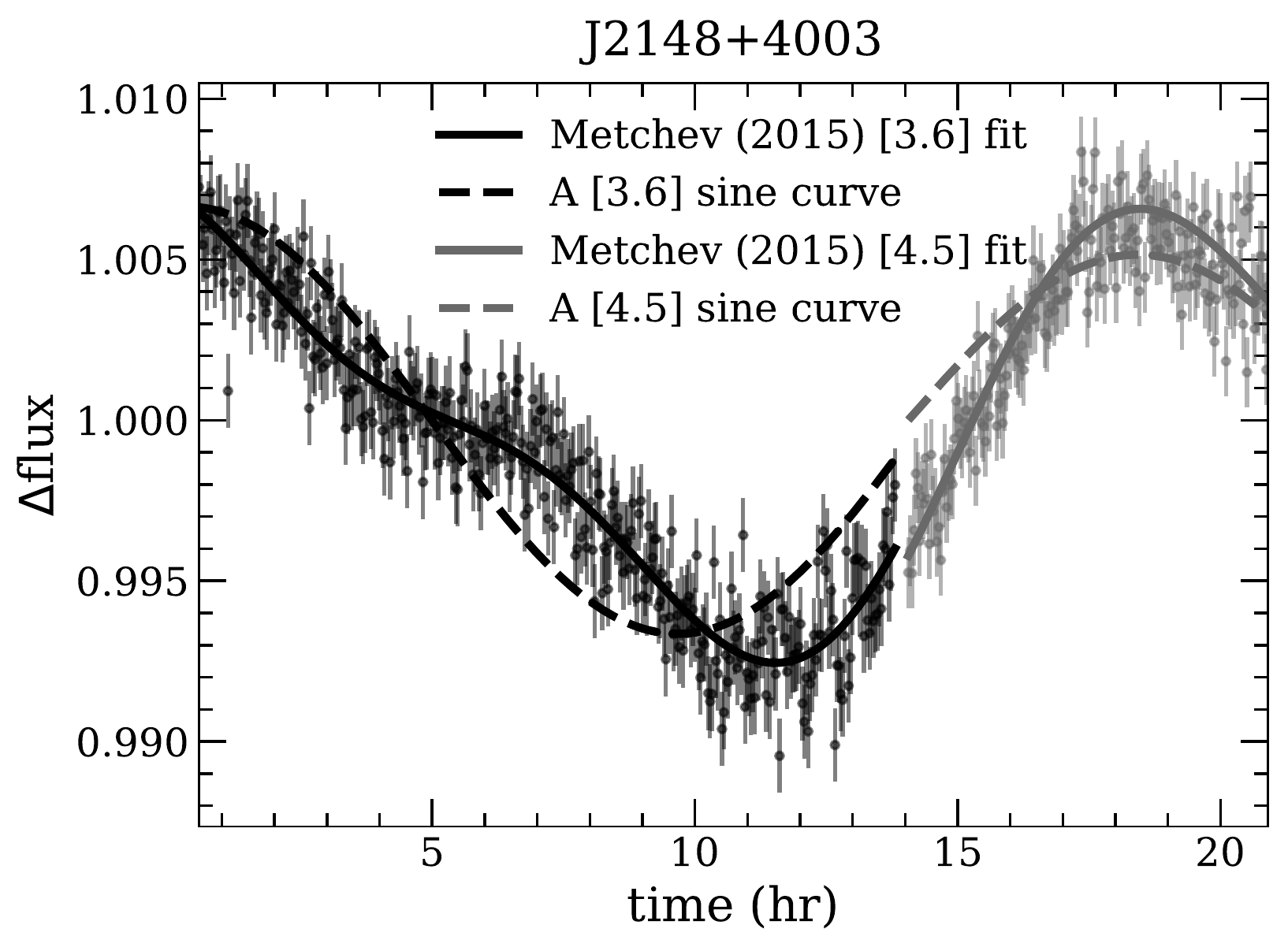}
    \includegraphics[width=1\linewidth]{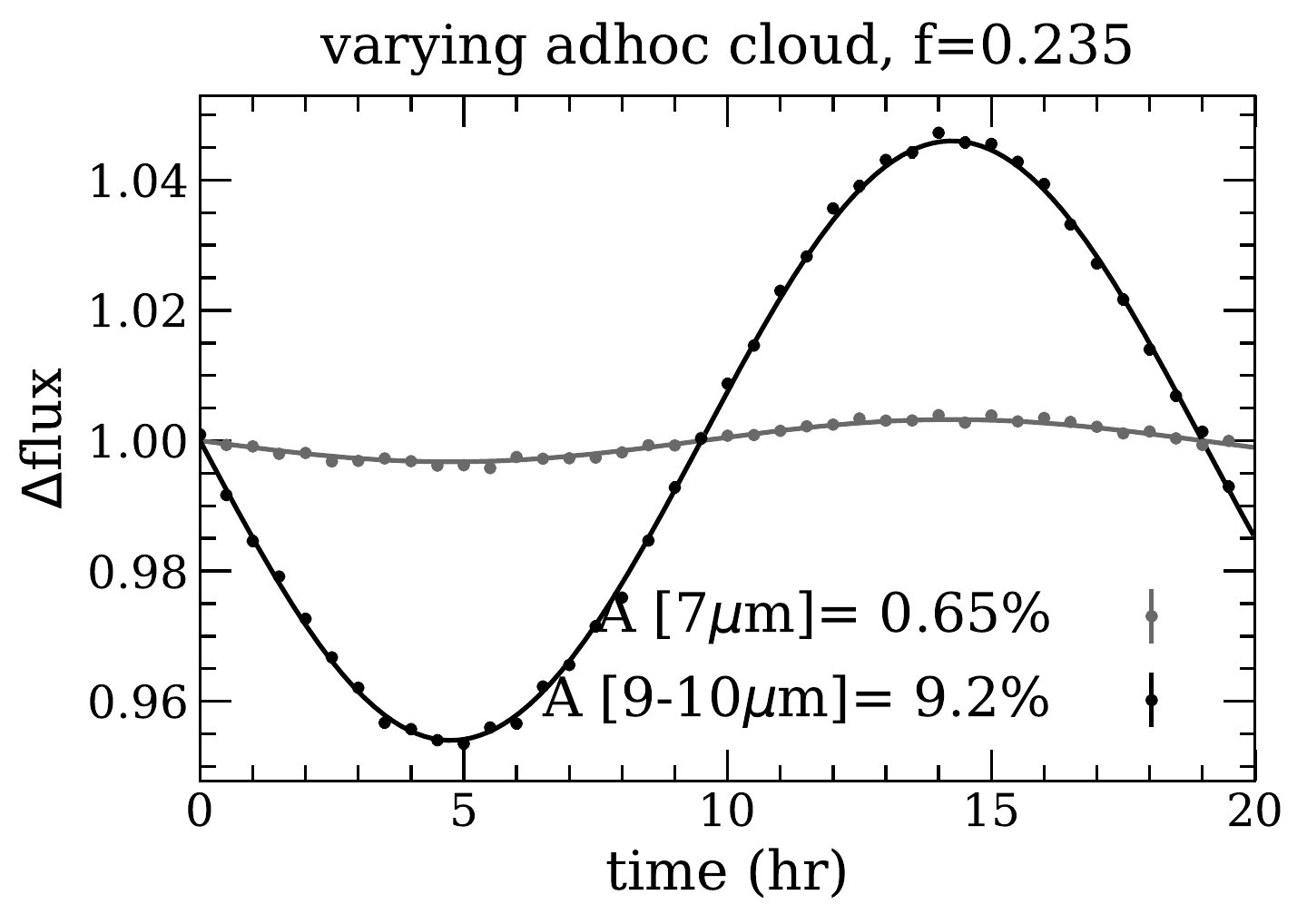}
    \caption{Same as Figure~\ref{fig:J1507fluxvstime} for J2148+4003. \label{fig:2148fluxvstime} }
\end{figure}

We were able to reproduce the observed amplitudes of variability for this object only by varying the ad hoc cloud alone. The fraction f $=$ 0.235 gives A [3.6 $\mu$m]$=$ 1.25$\%$ and A [4.5 $\mu$m]$=$ 1.14$\%$. We show the predicted mid-IR variability in the bottom panel of Figure~\ref{fig:2148fluxvstime}. The variability inside the silicate feature is $\sim$ 11 times the variability outside the feature.

%%%%-----------DISCUSSION SECTION----------------------------------------------------
\section{Discussion \label{sec:discussion}} 
\subsection{Small Particles in Cloud Models}
Our results show that small particles at low pressures are needed to produce the silicate feature. Simple advective-diffusive balance cloud models (eg. \citetalias{AM01}) typically have larger particle sizes (eg. r= 10 - 100 $\mu$m) and higher pressures (P= 0.1 - 1 bar) for typical $\sim$ 1800~K L dwarfs. At least a subset of brown dwarfs have small particles in their clouds that remain uncaptured by these models.  

Our results agree with \citet{Hiranaka16}, who showed that small particles above the main cloud deck could successfully reproduce the NIR colors of red brown dwarfs. Microphysical models do produce substantially small particles. For example, using the CARMA model, \citet{Powell18} found small particles can significantly dominate the cloud opacity when taking into account the full size particle distributions. In the era of \emph{JWST}, forward models and retrievals alike should be modified such that they can capture absorption from these small particles. 

Current retrievals of brown dwarfs often simplistically parameterize clouds by approximating the cloud as gray, power law, or not including clouds at all \citep{Line15,Burningham17}. Our work suggests that future retrieval work for brown dwarfs could retrieve information about particle size, composition, and crystalline structure (See \citet{burningham21}).

\subsection{Determining the Physics of Cloud Condensation \label{sec:physicsofclouds}} 
Determining the cloud condensation sequence in brown dwarfs will allow us to provide constraints on derived abundances. Clouds remove limited gaseous species (e.g., oxygen) from their atmospheres. The dominant silicate species for L dwarfs will result in more or less oxygen remaining in cooler objects; if most silicates form as \enst, 3 oxygen atoms are removed per magnesium atom, whereas if they form as \forst, 2 oxygen atoms are removed per magnesium atom. This impacts our measurements of both brown dwarfs and exoplanets for which we measure molecular abundances and C/O ratios. \emph{JWST}/MIRI observations can reveal the dominant silicate cloud mineral. One possibility is that brown dwarfs will follow rainout equilibrium calculations and condense predominantly forsterite. Another possibility is that enstatite will dominate the spectrum since it may be at higher altitudes than forsterite. Similarly, the cloud absorption features observed will shed light on the crystalline structure. Our results from fitting \emph{Spitzer} IRS spectra suggest that many of these objects may have predominately enstatite clouds with optical constants consistent with amorphous, Mg-rich crystal structures.

We find that 2MASS J2224-0158 is marginally better fit by a SiO cloud than either \enst~or \forst. Here we discuss the plausibility of this cloud species. SiO is the most abundant silicon gas species over a wide range of pressures and temperatures \citep{Visscher10}. SiO abundances can control silicate cloud formation (e.g. \citealt{Powell18,Helling&Woitke06}). However, many cloud models do not include direct SiO condensation. \citet{Helling06} proposed SiO$_2$ as the possible condensate responsible for the broad absorption seen in three brown dwarfs from \citet{Cushing06} using grain chemistry models. Furthermore, \citet{Helling08b} found that some SiO can form in the uppermost layers, broadly consistent with our fitted model containing SiO in the upper layers of the atmosphere. We suggest that future cloud modeling remains open to this possibility. 

Recently, \citet{burningham21} used the ``Brewster" retrieval framework to fit the 1-15 \um~\emph{Spitzer}/IRS spectrum of 2MASS~J2224-0158. They found that the data was best fit with enstatite and quartz slabs at low pressures and an Fe cloud deck deeper in the atmosphere. This is similar to our findings, where our best fitting model is an SiO cloud at low pressures and an Fe cloud at higher pressures (+Al$_2$O$_3$ and Mg$_2$SiO$_4$, see Figure~\ref{fig:compareobs_newmodels}). Additionally 2MASS J2224-0158 was inferred to have a high-metallicity atmosphere from the estimated Mg/Si and C/O ratios \citep{burningham21}. This is similar to 2MASS J2148+4003, whose goodness of fit values were significantly worse for 2MASS J2148+4003 than for any other object (see Table~\ref{tab:Gkvalues}). Future studies should aim to model atmospheres over a wide range of metallicities.

During \emph{Spitzer}'s Cold mission that ended in 2009, it observed 106 L/T dwarfs using \emph{Spitzer}/IRS. Since the end of its cold mission, hundreds of brown dwarfs have been discovered. Furthermore, brown dwarfs have since been discovered to be variable on the few percent level. \emph{JWST} will enable follow up spectroscopic observations of faint brown dwarfs to continue making strides in understanding the nature of these objects.

Future \emph{JWST} programs that aim to observe a wide range of cloudy brown dwarfs will enable more observations of the silicate feature. Additionally, the larger spectral range of MIRI will enable discoveries of other mineral features at longer wavelengths (eg. Al$_2$O$_3$ at 12\um, CaTiO$_3$ at 16\um).

\begin{figure} 
    \centering
    \includegraphics[width=1\linewidth]{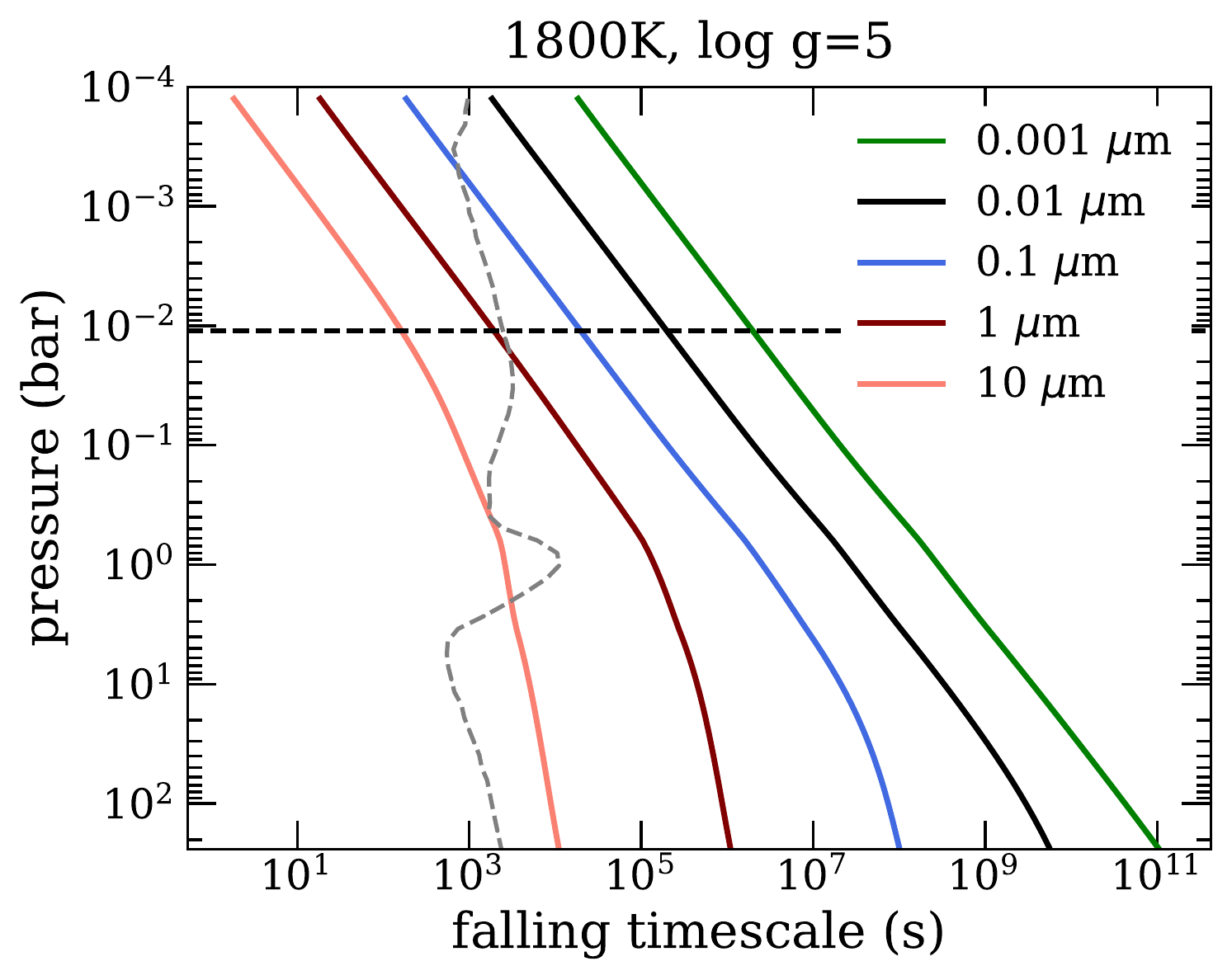}
    \caption{Falling timescales for cloud particles for an 1800 K, log g= 5 model. We show the $\tau \sim$1 pressure where the ad hoc cloud is placed (dashed black). The solid lines show the time it takes a particle to fall one pressure scale height for a range of particle sizes. The gray dashed line shows the falling timescale for the $K_{zz}$ of each pressure layer. }
    \label{fig:fallingtime}
\end{figure}

\subsection{Lofting of Particles by Vertical Mixing}
While silicates condense at pressures $\sim$ 0.1 - 0.2 bar in typical L dwarfs, to match the silicate feature they must be lofted to $\sim$ 0.01 bars. We can estimate how long the particles will stay lofted by comparing vertical mixing and settling timescales. 

We compare the lofting timescale for a given vertical eddy diffusion coefficient, $K_{zz}$, to the falling timescale. $K_{zz}$ is calculated within the atmosphere model, assuming mixing length theory in the convective region. Figure~\ref{fig:fallingtime} shows the time for a particle to fall one pressure scale height defined as, $H/v_{\rm fall}$,
where $H$ is the scale height, and $v_{\rm fall}$ is the particle falling velocity. We follow the appendix of \citetalias{AM01} to calculate the falling velocities assuming viscous flow. We find that the lofting timescale is shorter than the falling timescale for particles less than 1 $\mu$m, indicating that small particles could stay lofted high in the atmosphere with vigorous enough mixing. 

\subsection{Temperature-Dependent Optical Constants}
\begin{figure}
    \centering
    \includegraphics[width=1\linewidth]{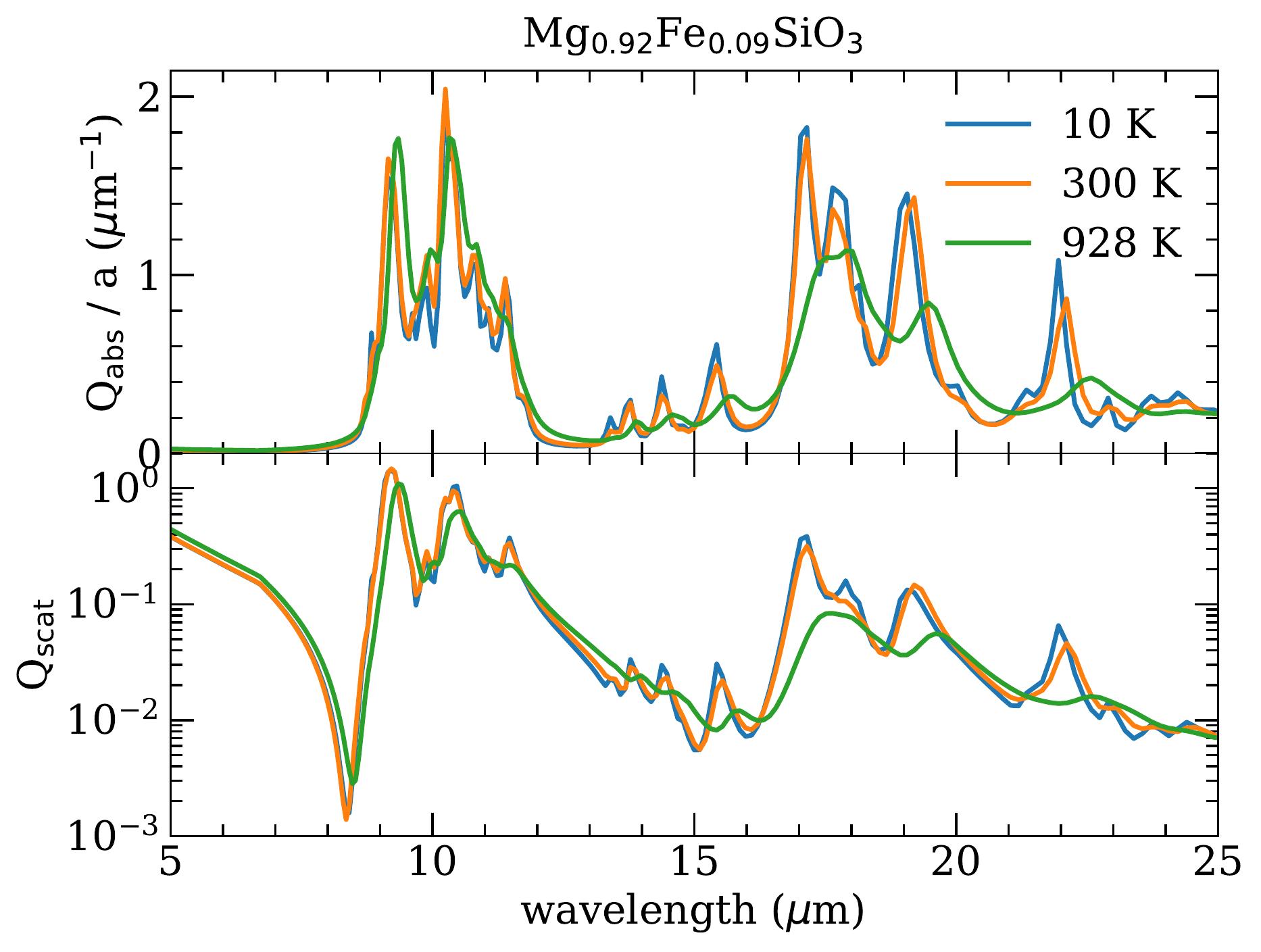}
    \caption{The absorption and scattering efficiencies for temperature-dependent orthoenstatite (Mg$_{0.92}$Fe$_{0.09}$SiO$_3$) at 10 K, 300 K and, 928 K for 1 \um~particles. }
    \label{fig:Qabstemps}
\end{figure}

\begin{figure}
    \centering
    \includegraphics[width=1\linewidth]{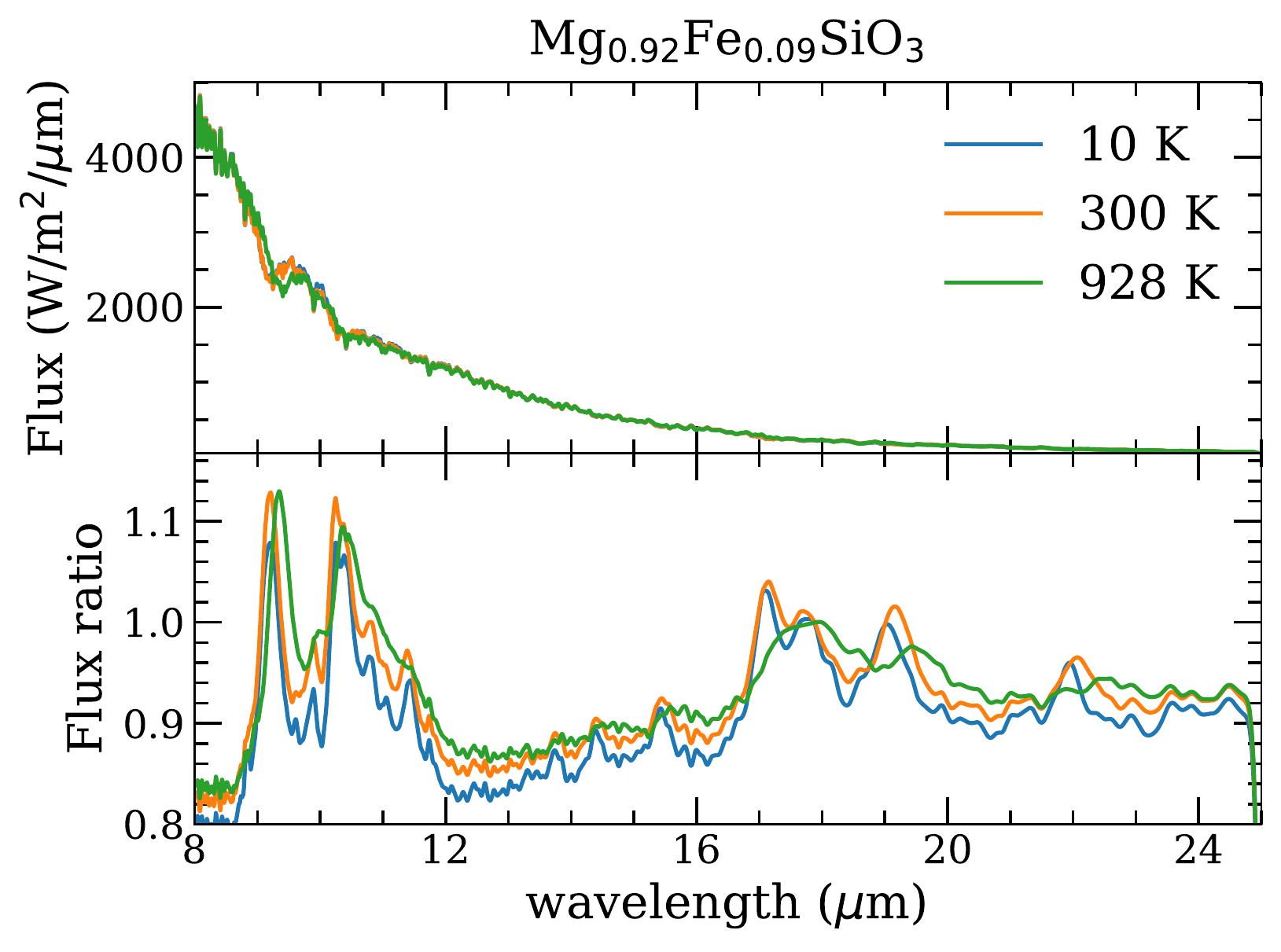}
    \caption{Thermal emission spectra and flux ratios for temperature-dependent optical constants of orthoenstatite (Mg$_{0.92}$Fe$_{0.09}$SiO$_3$) at 10 K, 300 K and 928 K. We use an 1800 K, log g$=$ 5 atmosphere with 1 $\mu$m~particles. }
    \label{fig:spectra_temps}
\end{figure} 

The majority of optical constants used for silicates are measured at room temperature. This is much lower than the temperatures of L dwarfs. We explored the differences in the thermal emission spectra for temperature-dependent optical properties for orthoenstatite (Mg$_{0.92}$Fe$_{0.09}$SiO$_3$) at 10 K, 300 K and, 928 K \citep{Zeidler15}. The crystalline orthoenstatite was measured on 3 crystallographic axes and we combine the information from each axis as described in Section \ref{sec:miescattering} and shown in Figure~\ref{fig:Qabstemps}. Using an 1800 K, log g$=$ 5 model with 1 \um~particles we computed model thermal emission spectra for the temperature-dependent optical constants shown in Figure~\ref{fig:spectra_temps}. We find that as the temperature increases the absorption features become broader and shallower. Additionally, the peak in the absorption feature moves redder $\Delta\lambda \approx~$0.13 $\mu$m. 

Temperature-dependent optical constants were not available for other materials studied here. Future studies of optical constants should aim to explore a range of optical constants closer to the astrophysical environments being studied. 

\subsection{Application to Directly Imaged Exoplanets }
Observations and models of brown dwarfs provide a testbed to study directly imaged exoplanet atmospheres. Some directly imaged planets are in the same temperature range as brown dwarfs and thus may show the same cloud mineral features in their thermal emission spectrum. 

\citet{vos19} found that young brown dwarfs are more likely to be variable by studying the near-IR variability in the J band. Recently, \emph{Spitzer} IRAC observations of young low-gravity brown dwarfs revealed an increase in variability amplitudes for late-L dwarfs at 4.5 \um ~\citep{vos20}. \emph{JWST}/MIRI observations of young brown dwarfs could reveal the silicate feature in the mid-IR \citep{Danielski18,Brande20}. Some of these observations may also be possible for directly imaged exoplanets with \emph{JWST}. Understanding the atmospheres of these young brown dwarfs is imperative for advancing knowledge of directly imaged exoplanets. 
%%%%------------------------------------------
%%%%-----------Conclusion SECTION-------------
\section{Conclusion \label{sec:conclusion}} 
We presented models for silicate and refractory clouds in warm brown dwarf atmospheres. These models include small particles in the upper layers of the atmosphere and have unique absorption features in the mid-IR.

\emph{JWST}/MIRI will potentially allow us to determine the cloud compositions, particle sizes, and mineral structures for warm brown dwarfs that have a sufficiently strong silicate feature. Soon, 2MASS J2148+4003 will be observed with \emph{JWST}/MIRI in cycle 1 under program ID: 2288 (PI: Joshua Lothringer, Co-PI: Jeff Valenti), and its spectrum will shed light on the nature of clouds in this brown dwarf. Time-series spectroscopy would allow us to establish patchy clouds as the cause for brightness variations seen in brown dwarfs and provide information about the inhomogeneity of different cloud layers. These measurements will allow us to empirically establish the condensation sequence for substellar atmospheres and test models of cloud physics and chemistry, with implications for studies of cooler brown dwarfs and exoplanets. 

\acknowledgments
We thank the reviewer for their thoughtful and helpful comments, which have improved the manuscript. This material is based upon work supported by the National Science Foundation Graduate Research Fellowship Program under Grant No. DGE-1610403. Any opinions, findings, and conclusions, or recommendations expressed in this material are those of the authors and do not necessarily reflect the views of the National Science Foundation. J.L. and C.V.M. acknowledge the National Science Foundation, which supported the work presented here under Grant No. 1910969.

This work benefited from the 2019 Exoplanet Summer Program in the Other Worlds Laboratory (OWL) at the University of California, Santa Cruz, a program funded by the Heising-Simons Foundation. We thank Stanimir Metchev for providing the reduced \emph{Spitzer} light curves. JL thanks Brittany Miles for insights using the \emph{JWST} ETC for brown dwarf observations. JL also thanks Jeremy Ritter for help with python data frames and useful UNIX code used to create plots made in this paper.

\software{astropy \citep{Astropy2013}, spectres \citep{Carnall17}}

\appendix

\startlongtable
\begin{deluxetable*}{ccccc}
\tablecolumns{8}
\tablewidth{0pt}
\tablecaption{ Goodness of fit values for all model fits for each brown dwarf. \label{tab:Gkvalues}}
\tablehead{
\colhead{Object} & \colhead{Condensate} & \colhead{T(K), log g, r(\um), $\tau_{\rm{cloud}}$}&  \colhead{$\chi^2$}& \colhead{Reject ?} \\ \vspace{-0.2cm}}
\startdata
J0036159$+$182110  & \enst, amorphous & 1800, 5, 1, 0.67 & 3078 & N \\
J0036159$+$182110  & \enst, amorphous   & 1800, 5,0.1, 0.67 &  3082 &  N  \\
J0036159$+$182110  & \forst, amorphous &  1800, 5, 0.1, 0.67 & 3305 & \textbf{Y}  \\
J0036159$+$182110  & SiO$_2$, crystalline &  1800, 5, 1, 0.2 & 3439 & \textbf{Y} \\
J0036159$+$182110  & SiO, amorphous & 1800, 5, 0.1, 0.67 & 3538 &\textbf{Y} \\
J0036159$+$182110  & \forst, amorphous &  1800, 5, 1, 0.67 & 3525 &\textbf{Y} \\
J0036159$+$182110  & \enst, crystalline & 1800, 5, 0.1, 0.2 & 3796 &\textbf{Y} \\
J0036159$+$182110  & SiO, amorphous & 1800, 5, 1, 0.67 &  3833 & \textbf{Y}\\
J0036159$+$182110  & \enst, crystalline & 1800, 5, 1, 0.2 & 3954 & \textbf{Y} \\
J0036159$+$182110  & \forst, crystalline & 1800, 5, 1, 0.2 & 4823&\textbf{Y} \\
J0036159$+$182110  & \forst, crystalline & 1800, 5, 0.1, 0.2 & 4799&\textbf{Y} \\
J0036159$+$182110  & SiO$_2$, crystalline &1800, 5,0.1, 0.2 & 4865 & \textbf{Y}\\
J0036159$+$182110  & SiO$_2$,amorphous & 1800, 5,1, 0.67 & 4920 &\textbf{Y}\\
J0036159$+$182110  & SiO$_2$,amorphous & 1800, 5,0.1, 0.67 & 6683 &\textbf{Y}\\ \hline
J1507476-162738 &  \enst, amorphous  & 1700, 4, 1, 0.67  & 3518 & N \\
J1507476-162738 &  \enst, amorphous  & 1700, 4, 0.1, 0.67  & 3609 & \textbf{Y} \\
J1507476-162738 &  \forst, amorphous  &  1700, 4, 0.1, 0.67  & 3903 &\textbf{Y} \\
J1507476-162738 &  \enst, crystalline  & 1700, 4, 0.1, 0.2  & 3955&  \textbf{Y}\\
J1507476-162738 &  \enst, crystalline  & 1700, 4, 1, 0.2  & 4050 & \textbf{Y}\\
J1507476-162738 &  \forst, amorphous  & 1700, 4, 1, 0.67  & 4000 & \textbf{Y} \\
J1507476-162738 &  SiO$_2$, crystalline  & 1700, 4, 1, 0.2  & 4344 &\textbf{Y} \\
J1507476-162738 &  SiO, amorphous  & 1700, 4, 0.1, 0.67  & 4838 & \textbf{Y} \\
J1507476-162738 &  SiO, amorphous  & 1700, 4, 1, 0.67  &  4908 &\textbf{Y}  \\
J1507476-162738 &  \forst, crystalline  & 1700, 4, 1, 0.2  & 5138 & \textbf{Y} \\
J1507476-162738 &  \forst, crystalline  & 1700, 4, 0.1, 0.2  & 5222 & \textbf{Y} \\
J1507476-162738 &   SiO$_2$, crystalline & 1700, 4, 0.1, 0.2  & 6170 &\textbf{Y} \\
J1507476-162738 &  SiO$_2$, amorphous  & 1700, 4, 1, 0.67  & 7376 & \textbf{Y} \\
J1507476-162738 &  SiO$_2$, amorphous  & 1700, 4, 0.1, 0.67  & 9236 & \textbf{Y} \\ \hline
J18212815$+$1414010 & \enst, amorphous    & 1800, 5, 1, 1 & 2999 & N \\
J18212815$+$1414010 & \enst, amorphous   & 1800, 5, 1, 0.67,  & 3063 &\textbf{Y} \\
J18212815$+$1414010 & \enst, amorphous   & 1800, 5, 0.1, 0.67 & 3315 &\textbf{Y} \\ 
J18212815$+$1414010 & SiO, amorphous   & 1800, 5, 0.1, 0.4 & 3329 & \textbf{Y}\\ 
J18212815$+$1414010 & SiO, amorphous   & 1800, 5, 0.1, 0.3 & 3318 & \textbf{Y}\\ 
J18212815$+$1414010 & \enst, amorphous   & 1800, 5, 0.1, 1 &3836 &\textbf{Y} \\ 
J18212815$+$1414010 & SiO$_2$, crystalline   & 1800, 5, 1, 0.2 & 3554 &\textbf{Y} \\
J18212815$+$1414010 & SiO, amorphous   & 1800, 5, 1, 0.4 & 3661 & \textbf{Y}\\ 
J18212815$+$1414010 & \forst, amorphous   & 1800, 5, 0.1, 0.67 & 3962 & \textbf{Y}\\ 
J18212815$+$1414010 & SiO, amorphous   & 1800, 5, 1, 0.3 &3719 &\textbf{Y} \\
J18212815$+$1414010 & SiO, amorphous   & 1800, 5, 0.1, 0.4 & 4133 & \textbf{Y}\\ 
J18212815$+$1414010 & \forst, amorphous   & 1800, 5, 1, 0.67 & 4018 & \textbf{Y}\\ 
J18212815$+$1414010 & \enst, crystalline   & 1800, 5, 0.1, 0.2 & 4168 &\textbf{Y}\\ 
J18212815$+$1414010 & SiO, amorphous   & 1800, 5, 1, 0.67 & 4044 &\textbf{Y} \\ 
J18212815$+$1414010 & \enst, crystalline   & 1800, 5, 1, 0.2 & 4478 & \textbf{Y}\\ 
J18212815$+$1414010 & SiO$_2$, crystalline   & 1800, 5, 0.1, 0.2 & 5645 &\textbf{Y} \\
J18212815$+$1414010 &  SiO$_2$, amorphous    & 1800, 5, 1, 0.67 & 5831 &\textbf{Y} \\
J18212815$+$1414010 & \forst, crystalline   & 1800, 5, 0.1, 0.2 & 6149 & \textbf{Y}\\ 
J18212815$+$1414010 & \forst, crystalline   & 1800, 5, 1, 0.2 & 6316 & \textbf{Y}\\ 
J18212815$+$1414010 & \forst, amorphous   & 1800, 5, 0.1, 0.67 & 8150 & \textbf{Y}\\  \hline
J21481628+4003593 &  \enst, amorphous    & 1800, 5, 0.1, 1  & 6000  & N \\
J21481628+4003593 & \enst, amorphous  & 1800, 5, 1, 1  &  6067 & \textbf{Y} \\
J21481628+4003593 & SiO, amorphous   & 1800, 5, 0.1, 0.67  & 6348  & \textbf{Y}\\
J21481628+4003593 & \enst, amorphous   & 1800, 5, 0.1, 0.67  & 7050  & \textbf{Y}\\
J21481628+4003593 & SiO, amorphous   & 1800, 5, 0.1, 0.4  & 7099  & \textbf{Y}\\
J21481628+4003593 & \enst, amorphous   & 1800, 5, 0.1, 0.67  & 7536  &\textbf{Y} \\
J21481628+4003593 & SiO, amorphous   & 1800, 5, 0.1, 0.3  & 8178  & \textbf{Y}\\
J21481628+4003593 & SiO$_2$, crystalline   & 1800, 5, 0.1, 0.2  & 8376 &\textbf{Y} \\
J21481628+4003593 & SiO$_2$, amorphous   & 1800, 5, 0.1, 0.67  & 8158  &\textbf{Y} \\
J21481628+4003593 & \forst, amorphous   & 1800, 5, 0.1, 0.67  & 8881  & \textbf{Y}\\
J21481628+4003593 & SiO, amorphous   & 1800, 5, 0.1, 0.67  & 9181  &\textbf{Y} \\
J21481628+4003593 & SiO, amorphous   & 1800, 5, 1, 0.4  & 9551  & \textbf{Y}\\
J21481628+4003593 & SiO, amorphous   & 1800, 5, 1, 0.3  & 10314  & \textbf{Y}\\
J21481628+4003593 & \forst, amorphous   & 1800, 5, 1, 0.67  & 10634 & \textbf{Y}\\
J21481628+4003593 & \enst, crystalline   & 1800, 5, 0.1, 0.2  & 11410  & \textbf{Y}\\
J21481628+4003593 & \enst, crystalline   & 1800, 5, 1, 0.2  & 12802  & \textbf{Y}\\
J21481628+4003593 & SiO$_2$, crystalline   & 1800, 5, 0.1, 0.2  & 13727  & \textbf{Y}\\
J21481628+4003593 &  SiO$_2$, amorphous   & 1800, 5, 0.1, 0.67  & 13873  &\textbf{Y} \\
J21481628+4003593 & \forst, crystalline  & 1800, 5, 0.1, 0.2  & 17863  & \textbf{Y}\\
J21481628+4003593 & \forst, crystalline   & 1800, 5, 1, 0.2  & 18757  & \textbf{Y}\\ \hline
J2224438-015852  & SiO, amorphous  & 1800, 5, 0.1, 0.67 & 1759 & N\\
J2224438-015852  & \enst, amorphous  & 1800, 5, 0.1, 0.67 & 2002 & \textbf{Y} \\
J2224438-015852  & \forst, amorphous  & 1800, 5, 0.1, 0.67 & 2139 & \textbf{Y} \\
J2224438-015852  & \enst, amorphous  & 1800, 5, 1, 0.67 & 2491 & \textbf{Y} \\
J2224438-015852  & SiO$_2$, crystalline  & 1800, 5, 1, 0.2 & 2253 & \textbf{Y} \\
J2224438-015852  & SiO$_2$, amorphous  & 1800, 5, 1, 0.67 & 2326 &  \textbf{Y}\\
J2224438-015852  & SiO, amorphous  & 1800, 5, 1, 0.67 & 2723 & \textbf{Y} \\
J2224438-015852  & \forst, amorphous & 1800, 5, 0.1, 0.67 & 2736 &\textbf{Y}  \\
J2224438-015852  & \enst, crystalline  & 1800, 5, 1, 0.2 & 3258 & \textbf{Y} \\
J2224438-015852  & \enst, crystalline  & 1800, 5, 0.1, 0.2 & 3356 & \textbf{Y} \\
J2224438-015852  & SiO$_2$, crystalline  & 1800, 5, 0.1, 0.2 & 3179 &\textbf{Y} \\
J2224438-015852  & SiO$_2$, amorphous  & 1800, 5, 0.1, 0.67 & 2957 &\textbf{Y} \\
J2224438-015852  & \forst, crystalline  & 1800, 5, 0.1, 0.2 & 3862 & \textbf{Y}\\
J2224438-015852  & \forst, crystalline  & 1800, 5, 1, 0.2 & 4050 & \textbf{Y}\\
\enddata
\tablecomments{Chi-square values for each model.}
\end{deluxetable*}

\bibliography{BD_cloud_mineralogy}
\bibliographystyle{aasjournal}

\end{document}